# Thermal Degradation Mechanisms and Stability Enhancement Strategies in Perovskite Solar Cells: A Review

Arghya Paul[1], Kanak Raj[2*], Prince Raj Lawrence Raj[2], Pratim Kumar[2]

[1] Department of Mechanical Engineering, Indian Institute of Science, Malleshwaram, Bengaluru, Karnataka, 560076, India

[2] Department of Aerospace Engineering and Applied Mechanics, Indian Institute of Engineering Science & Technology, Shibpur, Howrah 711103, India

[*]Corresponding author: kanakrajkota@gmail.com

## Abstract

Perovskite Solar Cells (PSCs) have garnered global research interest owing to their superior photovoltaic (PV) performance. The future of photovoltaic technology lies in PSCs since they can produce power with performance on par with the best silicon solar cells while being less expensive. PSCs have enormous potential; in just ten years, their efficiency increased from 3.8% to 25.2%, and research into new developments is still ongoing. Thermal instability is PSCs' main disadvantage, despite their high efficiency, flexibility, and lightweight nature. This paper looks at how temperature affects the ways that hole transport layers (HTLs) like spiro-OMeTAD and perovskite layers, especially $MAPbI_3$, degrade. Elevated temperatures cause $MAPbI_3$ to degrade into $PbI_2$, $CH_3I$, and $NH_3$, with decomposition rates affected by moisture, oxygen, and environmental factors. Mixed cation compositions, such as Cs-MA-FA, have higher thermal stability, whereas $MA^+$ cations break-down faster under heat stress. HTLs deteriorate due to morphological changes and the hydrophilicity of dopant additions like Li-TFSI and t-BP. Alternative dopant-free HTMs, such as P3HT and inorganic materials including CuSCN, $NiO_x$, and $Cu_2O$, have shown improved thermal stability and efficiency. Hybrid HTLs, dopant-free designs, and interface tweaks are all viable solutions for increasing the stability of PSC. Addressing thermal stability issues remains crucial for the development of more reliable and efficient PSC technology.

## 1. INTRODUCTION

Despite significant advancements in energy technology, conventional fossil fuels such as coal, natural gas, and petroleum oil continue to dominate electricity generation in 2025 [1–8]. However, extensive research over the past few decades has highlighted the detrimental effects of fossil fuel consumption on both the environment and public health[9]. Additionally, widespread awareness campaigns emphasise the necessity of transitioning from conventional power. Despite significant advancements in energy technology, conventional fossil fuels such as coal, natural gas, and petroleum oil continue to dominate electricity generation in 2025 generation methods shifting to more sustainable alternatives[3]. Numerous studies predict the depletion timelines for coal, natural gas, and petroleum oil reserves, underscoring the urgency of adopting renewable energy sources [10,11]. The continued reliance on fossil fuels is primarily attributed to their cost-effectiveness and established infrastructure [12]. However, their environmental impact is severe, contributing to greenhouse gas emissions, including carbon oxides (COx), nitrogen oxides (NOx), and sulfur dioxide ($SO_2$), etc [13]. Particulate matter released from power plants significantly contributes to smog formation and respiratory illnesses. According to the International Energy Agency (IEA) report [14], coal alone accounts for 40% of global CO2 emissions. Consequently, the transition to renewable energy sources, particularly solar power, has become imperative in mitigating climate change and reducing dependence on fossil fuels [15–21].

In addition to environmental concerns, the long-term availability of fossil fuel reserves is becoming increasingly uncertain [22,23]. Several studies predict that the depletion of coal, natural gas, and petroleum reserves could occur within the next few decades[24,25], emphasising the urgency of transitioning to renewable energy sources. The growing demand for energy, coupled with concerns about energy security and sustainability, has driven global efforts to adopt cleaner and more sustainable energy solutions[26,27]. Among various renewable energy sources, solar power has emerged as one of the most promising alternatives due to its abundance, scalability, and rapid technological advancements[28–30]. Unlike wind, geothermal, and hydroelectric energy, solar power can be harnessed in nearly all geographical locations, making it a highly accessible energy source worldwide [31,32]. Furthermore, technological improvements in photovoltaic (PV) systems have significantly reduced production costs while increasing efficiency and durability [33]. Silicon-based solar panels, which have been the primary technology for solar energy conversion, now boast lifespans of up to 25 years, making them a viable long-term investment for both residential and industrial applications[34–36]. Figure 1 illustrates the global contribution of various renewable and non-renewable energy sources to electricity generation, emphasising the growing role of solar power in the global energy mix [37].

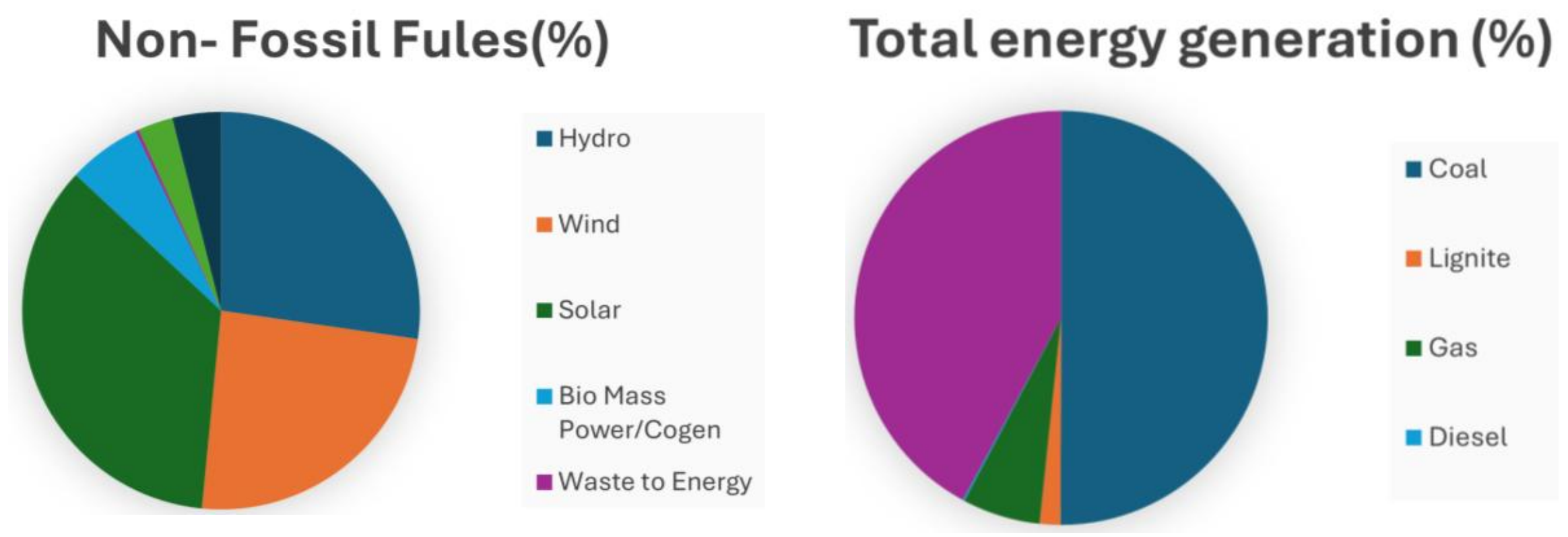

Figure 1. Generations of electricity from different sources as of December 2024.

Despite their advantages, silicon-based solar cells face several challenges that limit their widespread adoption[38]. The high manufacturing costs associated with silicon photovoltaics make them less economically viable, particularly for large-scale installations and residential use[39,40]. Additionally, their rigid structure limits their applicability in flexible and wearable electronics, restricting their potential in emerging energy-harvesting applications[41].Furthermore, silicon solar cells exhibit suboptimal performance under low-light conditions, reducing their efficiency in areas with limited sunlight exposure or frequent cloud cover[42]. To overcome these limitations, researchers have turned to alternative photovoltaic technologies that offer higher efficiency, lower production costs, and greater adaptability[15]. Among these emerging solutions, perovskite solar cells (PSCs) have gained significant attention due to their exceptional power conversion efficiency (PCE), ease of fabrication, and cost-effectiveness[15,43–46].

Since their first demonstration in 2009 by Kojima et al.[17], where methylammonium lead iodide ($MAPbI_3$) was used as a light-absorbing material in a dye-sensitized solar cell (DSSC) configuration, PSCs have undergone rapid improvements. Over the past decade, their efficiency has surged from 3.81% to over 25%, making them one of the most competitive photovoltaic technologies available today[47–50]. This remarkable progress is primarily attributed to key properties of perovskite materials, such as high absorption coefficients, long charge carrier diffusion lengths, and tunable band gaps, which enable efficient photon absorption and charge transport [51,52]. Figure 2 presents a historical timeline of solar cell development, outlining key advancements from Edmund Becquerel's discovery of the photovoltaic effect in 1839 to the development of silicon-based solar cells and the emergence of perovskite photovoltaics as a revolutionary alternative [53–55].

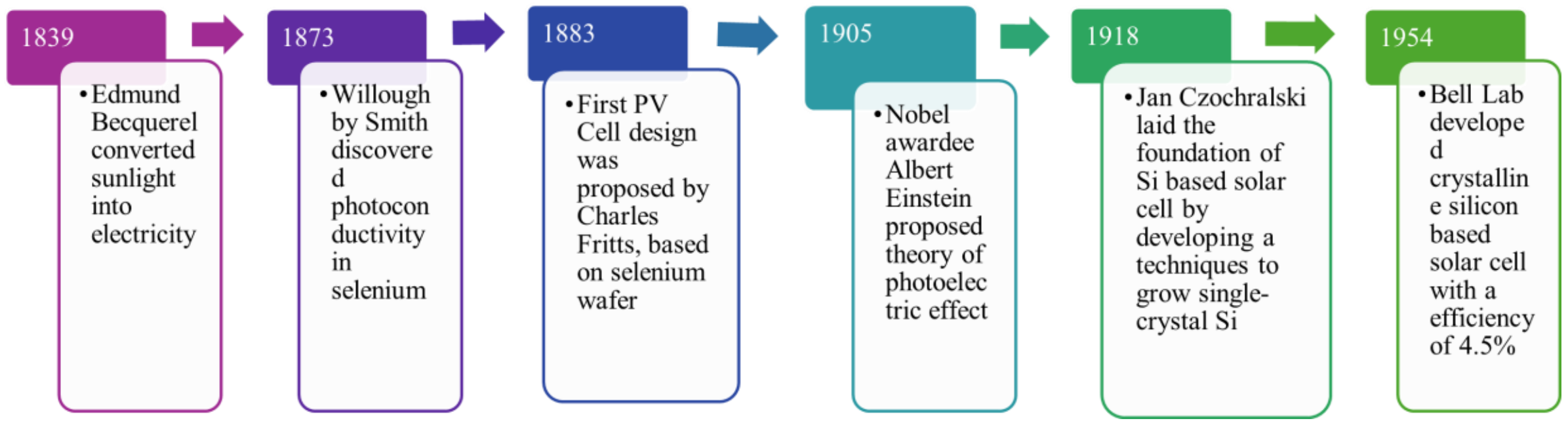

Figure 2. Historical timeline of solar cell development, highlighting key breakthroughs from early photovoltaic discoveries to modern advancements.

Researchers have worked to improve the stability, efficiency, and scalability of perovskite materials, which were initially suggested as promising photovoltaic materials [56–59]. Figure 3 outlines the timeline of perovskite solar cell development, showcasing key breakthroughs that have shaped the field, from the initial discovery of perovskite materials in 1839 to their first application in solar cells and subsequent stability enhancements [60–64]. Early perovskite solar devices demonstrated high efficiency but suffered from severe instability, with rapid degradation occurring within seconds to minutes [65,66]. Addressing these challenges has been a major research priority, leading to substantial advancementsin material composition and device architecture. One breakthrough involves the replacement of dimethyl sulfoxide (DMSO) with formalidinium iodide (FAI) in perovskite film processing, which has resulted in enhanced crystallinity and stability, allowing PSCs to achieve efficiencies exceeding 20% [53]. However, the degradation of perovskite layers remains a critical issue, as $MAPbI_3$-based devices are highly susceptible to moisture, oxygen, and elevated temperatures.Studies indicate that exposure to high humidity (80% RH) and temperatures above 85°C accelerates perovskite decomposition, leading to the formation of hydroiodic acid and other degradation by products [60].

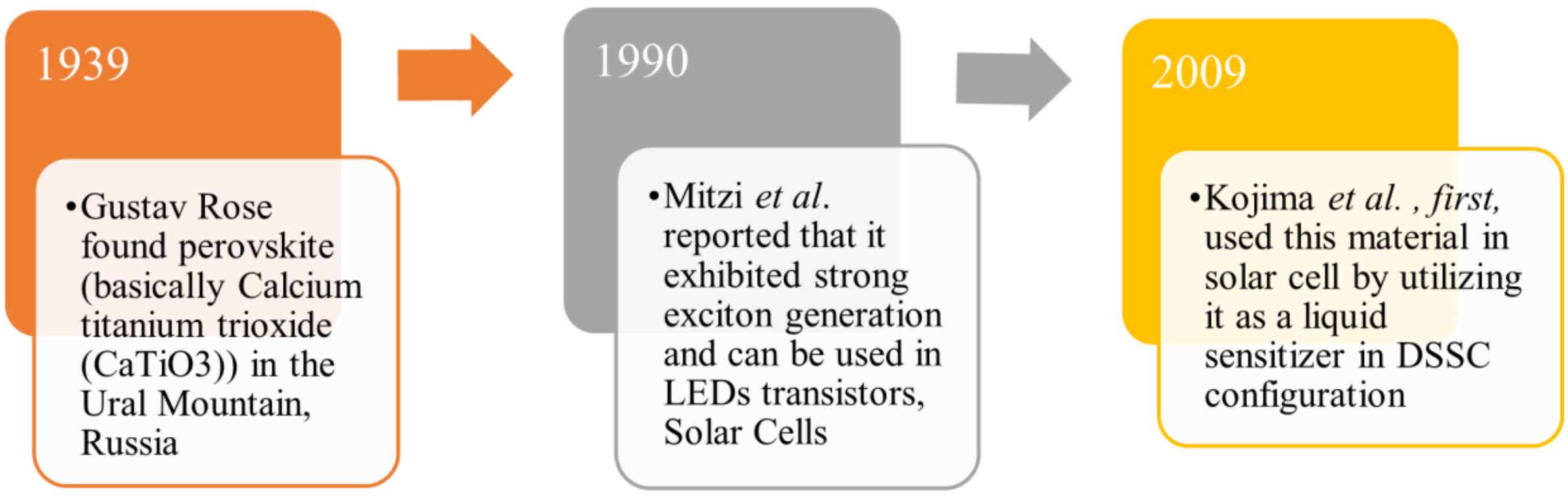

Figure 3. Timeline of perovskite solar cell development, highlighting key advancements in material stability and efficiency.

To combat these challenges, researchers have explored compositional engineering to enhance moisture and thermal stability. The partial substitution of cesium ($Cs^+$) for formamidinium ($FA^+$) and methylammonium ($MA^+$) in hybrid perovskite structures has significantly improved device stability and efficiency [61]. Additionally, incorporating ultrathin electron transport layers (ETLs) via atomic layer deposition (ALD) has enabled PSCs to maintain high efficiency under thermal stress, with operational stability exceeding 10 hours at 100°C [62]. Ono et al. [67] investigated the behaviour of spiro-OMeTAD as a hole transport material (HTM) under different environmental conditions and found that gas molecule incorporation reduced charge mobility, leading to p-type doping without significant oxidation of spiro-OMeTAD$^+$. Several advancements have been made to enhance perovskite solar cell (PSC) efficiency and stability [63,68]. A study integrating potassium cations ($Cs_{0.925}K_{0.075}PbI_2Br$) and rubidium cations into PSC devices improved stability, with one achieving 95% of its initial performance after 500 hours at 85°C under full illumination [64]. Another breakthrough introduced a dopant-free triphenylamine-based HTL, which outperformed conventional doped spiro-OMeTAD, reaching a PCE exceeding 16% with enhanced photostability [69]. Thermal degradation was also explored, where perovskite films exposed to 100 suns remained stable at room temperature but degraded at 45–55°C, indicating that thermal and light-induced stability is composition-dependent [70]. Furthermore, 2-pyridylthiourea additives improved crystal grain uniformity and moisture resistance, resulting in 18.2% PCE retention after 30 days in 55 ± 5% RH at 65°C [71].

To combat thermal instability and ion migration, Bai et al. [72] incorporated ionic liquids into perovskite films, significantly improving stability, with PCE retention of 80% even after 5200 hours at 70–75°C. Wang et al. [73] introduced caffeine as an additive, forming a molecular lock with $Pb^{2+}$ ions, which enhanced film crystallization, reduced ion migration, and improved stability; their device achieved a PCE of 19.8%, maintaining over 85% efficiency at 85°C. Schloemer et al. [74] further optimized PSCs by replacing $MoO_x$ interlayers with $VO_x$, increasing PCE retention to 71% under continuous illumination at 70°C for 1100 hours. However, $MoO_x$ interlayers suffered from delamination and rapid current loss under prolonged thermal stress, highlighting the need for further optimizations in charge transport layers to achieve commercially viable, long-lasting PSCs. These advancements demonstrate the significant potential of perovskite solar cells and the ongoing need for material innovations to enhance thermal and operational stability [75].

The present review aims to analyse the thermal degradation mechanisms of perovskite solar cells (PSCs), with a particular focus on $MAPbI_3$ perovskite layers and hole transport layers (HTLs) such as spiro-OMeTAD. As PSCs have demonstrated high power conversion efficiencies (PCEs) and cost advantages over silicon-based photovoltaics, their thermal instability remains a significant challenge for commercialization. The study investigates how elevated temperatures influence the decomposition of $MAPbI_3$ into $PbI_2$, $CH_3I$, and $NH_3$, with degradation rates further influenced by moisture, oxygen, and environmental conditions. Additionally, this review explores the role of cation engineering in improving thermal stability, highlighting how mixed cation perovskites (Cs-MA-FA) offer superior resilience to thermal stress compared to MA-based perovskites. The instability of HTLs is examined, particularly how dopant additives like Li-TFSI and t-BP contribute to morphological degradation and moisture sensitivity. Alternative dopant-free HTMs (e.g., $P_3HT$, CuSCN, $NiO_x$, and $Cu_2O$) and interface engineering strategies are discussed as potential solutions to improve PSC longevity. By providing a comprehensive assessment of stability challenges and emerging mitigation strategies, this review aims to support the development of more thermally stable and commercially viable PSC technology.

## 2. Working Mechanism and Equivalent Circuit Representation of Solar Cells

Perovskite solar cells (PSCs) generate electricity through a sequence of fundamental physical processes, as depicted in Fig. 4, including charge generation, charge transport, charge recombination, and charge extraction. Understanding these mechanisms is vital for optimizing the photovoltaic (PV) efficiency and overall performance of PSCs. Boix et al. [76] have provided an extensive review covering these processes in detail. Initially, incident photons with energies greater than the bandgap of the perovskite material are absorbed, resulting in the generation of excitons or electron-hole pairs [77]. Perovskite materials exhibit advantageous optical and electronic properties, such as a high extinction coefficient ($>10^4$ $cm^{-1}$), tunable bandgap, and sharp optical absorption edge. These properties allow efficient absorption of sunlight within a thin active layer (typically 300–500 nm), enabling high power conversion efficiency (PCE).

After generation, electron-hole pairs must effectively separate and travel toward respective electrodes—a step termed charge transport. Perovskites display notably long diffusion lengths (approximately 5 μm) and charge carrier lifetimes (~1 μs), significantly reducing recombination probabilities and enhancing charge collection efficiency[76]. Compared to traditional organic solution-processed photovoltaics, metal halide perovskites possess notably longer diffusion lengths and carrier lifetimes, thus achieving superior device efficiency[78]. Nevertheless, the recombination of charge carriers remains a critical obstacle, adversely affecting overall PSC efficiency. Recombination can occur in several forms, including bulk recombination due to trap states within the perovskite material, interface recombination at the boundary between the perovskite and transport layers, and shunt recombination arising from direct contact between electron transport layers (ETLs) and hole transport layers (HTLs). Minimising recombination losses through the proper engineering of perovskite films, interfaces, and transport layers is essential to enhance the photovoltaic performance of PSCs[79,80].

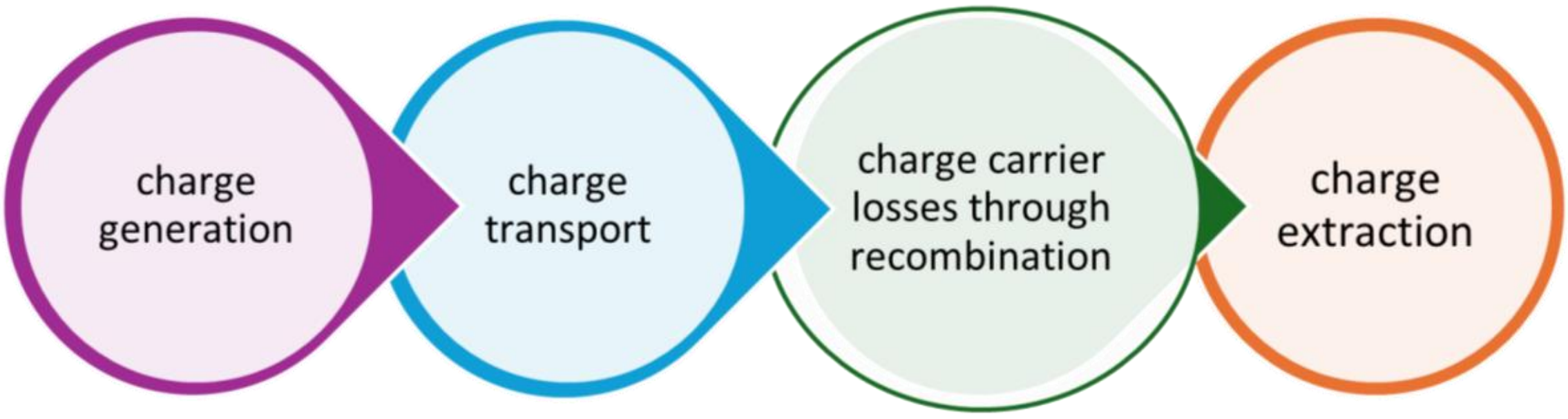

Figure 4. General mechanism of generating electricity in perovskite solar cell.

The final critical stage in PSC operation is charge extraction, wherein selective charge transport layers play a significant role. Selective contacts, or charge extraction materials, must facilitate unidirectional charge floweither electrons or holes, while preventing recombination at interfaces. Effective charge extraction layers exhibit high electrical conductivity, appropriate energy-level alignment, minimal parasitic light absorption, and robust stability under operational conditions. The extracted charges flow through conductive layers or metal electrodes, ultimately passing through an external circuit to produce electrical current[81,82].

In summary, understanding the fundamental physical processes involved in the operation of perovskite solar cells (PSCs), namely charge generation, transport, recombination, and extraction, is crucial for addressing their intrinsic degradation mechanisms, as depicted in Figure 5(a). The efficiency of PSCs strongly relies on key material properties such as a high absorption coefficient, long charge carrier diffusion lengths, and a tunable bandgap. These properties enable efficient photon absorption, effective charge separation, and improved overall photovoltaic performance. Nevertheless, PSCs experience performance losses due to intrinsic challenges like recombination processes, thermal instability, moisture-induced decomposition, and ion migration. For instance, thermal stress can trigger structural phase transitions and chemical decomposition, leading to by-products like $PbI_2$, $CH_3I$, and $NH_3$, consequently reducing both stability and efficiency. Moisture and oxygen exposure accelerate material degradation, while ion migration introduces instability at interfaces and within the active layer, further diminishing device lifespan.

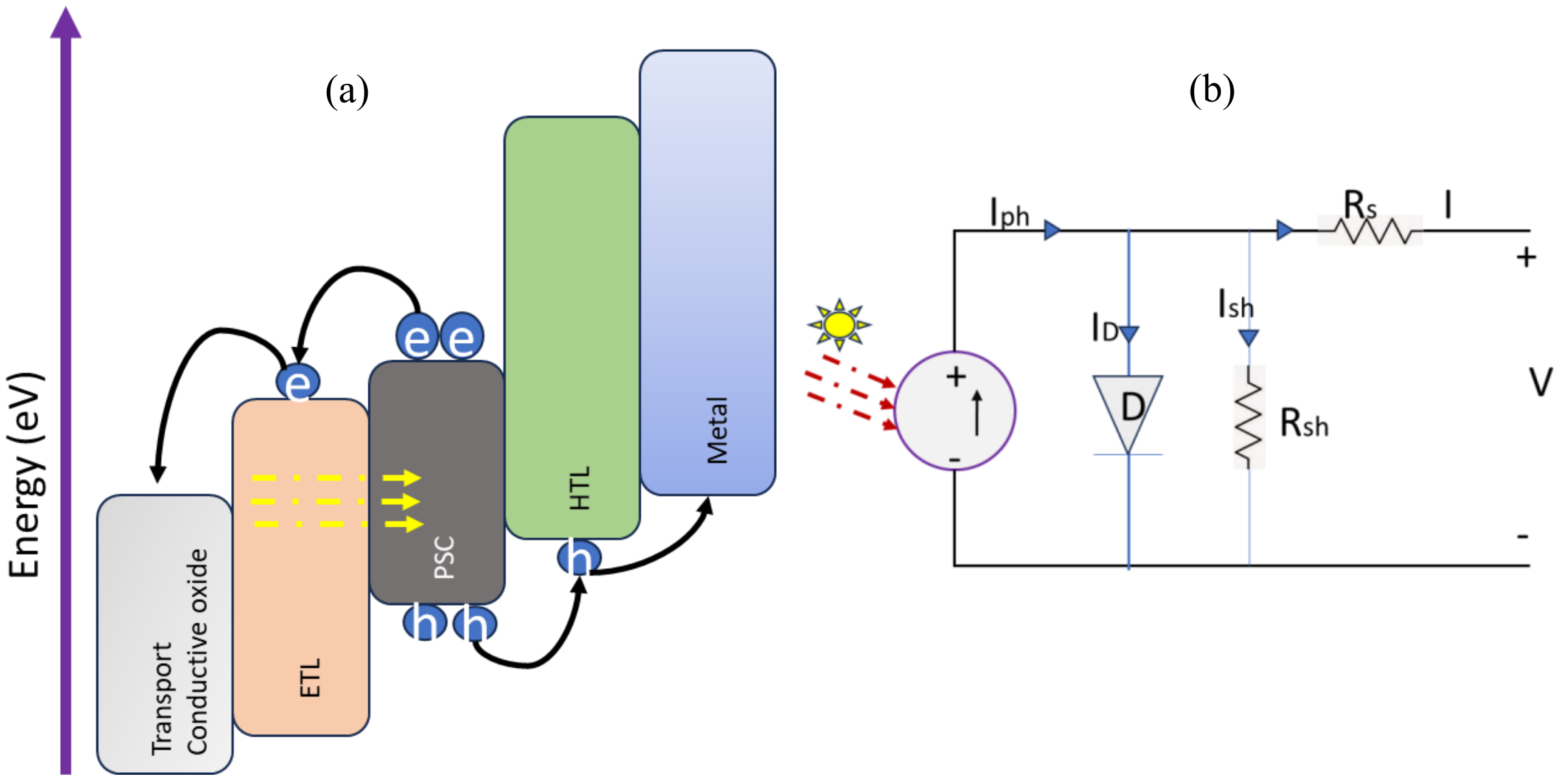

Figure 5. (a) Energy-level diagram illustrating the working mechanism of a typical perovskite solar cell (PSC) and (b) its equivalent circuit representation highlighting series and shunt resistances.

The simplest equivalent circuit representation of a photovoltaic (PV) solar cell consists of an ideal current source ($I_{ph}$) connected in parallel with an ideal diode (D) as shown in Figure 5(b). This fundamental configuration accurately describes the basic operational behaviour of a solar cell under illumination. Physically, the ideal current source represents the photogenerated current resulting from the absorption of photons within the semiconductor material. This photocurrent ($I_{ph}$) is directly proportional to the incident light intensity or photon flux. Concurrently, the diode symbolises the semiconductor junction's inherent properties, accounting for recombination and current leakage mechanisms within the cell. Under illumination, the diode allows current flow predominantly in one direction, characterising the solar cell's rectifying behaviour. Therefore, this simplified equivalent circuit effectively captures the essential physics underlying photovoltaic energy conversion processes [83–91].

(i) Current Flow in a Solar Cell

The output current from the equivalent circuit can be mathematically expressed by the Shockley diode equation [92]

$$I = -I_{ph} + I_0 \left[ \exp\left( \frac{qV}{\eta kT} \right) - 1 \right] \quad (1)$$

where:

$I_{ph}$ is the photocurrent generated by illumination, $I_0$ is the diode reverse saturation current, $q$ the electronic charge ($1.6*10^{-19} C$), $V$ is the voltage across the diode (terminal voltage), $\eta$ s the diode ideality factor, typically between 1 and 2, $k$ is Boltzmann's constant ($1.38*10^{-23}$ J/K). $T$ is the absolute temperature (K).

(ii) Electron-Hole Pair (EHP) Generation Rate[93]

The electron-hole pair generation rate ($G_{ph}$) at the illuminated surface of the solar cell decreases exponentially with the depth ($x$) inside the semiconductor material;

$$G_{ph}(x) = G_0 \exp(-\alpha x) \quad (2)$$

where:

$G_{ph}$ represents the generation rate at the surface, $\alpha$ is the absorption coefficient of the semiconductor material, For a small volume element within the solar cell, having cross-sectional area $A$ and infinitesimal thickness $dx$, the EHP generation rate ($\frac{dN}{dt}$) is given by:

$$\frac{dN}{dt} = G_{ph}(x) \cdot A \cdot dx \quad (3)$$

(iii) Carrier Diffusion Length and Active Region

The diffusion length, representing the average distance charge carriers travel before recombination, differs for electrons ($L_e$) and holes ($L_h$). It depends on their diffusion coefficients ($D$) and carrier lifetimes ($\tau$), expressed as:

$$L_e = \sqrt{2D_e \tau_e} \quad (4)$$

$$L_h = \sqrt{2D_h \tau_h} \tag{5}$$

The diffusion coefficient (D) is related to mobility ( $\mu$ ) by Einstein's relation:

$$D = \frac{kT}{q}\mu \tag{6}$$

The total active region width of a solar cell, where effective charge separation and collection occur, is defined as the sum of electron and hole diffusion lengths and the width of the depletion region (*W*):

$$Active\ Region\ = L_e + L_h + W \tag{7}$$

(iv) Dark I-V Characteristics and Illuminated Current

Under dark conditions (no illumination), a solar cell behaves as a standard diode with a current-voltage relationship given by the Shockley diode equation:

$$I_{\text{diode}} = I_0 \left[ \exp\left( \frac{qV}{\eta kT} \right) - 1 \right] \tag{8}$$

Two critical parameters characterize solar cell performance i.e., open-circuit voltage and short-circuit current. Open-circuit voltage $(V_{OC})$ is the maximum voltage available from the solar cell at zero current, occurring when terminals are disconnected from any load. At open-circuit conditions, the photocurrent $(I_{ph})$ generated by illumination precisely balances the forward-bias diode current. Short-circuit current $(I_{SC})$ is the maximum current produced by the solar cell when the terminals are directly shorted. Under this condition, nearly all photogenerated carriers are collected at the electrodes, resulting in the highest achievable current for the given illumination level.

## 3. Materials for Perovskite Solar Cell

The general chemical formula for perovskite materials is $ABX_3$, as shown in Fig. 6(a) where "A" typically represents an organic cation such as methylammonium ($CH_3NH_3^+$) or ethylammonium ($CH_3CH_2NH_3^+$), or an inorganic cation such as caesium ($Cs^+$), rubidium (Rb+), sodium ($Na^+$), potassium ($K^+$), or other rare-earth metal ions. "B" denotes a divalent metal cation (e.g., $Pb^{2+}$, $Sn^{2+}$, $Ge^{2+}$) with an octahedral coordination environment (coordination number of 6). "X" is commonly an anion, such as a halide ($Cl^-$, $Br^-$, $I^-$) or occasionally an oxide or nitride ion [94–99]. The structural stability and formation of perovskite crystals strongly depend on the ionic radii of these constituent ions. This dependency is quantified by the Goldschmidt tolerance factor (t)[100], defined by the ionic radii as:

$$t = \frac{r_A + r_X}{\sqrt{2}(r_M + r_X)} \tag{9}$$

For t = 1, the perovskite adopts an ideal cubic structure. If t < 1, the "A" cation radius is too small, leading to structural distortions. If t > 1, the "A" cation radius is too large, causing strain due to the excessive size. Typically, for stable halide perovskites, the tolerance factor lies in the range of 0.81 < t < 1.11 [101]. However, empirical observations have shown that even within the seemingly stable tolerance factor range of 0.8–0.9, perovskites may exhibit instability under certain conditions [102,103]. In addition to the tolerance factor, the octahedral factor (also called octahedral stability factor), ranging from 0.45 to 0.89, provides additional criteria for stability, particularly emphasizing the geometric compatibility between the divalent metal ion and the surrounding halide octahedra [104,105]. Understanding these factors is crucial for designing perovskite materials with enhanced stability and optimized performance in photovoltaic applications.[102]

Table 1. Effect of temperature on the structure of perovskite material

| **Temperature (K)** | **Structure** |
|---|---|
| T<160 | Ortho-rhombic |
| 162.2 < T < 327.4 | Tetragonal |
| T > 327.4 | Cubic |

Optical, unique electromagnetic, and thermal properties and cubic lattice-nested octahedral structures of perovskite materials have gained huge attention among researchers. The fundamental structure of perovskite material is $ABX_3$ composition, and the $CaTiO_3$ compound is the source of the perovskite material[106–112]. Well known characteristics of these materials are stated in Table 1 and Table 2.

Table 2. Properties of perovskite materials

| **Properties** | **Value Range** |
|---|---|
| Bandgap | 1.5–2.5 eV |
| Absorption coefficient | $10^4$ $cm^{-1}$ |
| Exciton binding energy | Less than 10 meV |
| Crystallization energy barrier | 56.6–97.3 kJ $mol^{-1}$ |
| PL quantum efficiency | 70% |
| Charge carrier lifetime | Greater than 300 nm |
| Relative permittivity | 3 |
| Carrier mobility | 800 $cm^2$/Vs |
| Exciton | Wannier type |
| Trap-state density | $10^{10}$ $cm^3$ (single crystals)<br>$10^{15}$–$10^{17}$ $cm^3$ (polycrystalline) |

These perovskite materials are very efficient in the absorption of solar energy. These exhibit high optical absorption coefficient, photoelectric properties, and low exciton binding energy. Effective transmission of electrons and holes from 100 nm to more than 1 μm [76]. Methylammonium lead iodide provides a desirable property in terms of electronic and optical absorption coefficients. The $CH_3NH_3PbI_3$ material-based PSC has an excellent electron mobility of $24.0 \pm 7$ $cm^2v^{-1}s^{-1}$, holes mobility of $105 \pm 35$ $cm^2v^{-1}s^{-1}$, and appropriate band gap of 1.55 eV with very good PCE of about 23.7% [113]. As indicated in the compound, this perovskite material contains lead (Pb), which is a toxic material, and this necessitates a lead-free perovskite material to be developed [114]. Other precautions may be taken, such as encapsulation, to reduce the detrimental effect on the environment of lead toxicity. Moreover, recycling of this perovskite material should carry forward to ensure the great lifetime of these materials. However, considering the stability of the Pb-based PSC in ambient conditions, it has very low stability [115–119], [120]

Group 14 materials, such as Ge or Sn, have the potential to become substitute materials as they exhibit a comparable electrical structure to Pb. However, transition metals, lanthanides, and alkaline earth metals can also be taken into consideration as lead replacements [120,121]. Other than this, inorganic perovskite materials have gained significant attention among researchers due to their outperformance in thermal stability. This is primarily because of the absence of weakly bonded organic compounds. $CsPbI_3$ is the top performer among the pool of inorganic compositions of $CsPbI_2Br_3$, $CsPbIBr_2$, and $CsPbBr_3$. Two significant problems limit the performance of $CsPbI_3$-based PSC. The low tolerance factor causes the preferred phase transformation from perovskite to a non-perovskite phase. The second is energy loss due to energy-level mismatch and unavoidable defects, which limits PCE [120]. Along with PSC material, it's worth taking note of conducting layer materials. CdS-based flexible PSC FTO/CdS/$CH_3NH_3PbI_3$/Spiro-OMeTAD/Ag, fabricated by Tong et al. [122], exhibited excellent performance in PSCs due to its enhanced hole mobility and reduced series resistance. Where the hole transport material (HTM) was Spiro-OMeTAD. This study documented that PCE is influenced by the thickness of the electron transport layer (ETL), which ranges from 30 to 120 nm. Maximum efficiency was obtained at 50 nm, while beyond 100 nm PCE decreased, likely due to the rise in series resistance and the reduced photon transmission to the perovskite layer. Abulikemu et al. (2017) [123] investigated $SnO_2$/CdS as an ETL in $CH_3NH_3PbI_3$-based PSCs. They quantitatively compared $SnO_2$ and $SnO_2$/CdS-based PSCs, which were fabricated using the spin-coating method [124]. The results demonstrated that CdS thin films, despite showing some instability under continuous illumination due to oxygen vacancies and other non-stoichiometric defects, emerged as a superior alternative to $TiO_2$ and ZnO for ETL in PSCs. Nonetheless, the highest efficiency was still achieved in PSCs utilizing $TiO_2$-based ETLs.

Another electron transport material (ETM), $Ti_3C_2Tx$ MXene, was investigated by Wang et al. [125] with the configuration ITO/MXene/Perovskite/Spiro-OMeTAD/Au. This two-dimensional ETM provides high transparency, conductivity, tunable binding energy, and excellent functional properties, but still loses the race with $TiO_2$ or $SnO_2$. The MXene-based PSCs showed promising results, achieving a current density of 21.5 mA/cm² and a power conversion efficiency (PCE) of 18.9%, along with enhanced device stability. Moreover, the configuration ITO/NiO/Perovskite/PCBM-$SnS_2$/ZnO/Ag was explored by Patil et al [126] where [6,6]-phenyl C61 butyric acid methyl ester (PCBM), a fullerene derivative, was used as an ETM. The PCE of this device was limited due to the challenges at the perovskite/PCBM interface, such as inefficient electron transport, large electron trap zones, poor film formation, and significant non-radiative recombination. To address these issues, a homogeneous blend of PCBM and $SnS_2$ was used as the ETM, which improved electron mobility and provided more favourable energy levels. Additionally, $SnS_2$'s higher relative permittivity reduced the electron capture radius from 0.62 nm to 0.22 nm, minimising leakage current and non-radiative recombination at the perovskite/PCBM-$SnS_2$ interface. As a result, the PSC with the PCBM-$SnS_2$ ETM achieved a PCE of 19.95%, outperforming the device with only PCBM, which had a PCE of 18.22%.

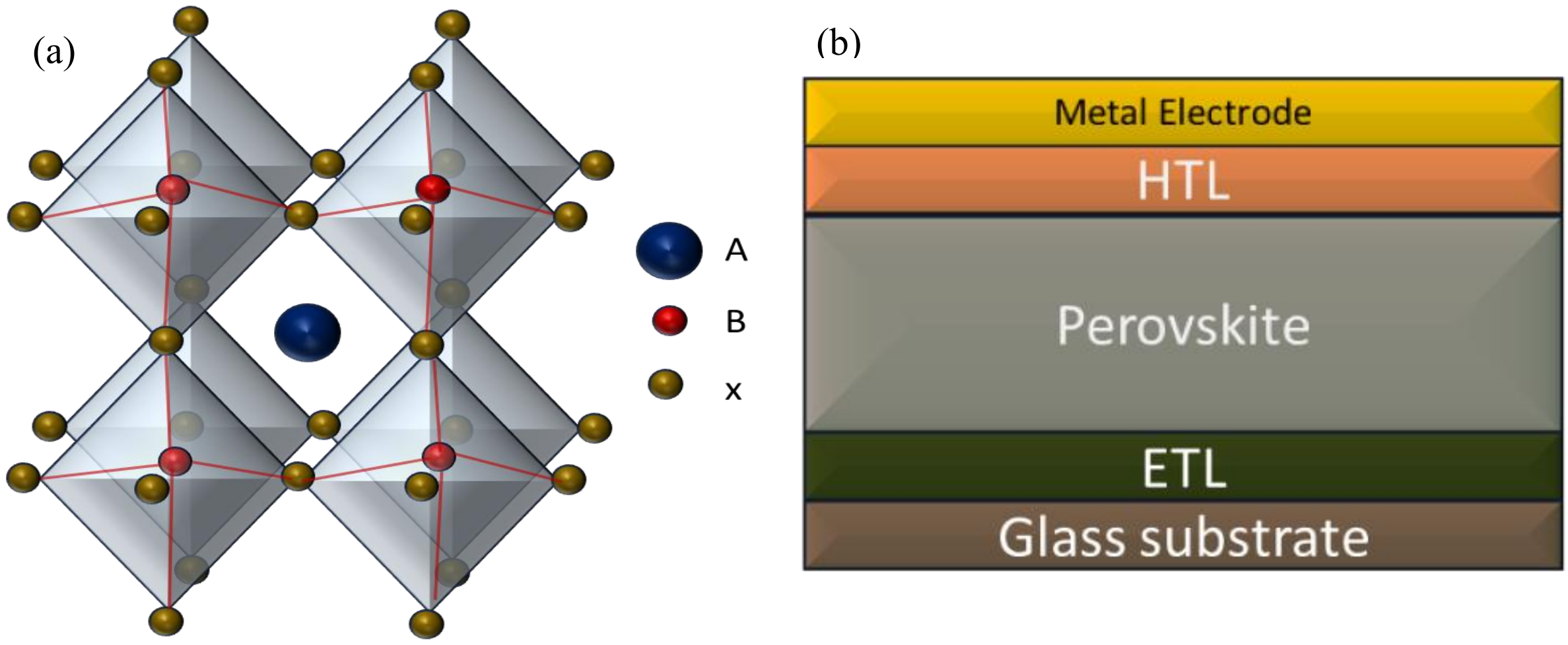

Figure 6. Layout of perovskite material with (a) crystal structure and (b) typical device structure.

## 4. Device Architectures of Perovskite Solar Cells

Since the first dye-sensitized perovskite solar cell (PSC) was introduced, device architecture has undergone substantial evolution to enhance efficiency, stability, and practicality. A typical PSC comprises several key components arranged sequentially: Transparent Conductive Oxide (TCO)[127], Electron Transport Layer (ETL) [128], Perovskite (absorber) [129], Hole Transport Layer (HTL) [130], and Metal Electrode (cathode) [131], as shown in Figure 6(b). Each of these layers has specific physical roles, such as facilitating selective charge extraction, preventing recombination, and optimizing optical absorption. The detailed functions, requisite properties, and commonly used materials for each component are summarized briefly in Table 3.

Table 3. Brief description of different layers in a perovskite solar cell

| Layer | Function | Required Properties | Examples |
|---|---|---|---|
| Hole Transport Layer (HTL) | Collect holes from absorber and transport toward cathode; block electrons | Higher HOMO energy level than the absorber | Spiro-OMeTAD, NiO, CuO, PTAA, CuI |
| Electron Transport Layer (ETL) | Collect electrons from absorber and transport toward anode; block holes | HOMO and LUMO levels higher than absorber; high UV-Vis transparency | $TiO_2$, $SnO_2$, $SiO_2$, ZnO |
| Metal electrode | Transport holes to an external circuit | Good electrical conductivity | Au, Ag |
| Perovskite Absorber | Generate excitons (electron-hole pairs) | High absorption coefficient, tunable bandgap | $MAPbI_3$, $FAPbI_3$ |
| Transparent Conductive Oxide (TCO) | Transport electrons to external circuit | High transparency, excellent conductivity | FTO, ITO |

The earliest PSC device architecture, proposed by Kojima et al. in 2009 [17], employed a DSSC configuration where perovskite materials ($MAPbI_3$ and $MAPbBr_3$) acted as sensitizers deposited on a mesoporous titanium dioxide (m-$TiO_2$) scaffold. This original design utilized fluorine-doped tin oxide (FTO) substrates, transparent conductive oxides, and a liquid electrolyte to facilitate charge transport. However, despite achieving initial power conversion efficiencies (PCE) up to 3.81% with $MAPbI_3$owing to its favourable lower bandgap (~1.5 eV) compared to $MAPbBr_3$ (2.3 eV), the liquid electrolyte significantly compromised the stability and longevity of the devices.

To address stability concerns inherent in liquid-based configurations, researchers transitioned toward solid-state architectures. In 2012, Kim et al. [132] developed the first solid-state mesoscopic

PSC, where the unstable liquid electrolyte was replaced by a solid organic hole transport material (HTM), specifically spiro-OMeTAD. This innovative approach significantly enhanced both the stability and efficiency of PSCs, achieving efficiencies of around 9.7%. Building upon these advancements, Snaith et al. further improved the structural integrity and performance by substituting the electron-conductive m-$TiO_2$ scaffold with an insulating mesoporous aluminium oxide (m-$Al_2O_3$) layer. Remarkably, this modification demonstrated efficient charge transport directly through the perovskite layer itself, increasing the device's PCE to approximately 10.9%. This evolution underscored the critical importance of optimizing charge transport dynamics and interface engineering in developing highly efficient and stable PSC technologies.

Building on these initial advancements, significant progress has been made to enhance the performance of perovskite solar cells (PSCs). The meso super structured configuration (MSSC) showed considerable promise, inspiring further innovations such as the "Regular Structure" introduced by Heo et al. in 2013 [133]. In this architecture, a pillared methylammonium lead iodide ($MAPbI_3$) structure was created by completely filling mesoporous titanium dioxide (m-$TiO_2$) pores with perovskite, followed by coating with a thin layer of polytriarylamine (PTAA) as a hole transport material (HTM), leading to an impressive power conversion efficiency (PCE) of approximately 12%. Subsequently, planar n-i-p heterojunction architectures, similar to traditional inorganic thin-film solar cells, eliminated the mesoporous scaffold to simplify fabrication. Although initial efforts by Snaith et al. in 2012 [134] resulted in low efficiencies caused by significant shunt paths and incomplete perovskite film coverage, these issues were effectively addressed by employing dual-source co-evaporation of $PbCl_2$ and methylammonium iodide (MAI). This advanced processing technique produced uniform and defect-free perovskite films, leading to an enhanced PCE of around 15%. Further advancements continued with the development of inverted planar structures aimed at better photon management and interface engineering. In 2018, Tang et al. [135] fabricated a highly efficient inverted planar PSC structure using nickel oxide (NiO) nanocrystals as the HTM. The introduction of NiO nanocrystals improved optical transparency and substantially reduced photon flux losses at interfaces, resulting in enhanced device performance. These continual refinements emphasize the importance of careful structural modifications in PSC design, highlighting that optimizing charge transport layers, interfacial engineering, and the precise compositional control of perovskite absorbers are crucial factors for achieving higher efficiencies and improved long-term operational stability.

The evolution of perovskite solar cell (PSC) architectures from their original dye-sensitized solar cell (DSSC) configuration has significantly improved both stability and efficiency. The early limitations associated with liquid electrolytes were effectively addressed by transitioning to solid-state mesoscopic structures utilizing organic hole transport materials, notably spiro-OMeTAD, leading to substantial increases in power conversion efficiency (PCE). Structural refinements, such as the introduction of insulating mesoporous $Al_2O_3$ scaffolds and subsequent advancements toward regular and planar n-i-p heterojunction architectures, have further enhanced performance by promoting efficient charge transport directly through the perovskite layer. Techniques such as dual-source co-evaporation have allowed researchers to fabricate uniform, defect-free perovskite films, yielding efficiencies exceeding 15%. More recently, the development of inverted planar architectures incorporating nickel oxide (NiO) nanocrystals as hole transport materials demonstrates the continued importance of optimizing interfaces and charge transport layers. These ongoing innovations underscore the critical role of careful architectural and compositional engineering in advancing the performance and practical viability of PSC technology.

Table 4. Evolution of perovskite solar cell architectures with key parameters, materials, and performance improvements

| Structures | Year | $V_{oc}$(V) | FF (%) | $J_{sc}$(mA/cm$^2$) | PCE (%) | Description | Image | Conclusion |
|---|---|---|---|---|---|---|---|---|
| DSSC | 2009 Kojima *et al* [17]. | 0.61 | 57 | 11 | 3.81 | ETL- compact thin $TiO_2$ (c-$TiO_2$ and a micron thick m-$TiO_2$) $MAPbI_3$ is used as a sensitizer over $TiO_2$ formed by spin-coating | 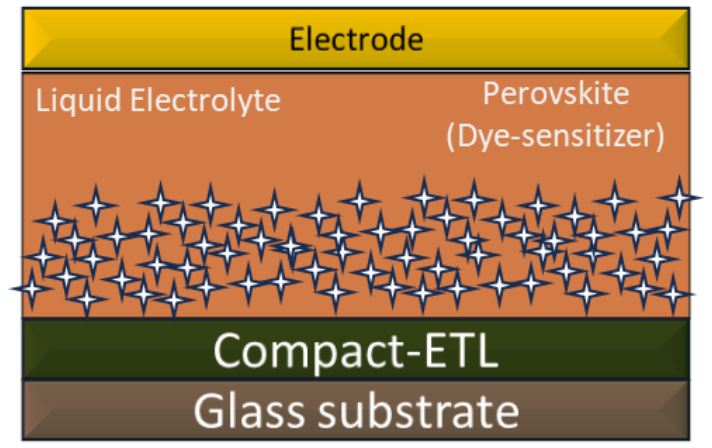  | Poor stability due to liquid electrolyte. |
| | 2011 Im*et al.* [136] | 0.706 | 58.6 | 15.82 | 6.54 | 2-3 nm of $MAPbI_3$ quantum dots in order to coat the 3.6 µm thick m-$TiO_2$ | | Significant performance degradation (80%) |
| Solid State Mesoscopic structure | 2012 Hui-Seon Kim et al. [133] | 0.88 | 62 | 17 | 9.7 | HTL- Spiro-OMeTAD $MAPbI_3$ used as a sensitizer over micron thick m-$TiO_2$ which is filled with pores | 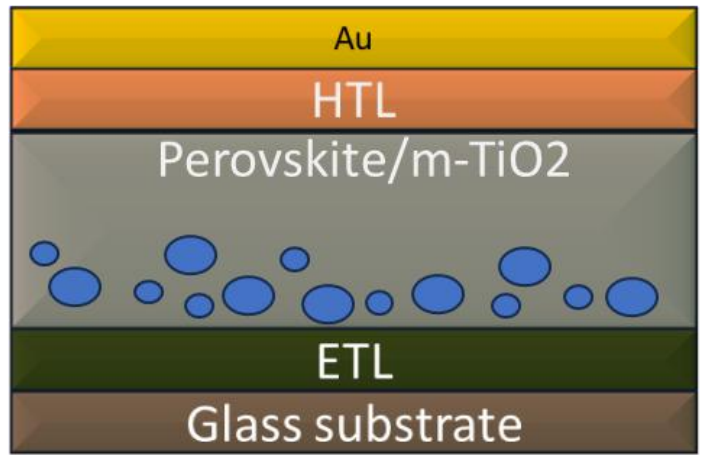  | Enhanced stability and efficiency |
| Meso-superstructure configuration (MSSC) | Tax *et al.* 2012 [134] | 0.98 | 63 | 17.8 | 10.9 | HTL- Spiro-OMeTAD $MAPbI_3$ used as a sensitizer and an insulating layer of m-$Al_2O_3$ which is filled with pores | 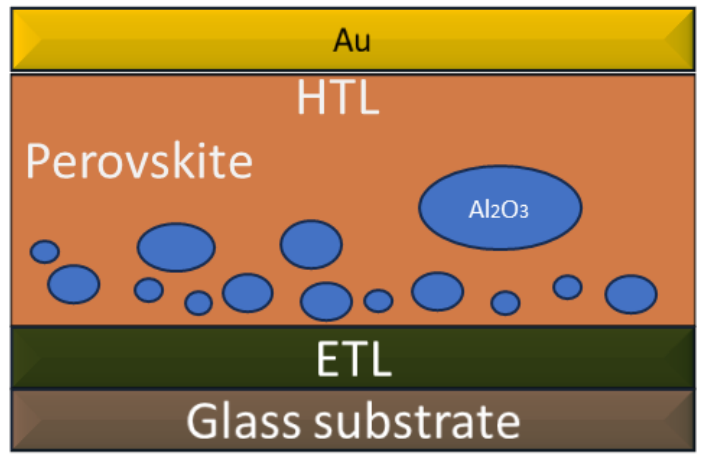  | Faster charge transport and higher efficiency |
| | Ball *et al.* 2013 [137] | 1.02 | 67 | 12.3 | 18 | $MAPbI_{3-x}Cl_x$ used as a sensitizer and lower temperature from 500 c to 150 c. | | Feasibility of low-temperature fabrication demonstrated |
| Regular | Heo *et al.* 2013 [133] | 0.99 | 72.7 | 16.5 | 12 | HTL- PTAA (polytriarylamine) $MAPbI_3$ used as a sensitizer over micron thick m-$TiO_2$ which is filled with pores | | Improved surface morphology required for higher efficiency |
| | NREL 2019 [4] | 1.056 | 74 | 21.64 | 17 | Used thicker perovskite film over 100nm m-$TiO_2$ with capping layer of 150nm using two step coating method | 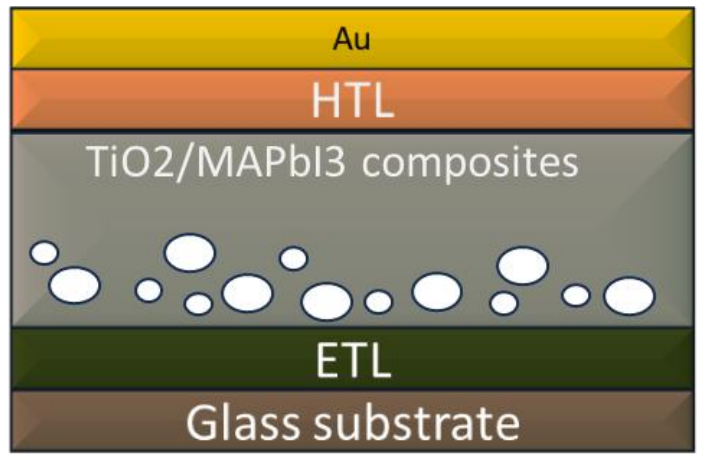 | High efficiency via improved absorber-layer deposition |
| Planar n-i-p heterojunction | Liu *et al.* 2013 [138] | 1.07 | 67 | 21.5 | 15 | "dual-source co-evaporation of PbCl2 and MAI and deposited the MAPbI3-XClx layer over c-TiO2 layer" [4]<br>p-type HTL, intrinsic perovskite layer, n-type ETL | | Eliminated shunting, improved film uniformity |
| | Yang *et al.* 2018 [139] | 1.11 | 79.2 | 24.55 | 21.6 | ETL- EDTA complex tin oxide<br>Absorber- $FAPbI_3$ with slight amount of Cs doping<br>HTL- Spiro-OMeTAD | 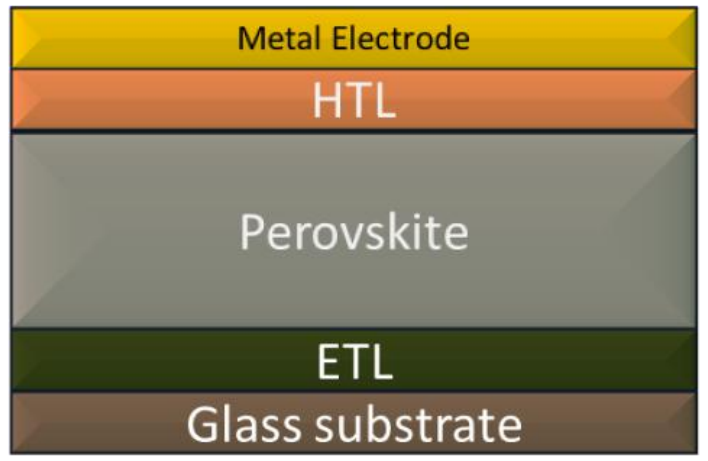 | Enhanced efficiency with improved interface properties |
| | Haider *et al.* 2018 [140] | 0.94 | 73 | 17.64 | 12.1 | HTL- NiPc (nickel pthalocyanine)<br>ETL - c-TiO2 | | Higher stability, retained 80% PCE after 38 days |
| Inverted planar | Tang *et al.* 2018 [135] | 1.06 | 75.02 | 19.41 | 15.47 | HTL- NiO nanocrystal (prepared by solvothermal method)<br>ETL- PCBM/BCP<br>Absorber- $MAPbI_3$<br>Anode- Au<br>They obtained best performance of HTL at 55 nm thickness from varying range of 30 nm, 55 nm, 70 nm, 100 nm, 170 nm | 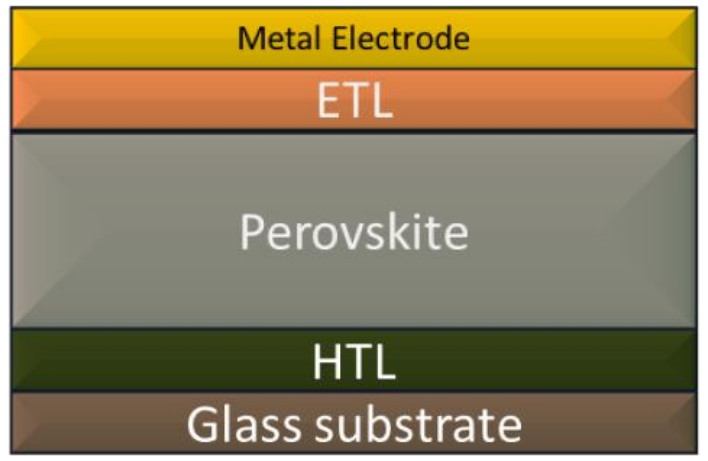 | Enhanced transparency and reduced photon flux loss |

## 5. Thermal Degradation Mechanisms in Perovskite Solar Cells

Perovskite solar cells (PSCs) have attracted considerable interest owing to their elevated power conversion efficiency, mechanical flexibility, and lightweight composition, establishing them as formidable rivals to traditional silicon-based solar cells. Their ability to achieve remarkable efficiency by means of cost-effective solution processing has inspired research in this domain even more. Still, its insufficient long-term stability, especially at high temperatures it makes commercialization extremely difficult. Particularly sensitive to heat breakdown is the hole transport layer (HTL), which collects and moves positive charge carriers (holes). This issue is most clear in organic HTL materials, which degrade under heat increase resistance and reduce hole mobility, therefore compromising the general stability of PSCs. In practical uses, including solar panels under continuous temperature stress, the perovskite absorber layer deteriorates faster, greatly reducing the operating lifetime of the equipment. PSCs' natural sensitivity to environmental variables is the main obstacle preventing them from challenging commercially available solar systems. Fragile semiconductors with weak chemical bonds mostly supported by hydrogen bonding, van der Waals forces, and ionic interactions define perovskites [141]. Environmental factors, including humidity, more sunlight, higher temperatures, and electric fields, cause the perovskite structure to deteriorate and ion migration across many functional layers of the solar cell results. This process accelerates the degradation of the perovskite absorber, electron transport layer (ETL), hole transport layer (HTL), and other necessary components, hence lowering the device efficiency [142–146]. Formulating strategies to improve the long-term stability of PSCs depends on a complete awareness of these degradation mechanisms.

A major limitation of perovskite solar cells (PSCs) under thermal stress is the instability of the hole transport layer (HTL), which is essential for charge transport. Organic HTLs, especially spiro-OMeTAD, don't hold up well at high temperatures because the dopants they contain, Li-TFSI (Lithium bis(trifluoromethanesulfonyl)imide) and tBP (tert-butylpyridine). It soaks up water and breaks down at temperatures above 70° and due to this decline, there is more electrical resistance and less hole extraction efficiency, which lowers the power conversion efficiency (PCE) and makes the device less reliable in general. Inorganic hole transport layers (HTLs), especially copper(I) thiocyanate (CuSCN) and nickel oxide (NiO), [147–151] have been seen as possible alternatives to address this issue because they are more stable at high temperatures. On the other side, we need to solve a few big problems before these materials can be widely used in perovskite solar cells. These problems are interface differences and energy level imbalance.[152]

Temperature-induced degradation affects the electron transport layer (ETL), metal electrode, and transparent conductive oxide (TCO) along with other layers in perovskite solar cells (PSCs). At high temperatures, materials like as titanium dioxide ($TiO_2$) can undergo oxygen vacancy formation and anatase-to--rutile phase transitions in ETLs, hence increasing charge recombination and reducing efficiency. Enhanced thermal stability made possible by tin oxide ($SnO_2$) makes it a better fit for high-temperature use [153,154]. Similarly, despite gold (Au) has better thermal resistance, metal electrodes, especially silver (Ag), may undergo thermal diffusion into adjacent layers, leading in interfacial instability [155,156]. Furthermore, showing little degradation under typical perovskite solar cell (PSC) operation settings are transparent conductive oxides (TCOs) such as fluorine-doped tin oxide (FTO) and indium tin oxide (ITO) [157]. This work mostly investigates the significant temperature-induced degradation mechanisms affecting the perovskite absorber and HTL layers. While simultaneously advancing thermally stable materials, interface engineering methods [158] and sophisticated encapsulation [159–162] techniques to improve the durability and market viability of perovskite solar cells for practical applications, it is imperative to fully understand thermal breakdown mechanisms to address these challenges.

**Perovskite film**

Perovskite solar cells (PSCs) have garnered significant attention owing to their superior efficiency, lightweight construction, and flexibility, positioning them as a formidable alternative to traditional silicon-based solar cells. PSCs have a major disadvantage, though, which limits their advantages: thermal instability. Solar panels turn sunlight into power; hence they are naturally hot. Research shows that the usual operating temperature for terrestrial solar cells is between 40°C to 85°C[163], which raises a fundamental question: Over extensive temperature exposure, can perovskite solar cells have constant performance? Strong power conversion efficiency but low thermal stability characterizes a common perovskite chemical used in solar cells: methylammonium lead iodide ($MAPbI_3$). $MAPbI_3$ has low conductivity [164]. At high temperatures it becomes unstable; it disintegrates around 85°C [165] and significantly improves performance. Yusoff et al. (2016), [166] show that the main reason the organic cation in $MAPbI_3$ is unstable is discovered; this makes it particularly prone to thermal breakdown. Clarifying the breakdown processes in these cells has thus been the main focus of researchers aiming at enhancing PSC thermal stability. Researchers have diligently carried out controlled tests in order to probe this instability. Coining et al. investigated $MAPbI_3$ under many conditions, including pure dry oxygen, dry nitrogen, and ambient air with 50% relative humidity. Following 24 hours of dark sample storage, they discovered indications of perovskite deterioration [165], therefore indicating the material's environmental sensitivity. Kim et al. [167] added new data on the deterioration of $MAPbI_3$ in air-free circumstances by heating perovskite films at 100 °C for 20 minutes and at 80 °C for more than an hour. Applying sophisticated techniques like Grazing Incidence Wide Angle X-ray Diffraction (GIWAXD) and High-Resolution X-ray Photoelectron Spectroscopy (HR-XPS), they proved that $MAPbI_3$ breaks down to methyl iodide ($CH_3I$), ammonia ($NH_3$), and lead iodide ($PbI_2$). This chemical breakdown alters the surface structure, therefore jeopardizing the integrity of the layer of perovskite. These findings showed really strong evidence that $MAPbI_3$ lacks the thermal stability required for extended solar cell operation.

Abdelmageed et al. [168] reported that the film underwent degradation at a temperature of 75 °C under a light environment, and after degradation, both metallic Pb and $PbI_2$ were found, but when the experiment environment was changed to a dark environment and temperature at 85°c, the film degraded to only $PbI_2$. This experiment was conducted by them in a light-induced inert atmosphere with $N_2$ (without oxygen and humidity) environment. They monitored the stability of perovskite film at two different temperatures (55 °C ± 2 °C and 75 °C ± 2 °C) for a duration of 4 days with a light exposure of 360 ± 10 mW/cm$^2$. Consequently, it was observed that at 55 °C, the film exhibited no visible degradation; however, at 75 °C, visible degradation was evident. The degraded film presented a distinctive grey-yellow hue post degradation, which is atypical and does not correspond to the bright yellow color characteristic of PbI2 films, the usual residual material following the degradation of $MAPbI_3$.The results showed that the film had no obvious degradation at 55°C and considerable degradation at 75°C. Atypical and unlike the vivid yellow $PbI_2$ films typically seen following the breakdown of $MAPbI_3$, the damaged film had a striking grey-yellow colour. Without light, a separate experiment was carried out at 75°C, 85°C, and 95°C where perovskite films were arranged on a heated plate inside a nitrogen-filled environment. The results showed notable variations: whereas complete film degradation happened at 95°C, degradation started after one day at 85°C. Under these circumstances, no clear evidence of deterioration was found at 75°C. These discoveries led to a suggested deterioration mechanism based on:

$$CH_3NH_3PbI_3 \rightarrow NH_3(g) + CH_3I(g) + PbI_2(s)$$

$$PbI_2(s) \rightarrow Pb(s) + 2I(g)$$

Brunetti et al. [169] conducted experiments using the same organic cation but different halide anions ($Cl^-$, $Br^-$, $I^-$) to evaluate the thermal stability of perovskite materials. The experiment was performed in a non-ambient reactor chamber with a controlled helium gas atmosphere in order to examine the behaviour of $MAPbCl_3$, $MAPbBr_3$, and $MAPbI_3$ within a temperature range of 130°C to 170°C. A gradual increment of 10°C in temperature and maintained at each stage for 10 hours before cooling the samples to room temperature; they observed a consistent degradation pattern across all three materials. The diffraction patterns as illustrated in Figure7, revealed that regardless of the halide composition, perovskite films underwent a similar thermal breakdown process as previously reported [168]. These results underlined even more how naturally limited perovskite materials are by the temperature-induced breakdown. It is not based on their specific halide composition. The results suggest that although altering the halide component could somewhat improve thermal stability it does not completely address the issue. Driven by this discovery, researchers have investigated new material compositions and structural modifications aimed at increasing PSC thermal resilience.

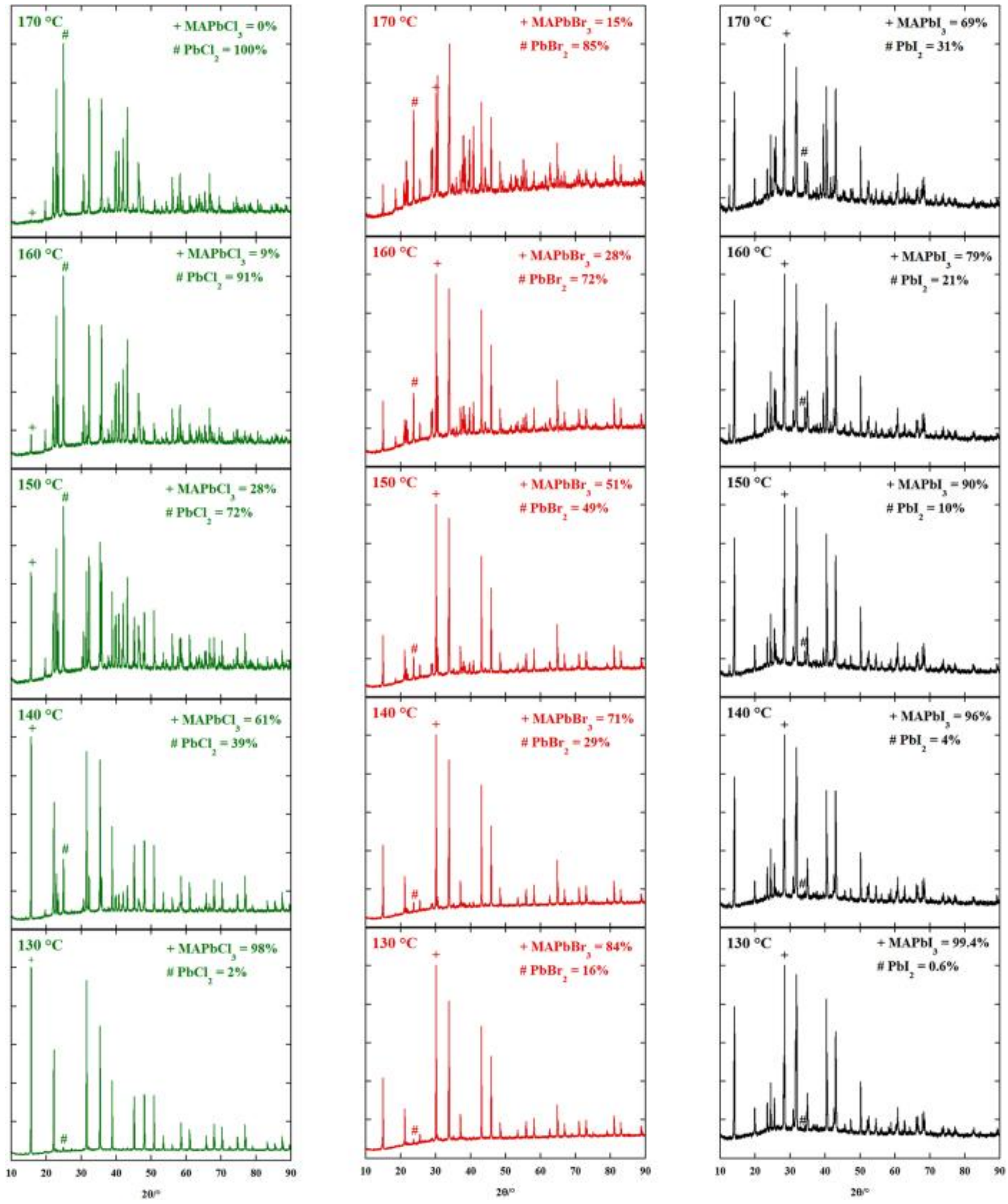

Figure 7.The XRD patterns of the $MAPbX_3$ compounds at various temperatures. The symbols "+" and "#" are used to direct the eye to recognize two characteristic reflections (one for each phase) and their relative variation after each thermal treatment[169].

Eperonet al., 2014 [161] gave a deep dive into the investigation of the effect of the size of cation in the PSCs. They chose three samples $Cs^+$, $MA^+$, $FA^+$ (effective ionic radius $Cs^+ < MA^+ < FA^+$). They prepared their perovskite film by using a single precursor solution, which was used to spin coat the substrate, followed by heating. They found that increasing the effective radius of the cation reduces the bandgap. They reported that $FAPbI_3$displayed a bandgap of 1.48eV, which is closer to the ideal single-junction solar cell (1.1 to 1.4eV). They also proved that tuning of bandgap can be done by mixing different halide anions, e.g., FAPbIyBr3-y, varying the value of y, and a range of bandgap can be obtained (1.48 to 2.23eV). They made a $FAPbI_3$ heterojunction perovskite solar cell which gave high $J_{sc}$ 23 mA/cm$^2$ and showed PCE 14.2%. They tested the sensitivity of their PSC, which made them conclude that FAPbI3 is more resistant to high temperature (withstood 150°C without any discolouration) than MAPbI3 (withstood up to 30min without discolouration), whereas for atmospheric conditions, both the cells had performed degradation at a same rate. The scanning electron microscope (SEM) image of the $FAPbI_3$ device structure is demonstrated in Figure 8.

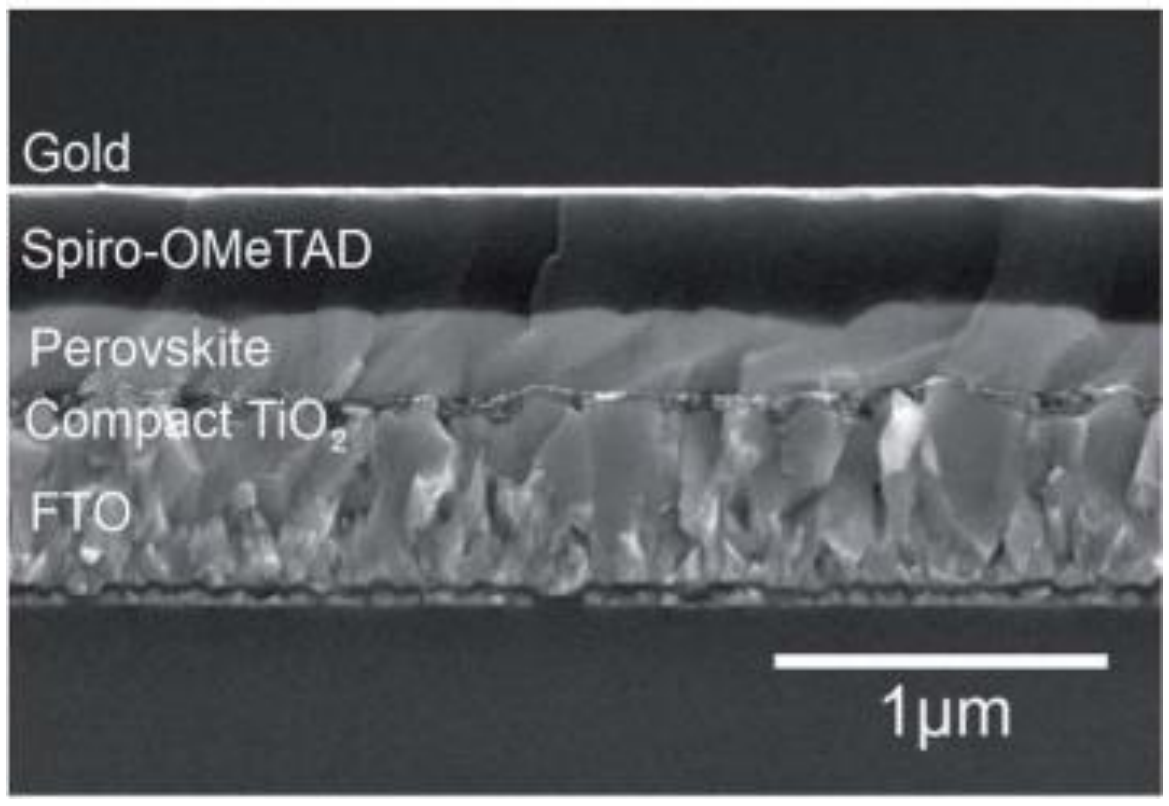

Figure 8. SEM image of $FAPbI_3$ PSC [170].

Table 5. Cs content in precursor and corresponding film composition determined by XPS

| Cs content in precursor (Cs/Cs + MA) (%) | Pb (%) | N (%) | I (%) | Cs (%) | Cs content in films (Cs/Cs+MA) (%) |
|---|---|---|---|---|---|
| 0 | 10.5 | 11.5 | 44.2 | NA | NA |
| 5 | 10.7 | 10.2 | 44.2 | 1 | 9 |
| 10 | 11 | 9 | 46 | 1.7 | 15.9 |
| 15 | 10 | 9.5 | 43.4 | 2.8 | 22.7 |
| 20 | 10 | 4.5 | 37.5 | 4 | 47 |
| 30 | 10 | 2.7 | 36.9 | 7 | 72 |
| 40 | 12 | 1.5 | 43 | 9.36 | 86 |

Wang et al.,[171] developed the mixed cation and single halide anion-based perovskite $Cs_xMA_{1-x}PbI_3$ to investigate the influence of temperature. The examinee was prepared using a one-step spin-coating method, followed by heating the entire solution at 60 °C. The findings revealed that the concentration of Cs in the film exceeded that of the precursor material. Their findings indicate that the phenomenon is attributed to the low solubility of precipitated Cs atoms. The data is also presented in a tabular format, as shown below in Table 5.

To investigate thermal stability, two conditions were selected: the first involved heating the film at 120°C for 3 hours in a nitrogen atmosphere, while the second condition consisted of heating at the same temperature for 3 hours in dry air ($V_{N2}$: $V_{O2}$ = 4:1). The specific concentrations of Cs (x) examined were 0, 0.09, 0.23, 0.47, 0.72, and 0.86. Based on the analysis presented in Figure9 (a, b), it can be concluded that following heat treatment, the relative absorption of the perovskite film with x=0.09% Cs exhibited superior performance compared to the other samples in dry air conditions. From the analysis of Fig 9(c) (dry air), it can be concluded that for x>0.23, the retained absorption at 700 nm was observed to be lower than that of the control sample (x=0). This unusual occurrence can be elucidated by the elevated concentration of Cs. The elevated concentration of Cs facilitated the segregation of $CsPbI_3$ from the film and allowed for a straightforward transition of the perovskite phase from cubic to orthorhombic. The black perovskite phase is identified as cubic, while the yellow non-perovskite phase is classified as orthorhombic. The yellow non-perovskite phase demonstrated no absorption at 700nm, attributed to the orthorhombic phase's band gap of 2.82 eV, indicating a high energy threshold[171]. Figure 9(d) presents the images of the perovskite film subjected to various atmospheric conditions and varying concentrations of Cs. It is evident that the film exhibited superior stability in a Nitrogen atmosphere compared to conditions in dry air. The team elevated the temperature to 150°C and maintained this condition for a duration of 3 hours, resulting in Figure 10 illustrating that the peak for x=0.09 is diminished when compared to x=0. Figure 11(a) illustrates that the films exhibited a flat morphology without pinholes. An increase in Cs content resulted in a reduction in grain size, likely due to the enhanced number of nucleation sites during the initial stages of film formation. Following thermal annealing at x=0, pits and pinholes were generated due to the loss of MAI or $CH_3NH_2$. Consequently, the film underwent recrystallization into $PbI_2$, resulting in a rugged texture. However, with the increase in Cs content, the film pinholes were effectively suppressed due to the significantly more stable cation site (Cs) present in the films. The analysis of the overall performance of PSC involved the selection of the $Cs_{0.09}MA_{0.91}PbI_3$ configuration, yielding effective results with Jsc at 22.57 mA/cm$^2$, Voc at 1.06 V, FF at 0.76, and an efficiency of 18.1%. This noteworthy outcome is attributed to the partial substitution of MA with Cs, which enhanced thermal stability. The reduced occupation of MA at the cation site resulted in diminished thermal loss, decreased oxidation, and a more compact crystal structure, ultimately enhancing stability. A positive outcome at x=0 indicates that the film exhibited a black coloration. The deterioration observed in dry air conditions can be attributed to the presence of oxygen and moisture, resulting in the oxidation of MA and subsequent yellowing. [171]

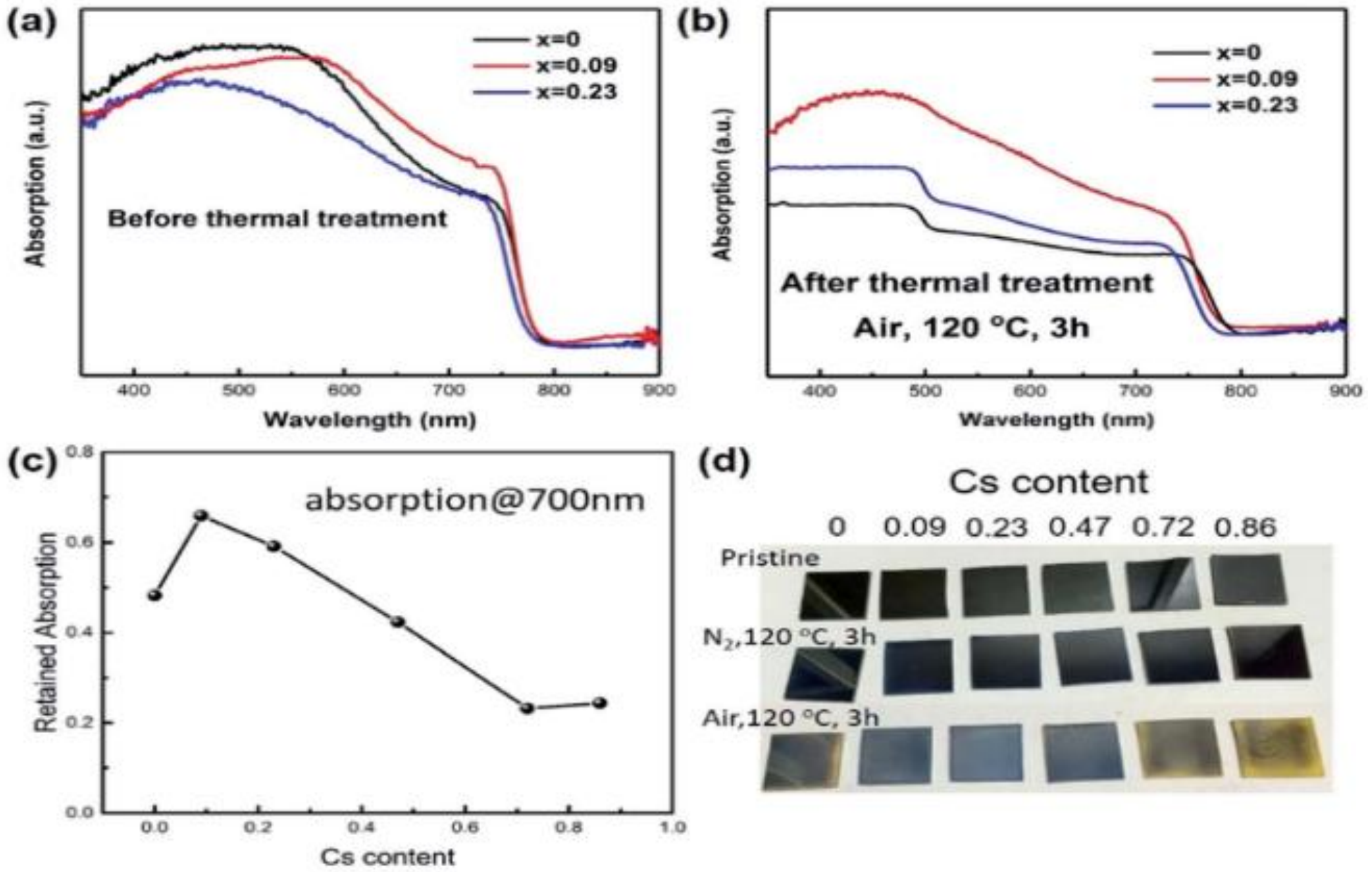

Figure 9. The absorption phenomenon of (a) pristine and (b) thermally aged perovskite CsxMA1-xPbI3, with varying Cs content represented by x. (c) demonstrates the retention of absorption of thermally aged perovskite at 700 nm as x ranges from 0 to 0.86. Images of perovskite films under varying atmospheric conditions [171].

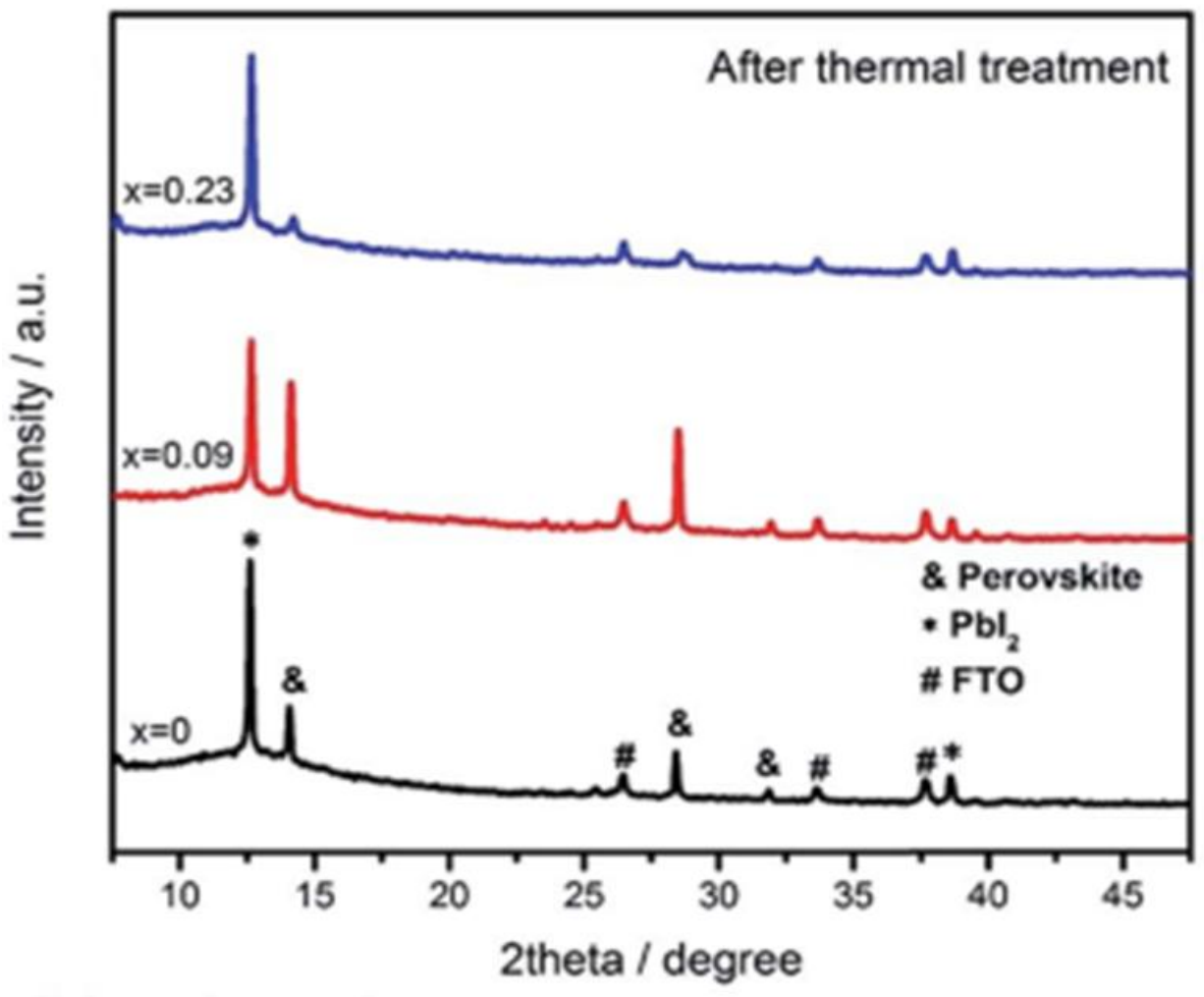

Figure 10. XRD patterns of $Cs_xMA_{1-x}PbI_3$ after thermal annealing [171]

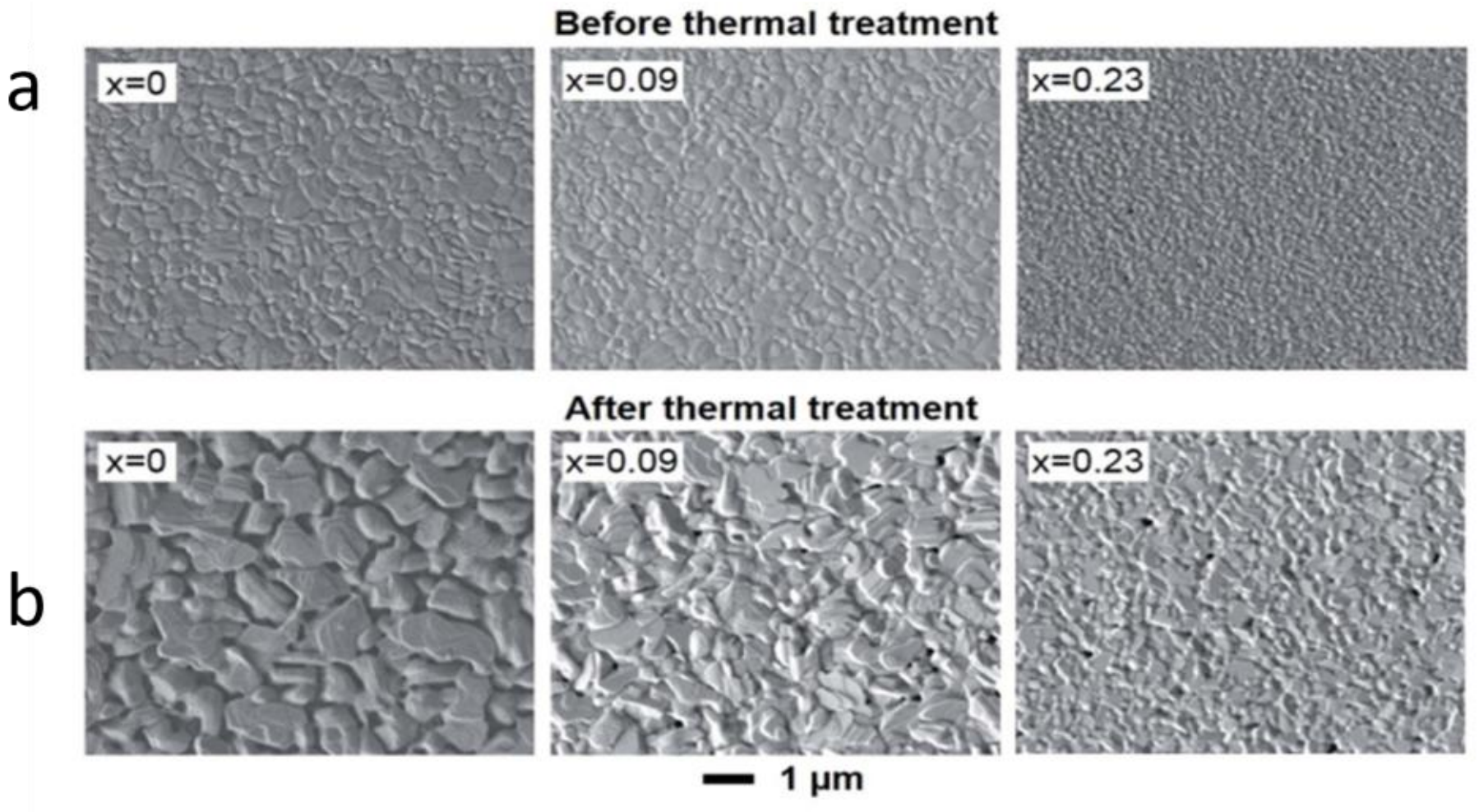

Figure11. SEM images of perovskite film[171].

Yang et al. [172] conducted an analysis of $Cs_{0.05}(MA_{0.17}FA_{0.83})_{0.95}Pb(I_{0.83}Br_{0.17})_3$ at an elevated temperature of 85°C, collecting data at 12 hours, 24 hours, and 48 hours. Their analysis indicated that following 12 hours of thermal annealing at 85°C, crystal grains exhibited growth, resulting in the observation of larger grains with a size of approximately 250 nm. This process persisted throughout the duration of thermal annealing, leading to the conclusion that this PSC device lacks thermal stability. Ultimately, they observed that their film exhibited irregularities, a rough texture, and an abundance of crystalline grains. The SEM images were provided to illustrate the process as given in Figure 12. Additionally, following a thorough examination of the crystal structure, charge carrier dynamics, and electronic configurations, the researchers concluded that the device $Cs_{0.05}(MA_{0.17}FA_{0.83})_{0.95}Pb(I_{0.83}Br_{0.17})_3$ can be interpreted as two distinct compounds: $(MAPbI_3)_{0.1615}$ and $Cs_{0.05}FA_{0.7885}Pb_{0.8385}I_{2.0055}Br_{0.51}$. The component $Cs_{0.05}FA_{0.7885}Pb_{0.8385}I_{2.0055}Br_{0.51}$ exhibits greater thermal stability compared to $(MAPbI_3)_{0.1615}$. The compound $(MAPbI_3)_{0.1615}$ adheres to the established degradation pathway, specifically $MAPbI_3 \rightarrow (-Ch_2-) + NH_3(g) + HI(g) + PbI_2$, where (-Ch2-) denotes the remaining hydrocarbon complex. The degradation occurred in a sequential manner, as illustrated in Figure 13(a). The authors outlined several factors contributing to this mode of degradation, including the presence of dangling bonds ($Pb^{2+}$ and $I^-$), and vacancies at grain boundaries. Collectively, these elements lead to a decrease in the energetic stability of the film surface, thereby initiating the degradation process. Identifying the cause of the degradation, they proposed and demonstrated that surface passivation could serve as a viable future approach for enhancing the thermal stability of PSC devices as shown in Figure 13(b).

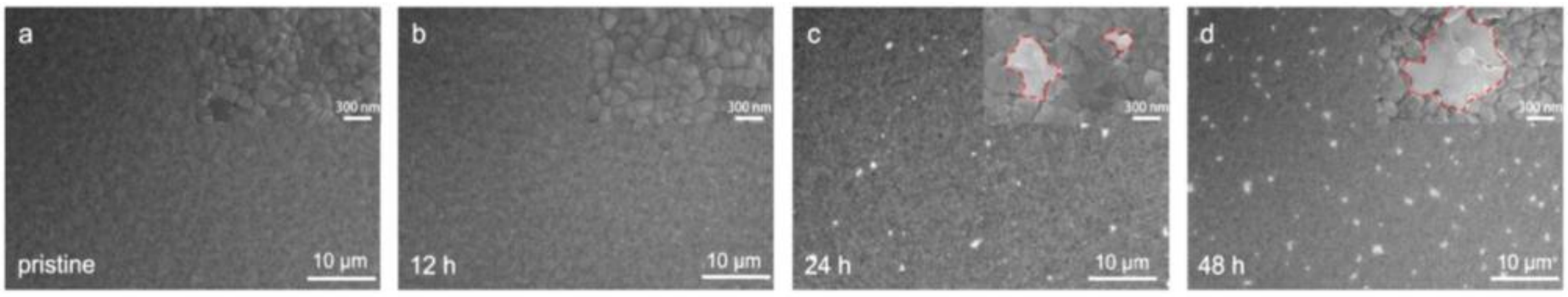

Figure 12. SEM images of perovskite film at 85oc for pristine, 12h, 24h, 48h [172]

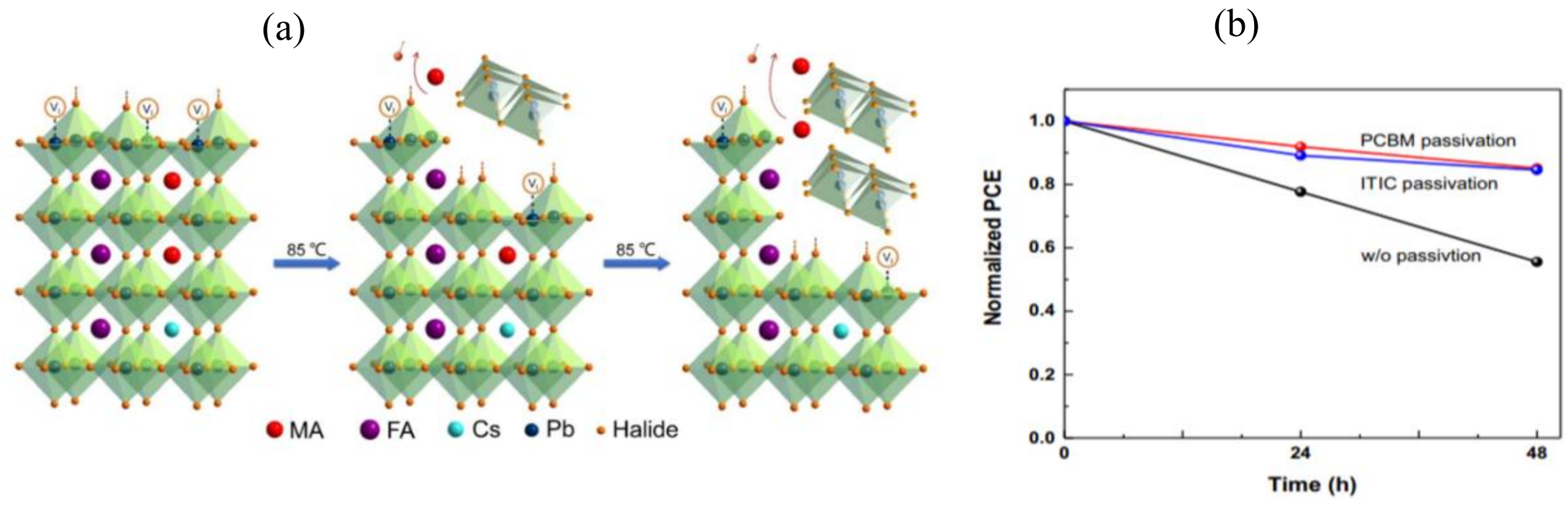

Figure 13. (a) layer-by-layer fashion degradation and (b) effect of passivation in PSC devices [172].

Tan and his team [173] elevated the temperature to 150°C and conducted an analysis of two PSC devices, $Cs_{0.17}FA_{0.83}Pb(I_{0.83}Br_{0.17})_3$ and $Cs_{0.05}(MA_{0.17}FA_{0.83})_{0.95}Pb(I_{0.83}Br_{0.17})_3$, under atmospheric conditions. The XRD analysis revealed that the decomposition kinetics adhered to zeroth order kinetics and demonstrated Arrhenius behaviour, with an activation energy of approximately 0.66 eV for $Cs_{0.17}FA_{0.83}Pb(I_{0.83}Br_{0.17})_3$ and 0.76 eV for $Cs_{0.05}(MA_{0.17}FA_{0.83})_{0.95}Pb(I_{0.83}Br_{0.17})_3$. Upon further investigation, it was determined that the device incorporating the $MA^+$ cation exhibited lower stability compared to the device that excluded the $MA^+$ cation. The device containing $MA^+$ exhibited a two-step decomposition process as shown in Figure 14. Initially, the decomposition of the $MA^+$ cation occurred rapidly, followed by the slower decomposition of $FA^+$.

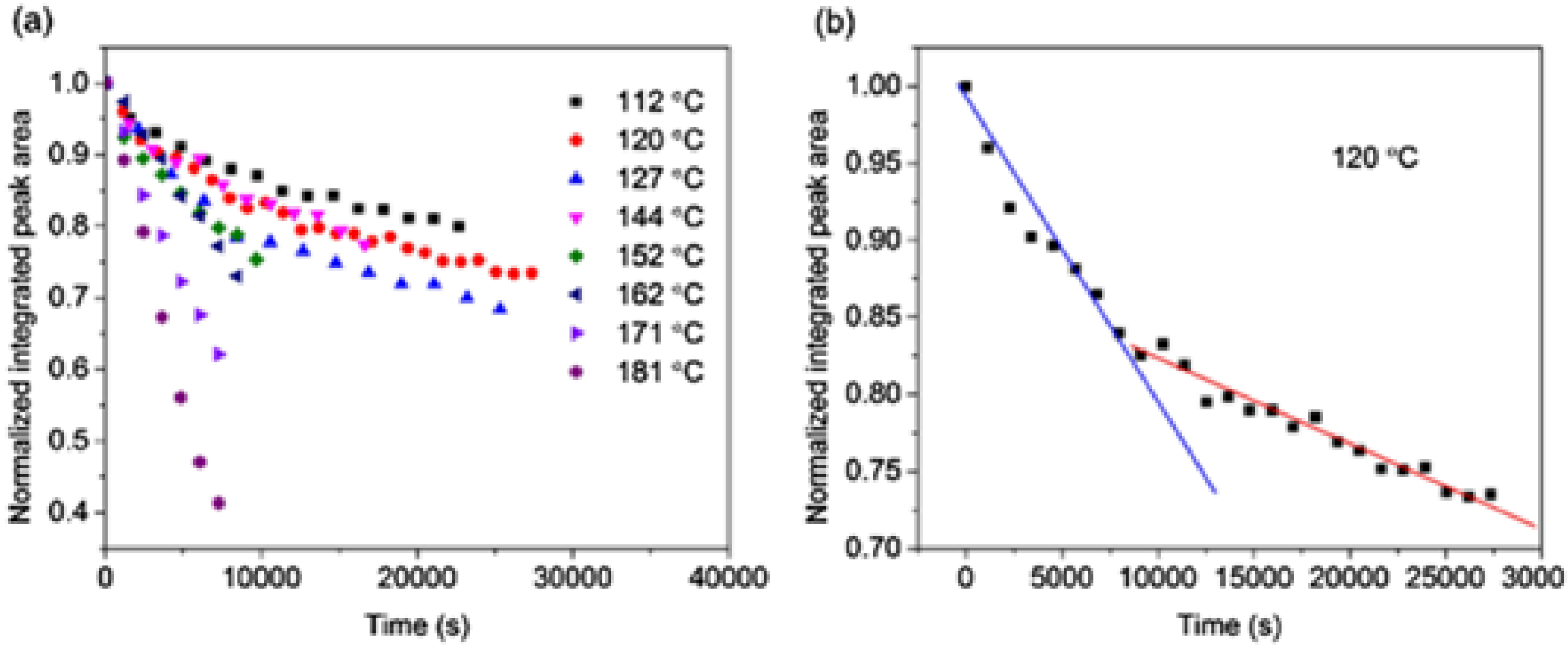

Figure 14. (a) kinetics of $Cs_{0.05}(MA_{0.17}FA_{0.83})_{0.95}Pb(I_{0.83}Br_{0.17})_3$ at different temperatures. (b) two stage degradation model of $Cs_{0.05}(MA_{0.17}FA_{0.83})_{0.95}Pb(I_{0.83}Br_{0.17})$ [173].

Dalal et al. [174] concentrated their efforts on developing a resilient PSC device designed for high-temperature environments. Consequently, the device FTO/In:CdS (In doped CdS)/ $CsPbI_xBr_{(3-x)}$/ P3HT/Au was deployed and subjected to thermal annealing at 200°C for durations of 24 hours and 72 hours within a nitrogen ambient environment, utilizing the vapor deposition method. After 24 hours of thermal aging, the device demonstrated no signs of compositional or phase degradation. Furthermore, when subjected to testing for 72 hours, the device maintained stable performance without any decline in power conversion efficiency or external quantum efficiency. The author indicates that the thermal stability and a bandgap of 1.87 eV position this material as a promising candidate for tandem solar cells when paired with Si or $Cu(Ga, In)Se_2$ in ambient conditions.

**Hole Transport Layer (HTL)**

The hole transport layer (HTL) is an important component of perovskite solar cells (PSCs) due to its ability to efficiently extract and transport positive charge carriers. Thermal stability is a major problem for HTL materials such as organic Spiro-OMeTAD. Thermal stress can induce chemical and structural degradation at a given temperature, impacting charge mobility and device performance.

Relevant research has indicated that Spiro-OMeTAD starts to crystallize at temperatures near 100°C, diminishing its conductivity and impacting the solar cell's efficiency. Further, with prolonged heating, chemical dopants within the HTL can undergo de-doping reactions, leading to charge carrier loss and causing deterioration in performance. There is a need to explore the thermal degradation mechanisms of HTLs in order to obtain more stable materials that will be capable of enhancing the long-term stability of PSCs [175–178].

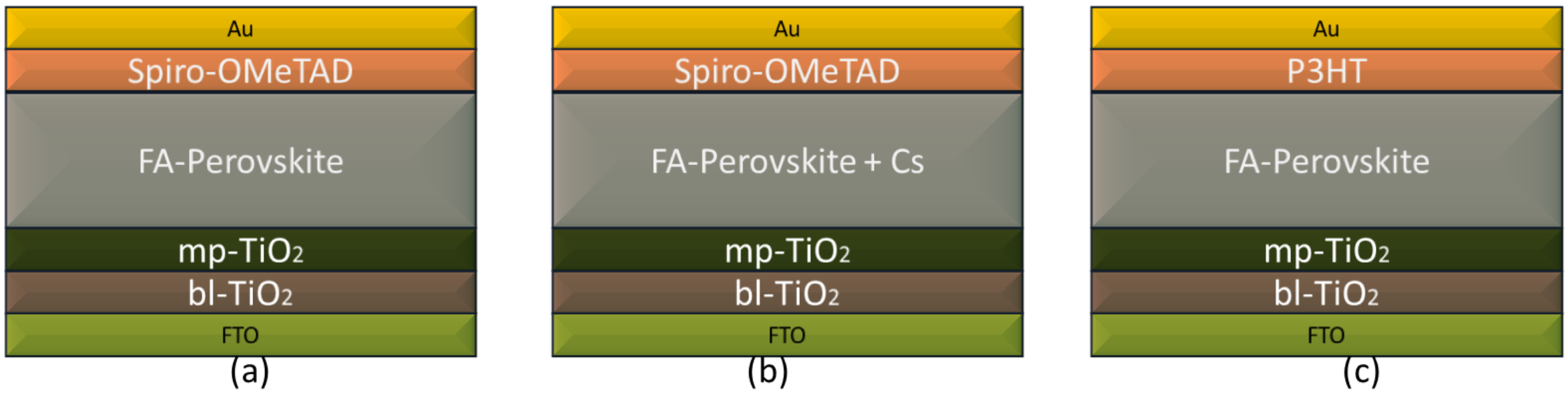

Figure 15: Schematic structure of perovskite solar cells explored in this study, (a) FA perovskite layer (spiro-OMeTAD as HTM with additives), (b) FA perovskite layer including $Cs^+$ (spiro-OMeTAD as HTM with additives), (c) FA perovskite layer (P3HT as HTM without additives).

A comprehensive investigation was carried out regarding the thermal stability of the hole transport layer (HTL) by Zheng et al. [179]. The degradation behaviour was examined under various temperature conditions. Three PSC samples were fabricated utilizing a perovskite composition of $(FAPbI_3)_{0.85}(MAPbBr_3)_{0.15}$, referred to as 'Fa' in Figure 15. The samples underwent thermal ageing tests in a meticulously controlled dark environment, maintaining relative humidity levels below 10%. The samples underwent exposure to three specific temperatures: 20°C, 60°C, and 85°C while being continuously illuminated by a 100 mW/cm² AM1.5G solar simulator. The degradation of PSC attributed to HTL was assessed by tracking normalized efficiency over a period of time [179]. Further investigations conducted by Zheng and his team[179] revealed that the instability of Spiro-OMeTAD, a widely used HTL material, primarily stems from the dopants Li-TFSI and t-BP, which exacerbate degradation when subjected to heat stress. To address this issue, they employed Spiro-OMeTAD without additives and evaluated its stability against that of P3HT, another hole transport layer material also used without additives. The findings indicated that P3HT showed enhanced thermal stability, positioning it as a more favourable option for high-temperature applications in PSCs. The schematic structures of the examined perovskite solar cells are presented in Figure 15, highlighting the analysis of various HTL and PSC layer configurations in relation to their thermal resilienceas illustrated in Table 6.

Table6. Thermal stabilityof different HTL and PSC layer configuration

| **Devices** | **Temperature (°C)** | | | **Experimental curve** |
|---|---|---|---|---|
| | 20 | 60 | 85 | |
| Unsealed FA PSC with spiro-OMeTAD as HTM with additives | 98% of initial PCE maintained after 300h storage | Curve declined to 70% of initial PCE after 300h storage | It had performed very poor after 100hof storage | 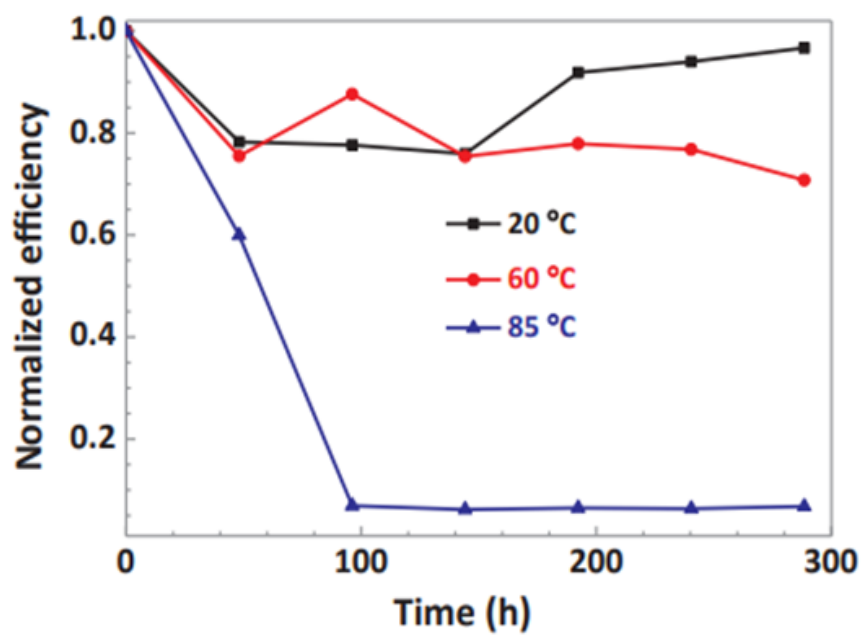 |
| Sealed FA PSC with spiro-OMeTAD as HTM with additives | PCE enhanced to over 100% after 300h storage | 75% of initial PCE had shown after 300h storage | PCE declined to 20% after 300h storage | 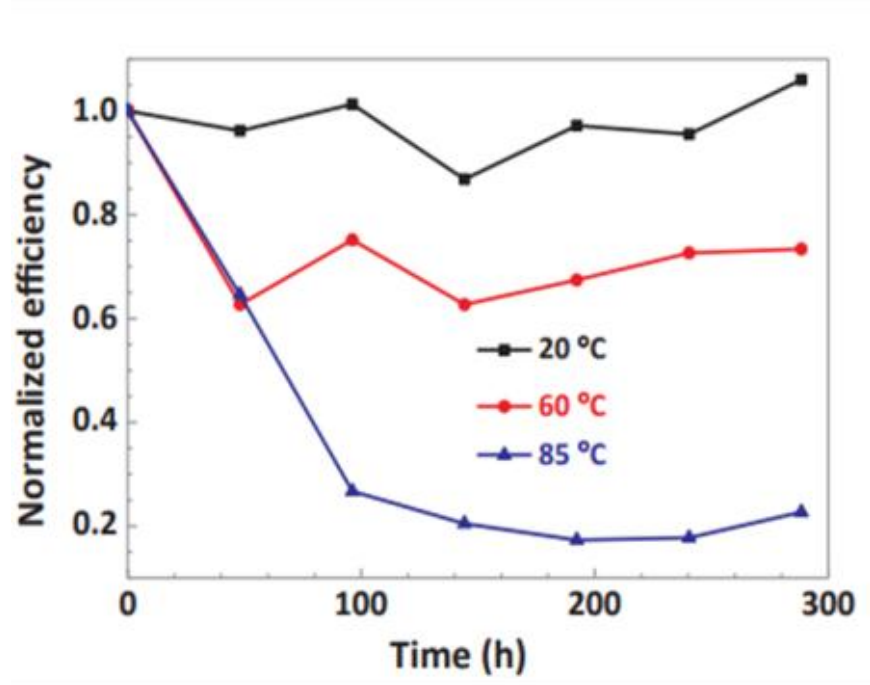 |
| Unsealed FA PSC with 10% $Cs^+$ cation with spiro-OMeTAD as HTM with additives | PCE decreased by 20% after 600h storage | PCE declined by 30% after 600h storage | PCE dropped by 100% after 600h storage | 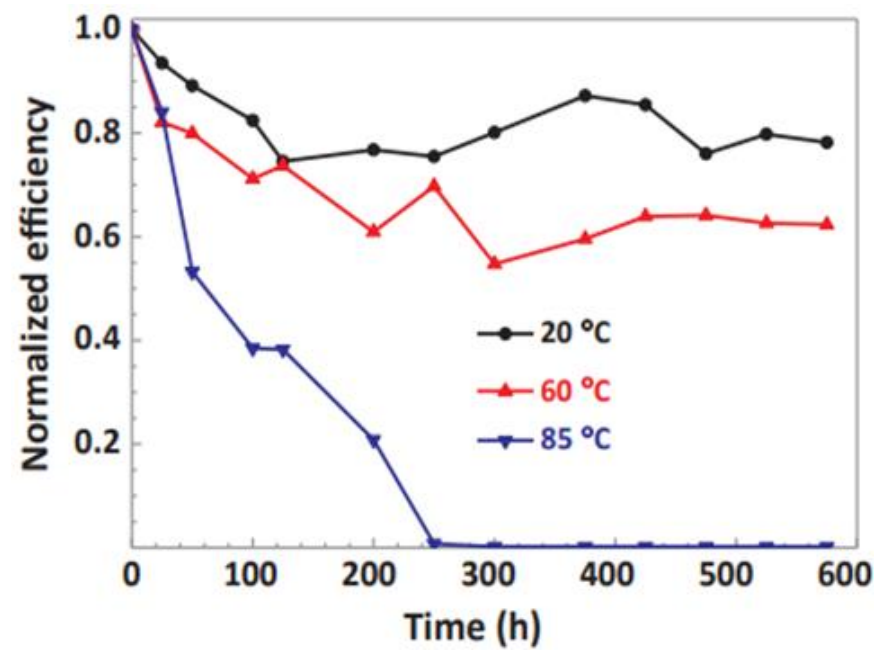 |
| Sealed FA PSC with 10% $Cs^+$ cation with spiro-OMeTAD as HTM with additives | PCE decreased by 20% after 600h storage | PCE decreased to 65% of initial PCE after 600h storage | PCE dropped by 80% after 600h storage | 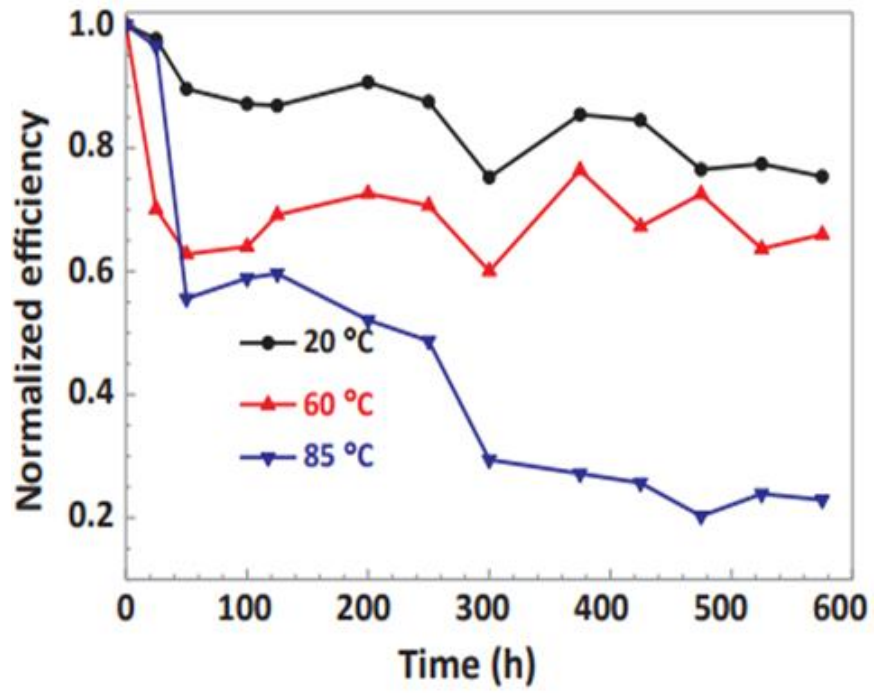 |
| Unsealed FA PSC with spiro-OMeTAD as HTM without additives | PCE enhanced over 100% after 300h storage up to 800h storage even it can be seen curve still going upward | PCE remained above 80% of initial value after 800h storage | PCE decreased to 0% after 800h storage | 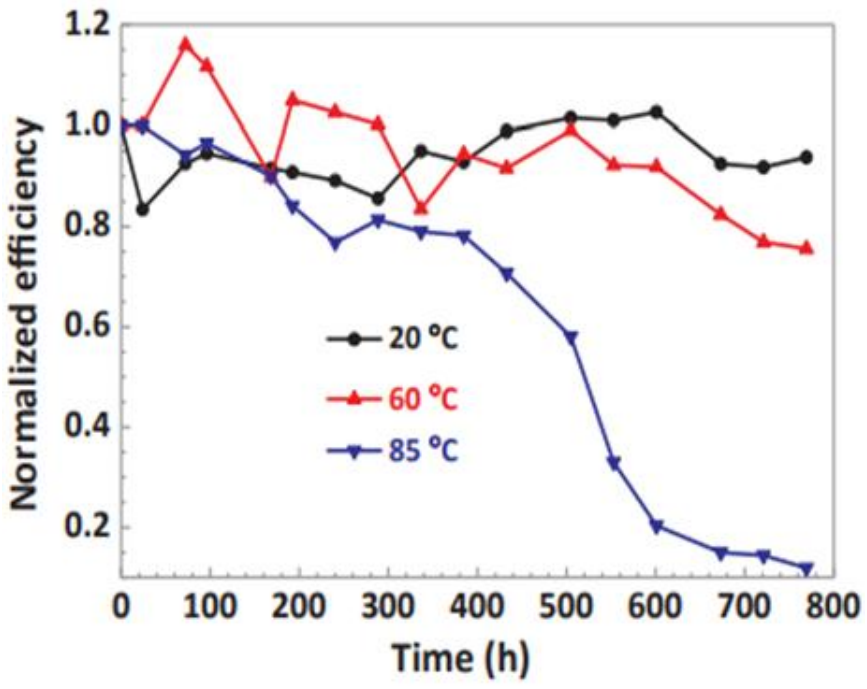  |
| Sealed FA PSC with spiro-OMeTAD as HTM without additives | PCE enhanced over 100% after 300h storage up to 800h storage even it can be seen curve still going upward | PCE remained above 80% of initial value after 800h storage | PCE decreased to 20% after 800h storage | 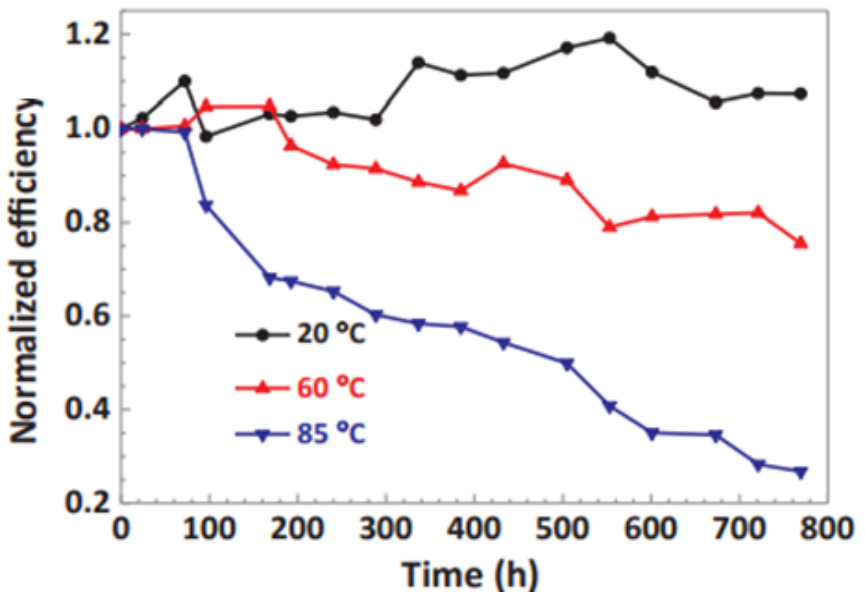  |
| Unsealed FA PSC with P3HT as HTM without additives | In all cases PCE had shown similar trends of retaining 80% of initial value after 800h storage | | | 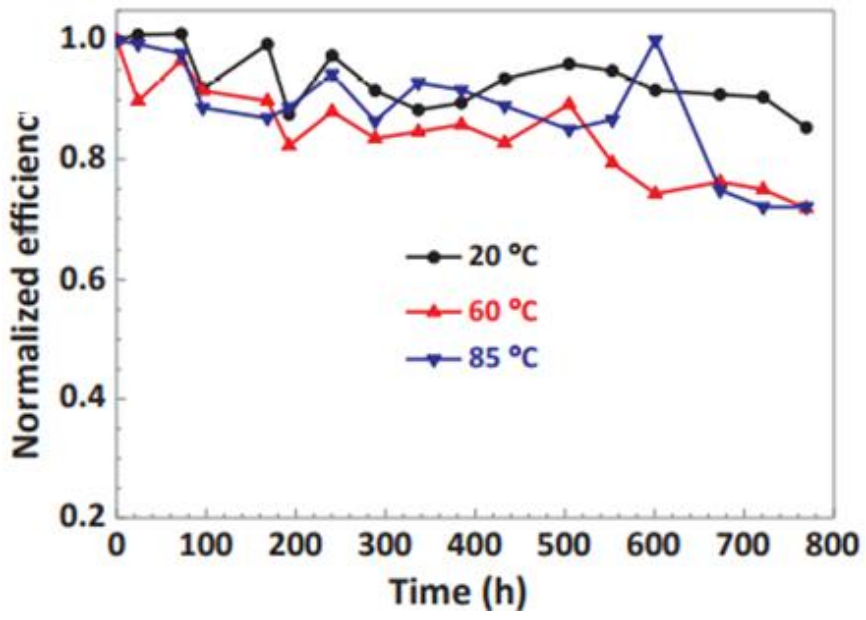  |
| Sealed FA PSC with P3HT as HTM without additives | In all cases PCE had shown similar trends of retaining 80% of initial value after 800h storage | Sealed FA PSC with P3HT as HTM without additives | In all cases PCE had shown similar trends of retaining 80% of initial value after 800h storage | 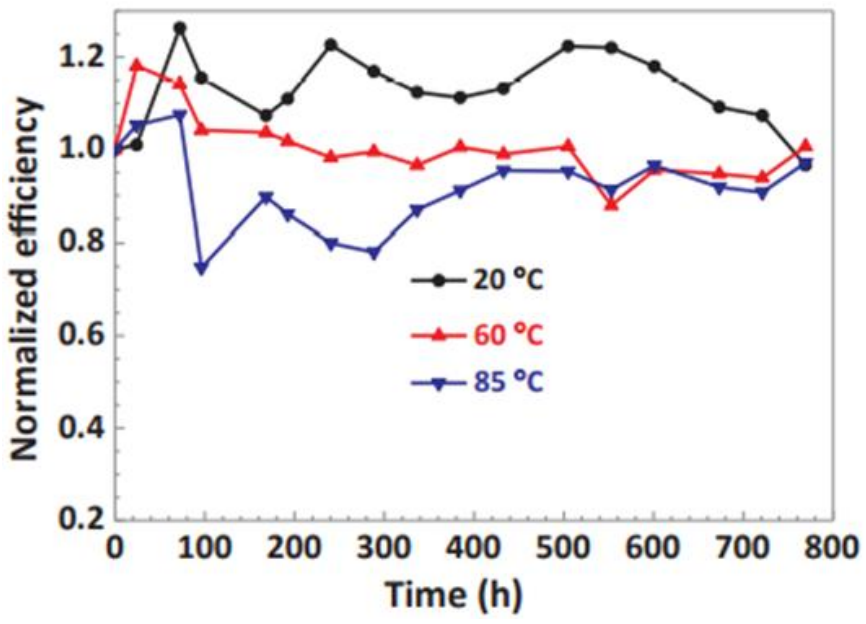  |

A.K. Jena and his team [180] indicate that the PSC device experiences thermal degradation as a result of HTL, leading to instability. In order to substantiate their hypothesis, they carried out a thermal treatment experiment on MA PSC (planar) utilizing spiro-OMeTAD as the hole transport material at various temperatures: 60°C, 80°C, 100°C, and 120°C for a duration of 1 hour in both dry conditions (less than 10% relative humidity) and ambient conditions (relative humidity between 30% and 50%). The findings indicated that the perovskite film remained stable up to 80°C, with minimal degradation observed at 100°C and 120°C. However, the spiro-OMeTAD experienced significant morphological deformation; substantial voids developed within the HTL at elevated temperatures, specifically in areas where the HTM was covered with the Au film. The researchers deliberately subjected the perovskite film to elevated temperatures to induce the formation of $PbI_2$, observing that the resulting power conversion efficiency (PCE) remained consistent at 16-17%, similar to that of the pristine perovskite without $PbI_2$. Another finding from their experiment was that the device with excess $PbI_2$ remaining from the precursor exhibited similar or slightly improved performance compared to the device without $PbI_2$. To investigate the cause of the degradation in the PSC device, they repeated the experiment incorporating the HTL. They observed that degradation occurred solely in the presence of the HTL, regardless of whether the Au film was applied to the spiro-OMeTAD. Observations indicated the formation of significant voids during the heating of the device, regardless of the presence of Au coating on spiro-OMeTAD. In both scenarios, the degradation occurred at a comparable rate [180].

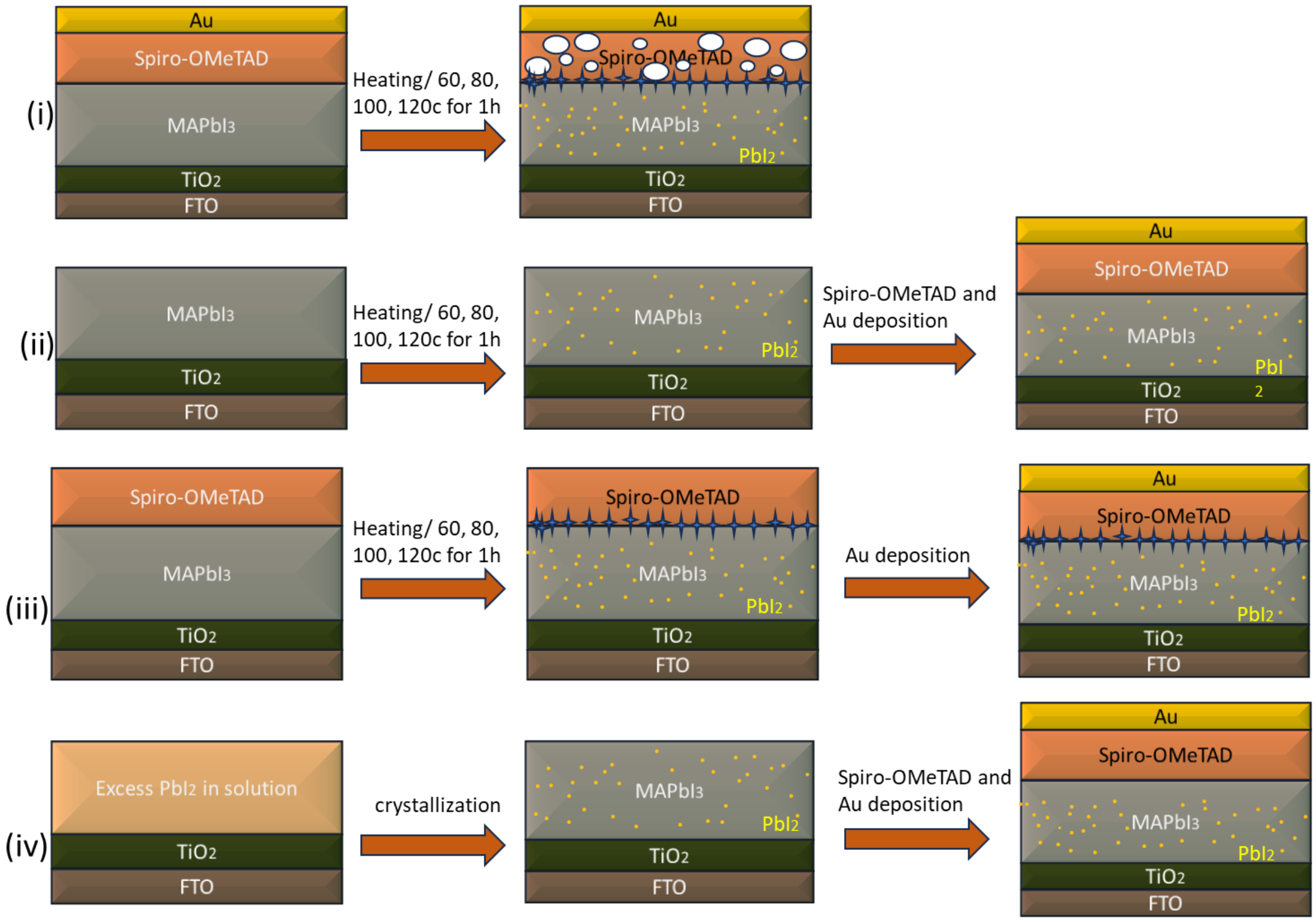

Figure 16. Different directions of coating different layers in PSC device

Consequently, the final conclusion indicated that thermal annealing resulted in certain chemical modifications at the perovskite/spiro-OMeTAD interface. This experiment outlines various methods for synthesizing $PbI_2$ and discusses the modifications in the layers induced by temperature variations as illustrated in Figure 16. It is evident from the preceding discussion that further investigation into HTL is essential for enhancing thermal stability. Consequently, M.R. Leyden and his team [181] conducted an analysis on the thermal degradation of spiro-OMeTAD (incorporating additives Li-TFSI and t-BP) in the context of PSC $Cs_{0.05}(FA_{0.85}MA_{0.15})_{0.95}Pb(I_{0.85}Br_{0.15})_3$. The experiment was carried out at two distinct temperatures, 25°C and 85°C, over a duration of 20 hours, and it was observed that crystallization did not occur. The observed phenomenon may be attributed to the chosen temperature being below the Glass Transition Temperature (tg). Following the acquisition of this result, the inquiry emerged regarding the rapid degradation of PSC at 85°C. To explore the underlying cause, they assessed the HOMO layers of spiro-OMeTAD at 25°C (under dark conditions) and 85°C. The findings indicated that at 25°C, there was no alteration in the HOMO layer value, remaining at -5.3 eV. However, at 85°C, the HOMO layer value increased to -5.8 eV, surpassing the HOMO layer of perovskite, which is -5.5 eV. Thus, variations in HOMO energy levels introduced barriers during the hole extraction process as shown in Figure 17, resulting in reduced stability and lower power conversion efficiency.

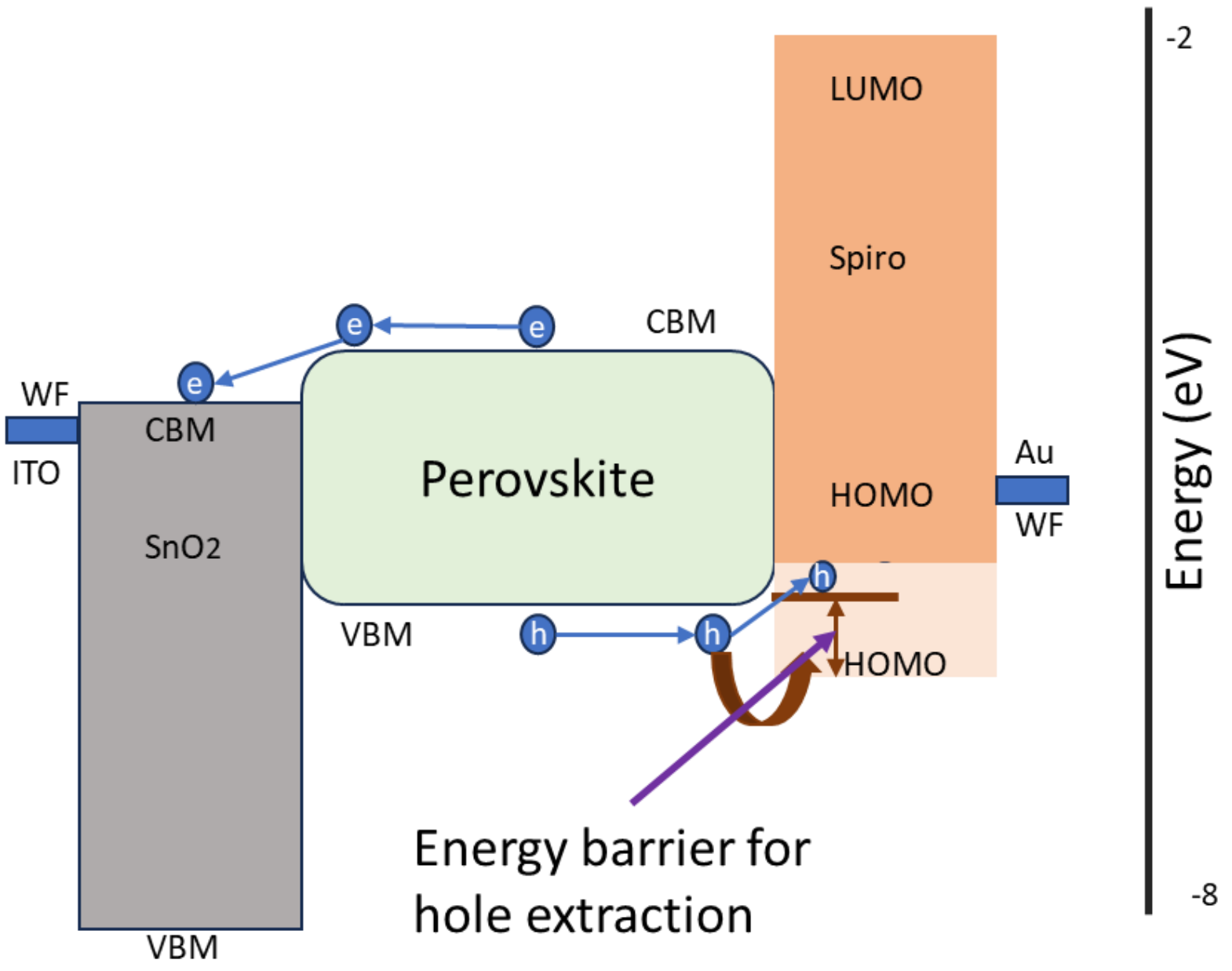

Figure 17. Energy level diagram of PSC at 85 °C

To comprehend the fundamental cause of spiro-OMeTAD degradation, Song et al. [182] performed various experiments and proposed hypotheses regarding the degradation of spiro-OMeTAD at elevated temperatures (85°C). The articles indicate that the additives used in spiro-OMeTAD contribute to degradation. Among the two primary additives, Li-TFSI and t-BP, which one is responsible for the degradation? In order to address this question, they conducted an analysis of 48 n-i-p PSC devices featuring a structure composed of glass/150 nm indium tin oxide (ITO)/20 nm $SnO_2$/3 nm phenyl-C61-butyric acid methyl ester (PCBM)/450 nm perovskite/250 nm spiro-OMeTAD/80 nm Au. The experiment was carried out in a dark nitrogen environment for 1032 hours at 85 degrees Celsius. From a total of 48 devices, they developed two sets, each containing 24 devices. Both additives were included in one set, while only Li-TFSI was added in another. The results of the experiment are presented below. Based on the data presented, it is evident that the use of both additives resulted in a significant decrease in PCE after 200 hours of thermal stressing. In contrast, when only the Li-TFSI additive was utilized, the initial PCE was slightly lower, but it gradually increased and remained relatively stable up to 1032 hours of thermal stressing. The POM and SEM images presented in figures 18 b, b1, and b2 indicate that when solely the Li-TFSI additive is utilized in the spiro-OMeTAD film, it retains an amorphous state, accompanied by some aggregates and voids. The images presented in Figures 18 c, c1, c2, c3, and c4 collectively indicate that the area capped with Au has undergone greater crystallization, resulting in the formation of empty trenches with depths reaching several micrometres. In contrast, the area without Au capping has retained an amorphous state. The data unequivocally demonstrates that the coexistence of t-BP and Au will result in the development of extensive domain crystallization and micro-meter-sized trenches. Their hypothesis suggests that the amorphous state of the non-capped area of Au may be attributed to the rapid evaporation of t-BP, which raises the film's Tg above 85°C. This condition decreases the likelihood of new crystal formation, leaving the non-capped area in an amorphous state. In contrast, while t-BP also evaporates in the capped area, it does so at a slower rate. The potential explanations are as follows: Firstly, the region is covered with Au, and secondly, pyridine has been adsorbed onto Au [183,184]. Three strategies were proposed to inhibit this phase transition and maintain the amorphous state: the first involves inserting an additional layer between the spiro-OMeTAD and the Au layer to prevent direct contact between the two. Secondly, minimize residual t-BP by preheating the bare spiro-OMeTAD film prior to the deposition of subsequent layers. Third, integrate both the first and second strategies. In addition to these findings, two factors were identified that contribute to the increase in series resistance (Rs) and subsequently lead to a decrease in the PCE of PSC: firstly, the contact area between the perovskite film and spiro-OMeTAD was diminished due to crystallization. Secondly, the phase transition of the spiro-OMeTADfilm has the potential to alter both the resistivity of the film itself and the contact resistance between neighbouring layers. This ultimately led to an increase in the Rs.

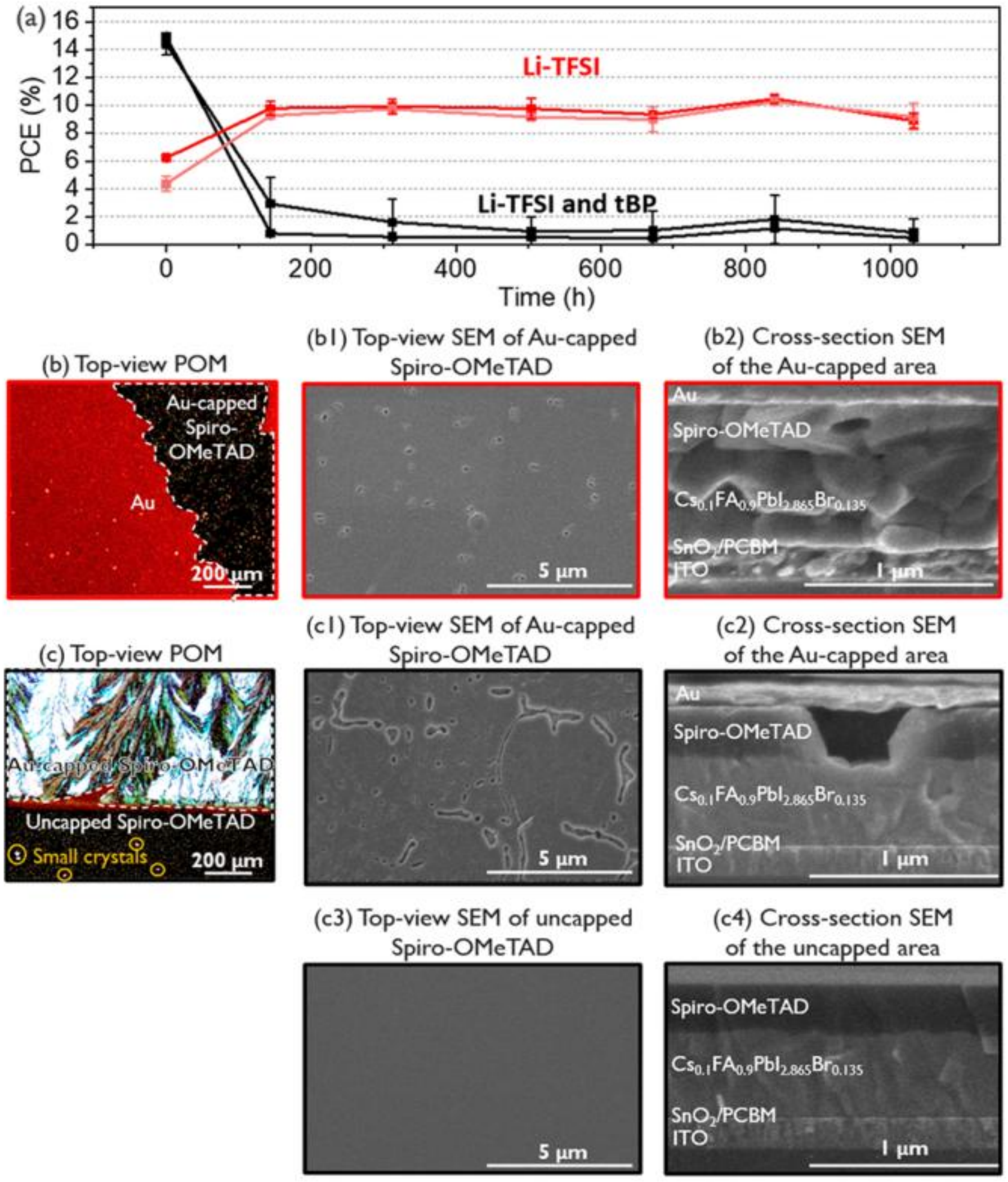

Figure 18. (a) PCE stability plots at 85°C (mean values from 12 PSCs). (b, b1, b2) Top-view POM, top-view SEM, and cross-sectional SEM images of stressed PSCs with Li-TFSI additive; white dashed area indicates initial Au coverage. (c, c1, c2) Corresponding images for PSCs with Li-TFSI and t-BP additives; dashed areas initially Au-covered; yellow circles mark small crystals. (c3, c4) SEM images of uncapped regions in PSCs containing Li-TFSI and t-BP.

**Dopant free HTM**

Having examined the impact of dopants on the degradation of PSC, a question naturally arises: why not develop an HTM for PSC that is free of dopants? Numerous studies have been conducted, and a significant number of articles have been published, paving the way for dopant-free HTM to enter the market. Before delving into the examples of dopant-free HTM, it is essential to comprehend the role of dopants. Notably, spiro-OMeTAD is widely recognized as the most effective hole transport material, attributed to its high efficiency and excellent hole mobility, among other properties. This presents an ideal alignment for PSC HTM; however, the device is experiencing delays attributed to the instability associated with HTM. Spiro-OMeTAD primarily consists of two dopants: Li-TFSI and t-BP. The electrochemical properties of TiO2 are influenced by the control exerted by Li+ ions via absorption and intercalation processes. Consequently, the introduction of Li-TFSI enhances the hole mobility of spiro-OMeTAD by improving disorder and broadening the density of states, while also partially passivating deep Coulomb traps, thereby reducing the charge transfer barrier[185,186]. Additionally, t-BP serves as a passivating agent for uncoordinated titanium atoms, leading to a decrease in charge recombination. Furthermore, both t-BP and Li-TFSI enhance the miscibility of spiro-OMeTAD[187]. However, alongside these advantages, these dopants also present the drawback of hydrophilic nature, which contributes to the degradation of devices. Recent studies indicate that the addition of Li-TFSI and tBP dopants to Spiro-OMeTAD enhances its hydrophilic characteristics, resulting in faster degradation in devices. The hygroscopic behaviour can be explained by the salt-like properties of the Li-TFSI dopants. As a result, several approaches utilizing undoped hole transport materials have been recently suggested to improve the stability of perovskite solar cells (PSCs) [188–190]. Numerous attempts have been undertaken to replace hygroscopic Li-TFSI dopants with more stable additives, including Cu(II) pyridine, F4-TCNQ, Co(III), CuSCN, and CuI. Nonetheless, these alternatives frequently entail intricate and expensive synthesis procedures and require preparation at elevated purity levels to ensure effective PSCs[191–194]. To effectively compete with spiro-OMeTAD, dopant-free HTMs must exhibit superior characteristics, including high purity and yields, elevated hole mobilities, and suitable bandgap properties for the HOMO/LUMO layers. It must exhibit stability, affordability, and efficacy as a high-throughput method. Sonar et al. have provided a compilation of dopant-free HTMs [195].

**Inorganic**

Setting aside all the organic molecules and polymers, inorganic materials exhibited an intriguing behaviour as hole transport materials in perovskite solar cell devices. A multitude of inorganic HTMs has been reported, with some demonstrating lower stability and others exhibiting competitive behaviour with spiro-OMeTAD. I am providing a limited number of examples of inorganic HTMs. In 2016, sputtered Cu2O was utilized as an HTM in the n-i-p heterojunction device structured with FTO/c-TiO2/MAPbI3-xClx/sputtered Cu2O/Au, demonstrating stability for one month while maintaining over 90% of its initial efficiency of 8.93% under ambient conditions of 25°C and RH = 28.2%. The experience was both thrilling and demanding when compared to traditional spiro-OMeTAD, particularly given that its lifespan was only 12 days at that time [195]. Guo et al., 2018 [196] achieved a highly promising result regarding stability with the use of an organic/inorganic hybrid HTL of FBT-BH4/CuOx in an environment with 70-80% relative humidity. The device, FTO/$SnO_2$/PCBM/$MAPbI_3$/FBT-BH4/$CuO_x$/Au, maintained 90% of its initial efficiency of 18.85% over a period of 500 hours without encapsulation.

Ito et al. [197] utilized a different inorganic hole transport material in an n-i-p device configuration, consisting of TCO/cp-$TiO_2$/mp-$TiO_2$/$MAPbI_3$/CuSCN/Au. In 2014, they initially set up this device using CuSCN HTM, achieving a power conversion efficiency of 4.85% as a result. CuSCN is now considered suitable for HTM because of its impressive hole mobility ranging from $10^{-2}$ to $10^{-1}$ $cm^2$ $V^{-1}$ $s^{-1}$. Jung and his colleagues [198] developed thermally stable perovskite solar cells utilizing CuSCN as a hole transport material, combined with the absorber composition of $(FAPbI_3)_{0.85}(MAPbBr_3)_{0.15}$. The report presented a significant outcome of 18% PCE, accompanied by a Jsc of 23.1 mA/$cm^2$, and demonstrated a retention of 60% of the initial PCE following thermal stress at 125°C, with 40% relative humidity in the ambient air environment. You et al., 2015 [199] demonstrated remarkable stability of 60 days in ambient conditions, maintaining 80% of the initial PCE of the device. $NiO_x$ was integrated as the hole transport material. In 2017, an additional experiment was conducted by Wu et al [200] involving a combination of mixed inorganic HTL and a conventional absorber layer. In that experiment, $Li_{0.05}Mg_{0.15}Ni_{0.8}O$ was selected as the hole transport layer, while $MAPbI_3$ served as the absorber layer. The device was exposed to AM1.5 irradiance for 1000 hours, during which it maintained 90% of its initial power conversion efficiency. During the testing conducted in a dark environment at 85°C for 500 hours, the device demonstrated a retention of 80% of its initial power conversion efficiency.

## 6. Perovskite Stability Enhancement

Long-term dependability and commercialization of perovskite solar cells (PSCs) depend on their thermal stability being improved. Several methods have been investigated to mitigate temperature-induced degradation: compositional engineering and interface changes among others. Among these techniques, the inclusion of appropriate additives has become clear as a successful means of reducing grain boundary defects, controlling crystallization, and improving the general thermal stability of PSCs. By stabilizing the perovskite structure against thermal stresses, several compounds have been found recently to greatly improve device performance.

A bifunctional additive featuring multiple Lewis base groups is biuret, which has the capability to regulate the crystallization process and reduce defects at grain boundaries. The findings presented by Shi et al., 2020 [201] indicated that the incorporation of a biuret additive in the $MAPbI_3$ device resulted in the retention of 94% of its initial power conversion efficiency (21.1%) following a 12-day thermal annealing process at 85°C in a nitrogen atmosphere. Biuret established a cross-link chemical bond involving uncoordinated $Pb^{+2}$ and iodide derived from iodoplumbate. The cross-link chemical bond enhanced the thermal stability. Additionally, the findings indicated a reduction in trap state density, an increase in grain size, and a suppression of trap-assisted non-radiative recombination. In a study conducted by Wu et al., 2019 [202], the additive bilateral alkyl amine (BAA) was analyzed, yielding a $V_{oc}$ of 0.35V under AM1.5G illumination and a PCE of 22.6% under 0.3 sun. The linking alkyl chain was exposed, effectively addressing the defect present on the perovskite surface. This led to an enhancement in thermal stability. Additional additives include poly-vinylpyridine (PVP) [203] and guanidine thiocyanate [204].

Smith et al., 2014 [205] demonstrated that 2-D perovskite has the potential to serve as a viable option for enhanced thermal stability. Consequently, the researchers investigated $(PEA)_2MA_2[Pb_3I_{10}]$ perovskite, revealing a notably low power conversion efficiency of 4.73%. However, it was observed that the material maintained over 95% of its initial efficiency following 200 hours of thermal stress at 85°C in a nitrogen atmosphere. Yao et al. (2015)[206] combined the additives with 2D perovskite to improve stability, achieving a power conversion efficiency exceeding 15% by utilizing polymer ammonium PEI.HI additive through a two-step spin-coating technique. Davy et al., 2021 [207] and Muhammed et al., 2015 [208] concluded that two-dimensional perovskite enhances power conversion efficiency and thermal stability, as it effectively passivates materials, minimizes surface defects, and mitigates recombination at the interface between the perovskite and the hole transport layer. The incorporation of additives into this perovskite may enhance the interface contact characteristics and mitigate the surface defects of the absorbing layer. The mixed dimensional perovskite, 3-D/2-D $MAPbI_3$-$PEA_2Pb_2I_4$, was investigated by Bai et al. in 2017 [209], revealing an improvement in thermal stability, with the device maintaining 60% of its initial power conversion efficiency at 83°C for a duration exceeding 30 days. The findings suggest that the inhibition of $I^-$ ion migration from the absorber layer to the $PC_{61}BM$ layer could be the underlying cause of this stable outcome. The study revealed that the graded surface of the device facilitated the formation of a barrier layer at the interface between the absorber layer and $PC_{61}BM$, effectively inhibiting the diffusion of $I^-$ ions.

Grancini et al., 2017 [210]developed a hole transport material-free perovskite solar cell device $[HOOC\ (CH_2)_4NH_3]_2PbI_4/MAPbI_3$, achieving a power conversion efficiency of 11.2% with no loss under AM1.5G illumination at 55 °C, over an area of 100 $cm^2$, demonstrating durability for 12,000 hours. Zhang et al., 2017a, 2017b[211,212], utilized a 3-D/2-D/0-D approach with the $CsPbBrI_2$ device, demonstrating that it maintained 80% of its initial PCE after being exposed to 60°C for 450 hours. This improvement can be attributed to enhanced hole extraction, increased conduction efficiency, and minimized charge recombination losses.Upon analyzing the articles, several questions arise regarding the presence of only three types of cations at the cation site of perovskite. The exploration of multication perovskites has primarily been limited to three common cations ($Cs^+$, $FA^+$, $MA^+$), but recent studies[213–216] suggest that incorporating four or five different cations (e.g., $Cs^+$, $FA^+$, $MA^+$ (minimal), $Rb^+$, $K^+$) could further enhance thermal stability and structural robustness. Investigations into Pb/Sn mixtures as partial substitutes for lead[217–222] demonstrate their ability to optimize stability and efficiency, addressing concerns about toxicity while maintaining perovskite performance. Furthermore, studies have explored the partial substitution of halide ions with oxides, nitrides, or carbon-based ions[223–229], revealing promising pathways to mitigate degradation and enhance PSC longevity.

Since the inception of spiro-OMeTAD, a traditional small-molecule-based hole transport material, it has maintained its dominance even after two decades. Ahmad et al. [230] indicate that the extensive use of spiro-OMeTAD is driven by several factors, including complex synthetic procedures, the necessity for high purity, sensitivity to UV light, low hole mobility, and the potential for crystallization under thermal stress. Numerous alternatives to spiro-OMeTAD have been documented, including small conducting polymers, inorganic p-type semiconductors [231–235]. The performance of these HTMs was satisfactory, yet it did not reach the level of enthusiasm generated by spiro-OMeTAD. Ahmad et al. [230] outlined several essential properties that a successful HTM must exhibit the following properties as given in Table 7.

Table 7. Properties for Hole Transport Material

| Property | Value | Reason |
|---|---|---|
| Hole Mobility | $10^{-3}$ to $10^{-4}$ $cm^2V^{-1}s^{-1}$ | Enhancement of structural stability through doping |
| Glass transition Temperature ($T_g$) | >120c | Prevents structural degradation under prolonged sunlight exposure. |
| Optical Transparency | transparent in the visible and near IR spectrum | Allows red light absorption in PSCs |
| Pinhole-Free Structure | Must be maintained | Direct contact between perovskite and metallic conductor is inhibited |
| HTM Thickness Optimization | Neither too thick nor too thin | Though thick layer prevents pinholes but excessive thickness increases **resistance**. |
| Additional Layer (MoOx, VOx, etc.) | Optional | Direct contact between perovskite and metallic conductor may be inhibited |

Exploring a novel alternative additive to t-BP, given that this article exclusively examines the thermal stability of PSC, or refining the doping and processing parameters of HTMs could represent a promising avenue for future investigation. In addition, numerous reports exist that provide a compilation of dopant-free HTMs [195] and small organic molecule-based HTMs [230].

## 7. Conclusion

Perovskite solar cells (PSCs) have emerged as promising next-generation photovoltaic devices due to their impressive power conversion efficiencies, cost-effective fabrication processes, flexibility, and lightweight properties. However, their large-scale commercialization remains constrained primarily by significant thermal instability issues. Elevated operating temperatures accelerate the decomposition of perovskite materials, particularly methylammonium lead iodide ($MAPbI_3$), leading to degradation into $PbI_2$, $CH_3I$, and $NH_3$, severely impacting device efficiency. Additionally, the organic hole transport layer (HTL), typically spiro-OMeTAD doped with additives such as Li-TFSI and tBP, suffers pronounced thermal degradation due to moisture absorption, increased electrical resistance, and morphological alterations at elevated temperatures. Therefore, addressing these thermal stability concerns is critical for enabling practical and durable PSC applications.

Compositional engineering of perovskite layers through cation and anion substitution, such as incorporating cesium ($Cs^+$), formamidinium ($FA^+$), or rubidium ($Rb^+$), has effectively enhanced thermal stability. Such mixed-cation perovskite compositions exhibit improved resilience against thermal decomposition, reduced volatility of organic components, and stable crystal phases, providing longer operational lifetimes. Moreover, additive engineering approaches using materials like biuret, polyvinylpyridine (PVP), and guanidine thiocyanate have demonstrated significant improvement in thermal resistance by suppressing non-radiative recombination, reducing trap-state density, and enhancing grain size. Similarly, two-dimensional (2D) perovskite structures and mixed-dimensional (2D/3D) perovskite systems have proven highly effective in mitigating ion migration and defect formation, contributing substantially to enhanced thermal and operational stability[236].

Future developments should focus on further optimizing the PSC structure through advanced interface engineering, exploring dopant-free and inorganic HTLs such as CuSCN and $NiO_x$, and improving encapsulation techniques to prevent thermal and environmental degradation. Such advancements would significantly improve PSC durability, paving the way for practical applications including rooftop photovoltaic systems, flexible electronics, portable power generation systems, and integration into building-integrated photovoltaics (BIPV). Achieving greater thermal stability and long-term reliability will enable PSC technology to effectively compete with conventional silicon-based solar cells, thus fostering broader commercialization and contributing substantially to global renewable energy initiatives.

**References:**

[1] El-Araby R. Biofuel production: exploring renewable energy solutions for a greener future. Biotechnol Biofuels Bioprod 2024;17:129. https://doi.org/10.1186/s13068-024-02571-9.

[2] Tsai I-C. Fossil energy risk exposure of the UK electricity system: The moderating role of electricity generation mix and energy source. Energy Policy 2024;188:114065. https://doi.org/10.1016/j.enpol.2024.114065.

[3] Hadi A, Budiman AH, Darmawan A, Primeia S, Salmahaminati. Fuels Characteristics for Thermal Power Plants, 2025, p. 35–68. https://doi.org/10.1007/978-981-97-9360-0_2.

[4] Su CW, Song XY, Dou J, Qin M. Fossil fuels or renewable energy? The dilemma of climate policy choices. Renew Energy 2025;238:121950. https://doi.org/10.1016/j.renene.2024.121950.

[5] Fanchi JR. Energy In The 21st Century: Energy In Transition. World Scientific; 2023.

[6] Holechek JL, Geli HME, Sawalhah MN, Valdez R. A Global Assessment: Can Renewable Energy Replace Fossil Fuels by 2050? Sustainability 2022;14:4792. https://doi.org/10.3390/su14084792.

[7] Gürsan C, de Gooyert V. The systemic impact of a transition fuel: Does natural gas help or hinder the energy transition? Renew Sustain Energy Rev 2021;138:110552. https://doi.org/10.1016/j.rser.2020.110552.

[8] Finkelman RB, Wolfe A, Hendryx MS. The future environmental and health impacts of coal. Energy Geosci 2021;2:99–112. https://doi.org/10.1016/j.engeos.2020.11.001.

[9] Khatibi SR, Moradi-Lakeh M, Karimi SM, Kermani M, Motevalian SA. Catalyzing healthier air: the impact of escalating fossil fuel prices on air quality and public health and the need for transition to clean fuels. Biofuel Res J 2024;11:2099–104. https://doi.org/10.18331/BRJ2024.11.2.4.

[10] Goodstein D. Out of gas: the end of the age of oil. WW Norton \& Company; 2005.

[11] Berners-Lee M, Clark D. The Burning Question: We can't burn half the world's oil, coal and gas. So how do we quit? Profile Books; 2013.

[12] Bartnik R. The modernization potential of gas turbines in the coal-fired power industry: thermal and economic effectiveness. Springer Science \& Business Media; 2013.

[13] Filonchyk M, Peterson MP. An integrated analysis of air pollution from US coal-fired power plants. Geosci Front 2023;14:101498. https://doi.org/10.1016/j.gsf.2022.101498.

[14] International Energy Agency (IEA). Global CO2 emissions rebounded to their highest level in history in 2021. Glob Energy Rev CO2 Emiss 2021 2022:1–3.

[15] Zuo F, Williams ST, Liang P, Chueh C, Liao C, Jen AK -Y. Binary-Metal Perovskites Toward High-Performance Planar-Heterojunction Hybrid Solar Cells. Adv Mater 2014;26:6454–60. https://doi.org/10.1002/adma.201401641.

[16] Noel NK, Stranks SD, Abate A, Wehrenfennig C, Guarnera S, Haghighirad A-A, et al. Lead-free organic–inorganic tin halide perovskites for photovoltaic applications. Energy Environ Sci 2014;7:3061–8. https://doi.org/10.1039/C4EE01076K.

[17] Kojima A, Teshima K, Shirai Y, Miyasaka T. Organometal Halide Perovskites as Visible-Light Sensitizers for Photovoltaic Cells. J Am Chem Soc 2009;131:6050–1. https://doi.org/10.1021/ja809598r.

[18] Sims REH. Renewable energy: A response to climate change. Sol Energy 2004;76:9–17. https://doi.org/10.1016/S0038-092X(03)00101-4.

[19] Hassan Q, Viktor P, J. Al-Musawi T, Mahmood Ali B, Algburi S, Alzoubi HM, et al. The renewable energy role in the global energy Transformations. Renew Energy Focus 2024;48:100545. https://doi.org/10.1016/j.ref.2024.100545.

[20] Akinwale Ishola. Global renewable energy transition in fossil fuel dependent regions. World J Adv Res Rev 2024;24:1373–80. https://doi.org/10.30574/wjarr.2024.24.1.3046.

[21] Gielen D, Boshell F, Saygin D, Bazilian MD, Wagner N, Gorini R. The role of renewable energy in the global energy transformation. Energy Strateg Rev 2019;24:38–50. https://doi.org/10.1016/j.esr.2019.01.006.

[22] Speirs J, McGlade C, Slade R. Uncertainty in the availability of natural resources: Fossil fuels, critical metals and biomass. Energy Policy 2015;87:654–64. https://doi.org/10.1016/j.enpol.2015.02.031.

[23] Abas N, Kalair A, Khan N. Review of fossil fuels and future energy technologies. Futures 2015;69:31–49. https://doi.org/10.1016/j.futures.2015.03.003.

[24] Sorrell S, Speirs J, Bentley R, Brandt A, Miller R. Global oil depletion: A review of the evidence. Energy Policy 2010;38:5290–5. https://doi.org/10.1016/j.enpol.2010.04.046.

[25] Maggio G, Cacciola G. When will oil, natural gas, and coal peak? Fuel 2012;98:111–23. https://doi.org/10.1016/j.fuel.2012.03.021.

[26] Adetomi Adewnmi, Kehinde Andrew Olu-lawal, Chinelo Emilia Okoli, Favour Oluwadamilare Usman, Gloria Siwe Usiagu. Sustainable energy solutions and climate change: A policy review of emerging trends and global responses. World J Adv Res Rev 2023;21:408–20. https://doi.org/10.30574/wjarr.2024.21.2.0474.

[27] Jayachandran M, Gatla RK, Rao KP, Rao GS, Mohammed S, Milyani AH, et al. Challenges in achieving sustainable development goal 7: Affordable and clean energy in light of nascent technologies. Sustain Energy Technol Assessments 2022;53:102692. https://doi.org/10.1016/j.seta.2022.102692.

[28] Hasan MM, Hossain S, Mofijur M, Kabir Z, Badruddin IA, Yunus Khan TM, et al. Harnessing Solar Power: A Review of Photovoltaic Innovations, Solar Thermal Systems, and the Dawn of Energy Storage Solutions. Energies 2023;16:6456. https://doi.org/10.3390/en16186456.

[29] Choudhary P, Srivastava RK. Sustainability perspectives- a review for solar photovoltaic trends and growth opportunities. J Clean Prod 2019;227:589–612. https://doi.org/10.1016/j.jclepro.2019.04.107.

[30] Breyer C, Khalili S, Bogdanov D, Ram M, Oyewo AS, Aghahosseini A, et al. On the History and Future of 100% Renewable Energy Systems Research. IEEE Access 2022;10:78176–218. https://doi.org/10.1109/ACCESS.2022.3193402.

[31] Foster R, Ghassemi M, Cota A. Solar energy: Renewable energy and the environment. CRC Press; 2009. https://doi.org/10.5860/choice.47-5672.

[32] Kabir E, Kumar P, Kumar S, Adelodun AA, Kim K-H. Solar energy: Potential and future prospects. Renew Sustain Energy Rev 2018;82:894–900. https://doi.org/10.1016/j.rser.2017.09.094.

[33] Allouhi A, Rehman S, Buker MS, Said Z. Up-to-date literature review on Solar PV systems: Technology progress, market status and R&D. J Clean Prod 2022;362:132339. https://doi.org/10.1016/j.jclepro.2022.132339.

[34] Ramírez-Cantero J, Pérez-Huertas S, Muñoz-Batista MJ, Pérez A, Calero M, Blázquez G. State of the art of end-of-life silicon-based solar panels recycling with a bibliometric perspective. Sol Energy Mater Sol Cells 2025;281:113312. https://doi.org/10.1016/j.solmat.2024.113312.

[35] Li N, Niu X, Chen Q, Zhou H. Towards commercialization: the operational stability of perovskite solar cells. Chem Soc Rev 2020;49:8235–86. https://doi.org/10.1039/D0CS00573H.

[36] Liang B, Chen X, Wang X, Yuan H, Sun A, Wang Z, et al. Progress in crystalline silicon heterojunction solar cells. J Mater Chem A 2025;13:2441–77. https://doi.org/10.1039/D4TA06224H.

[37] Ember. Yearly Electricity Data. 2024.

[38] Ceccaroli B, Ovrelid E, Pizzini S, editors. Solar Silicon Processes. CRC Press; 2016. https://doi.org/10.1201/9781315369075.

[39] Minemoto T, Murata M. Theoretical analysis on effect of band offsets in perovskite solar cells. Sol Energy Mater Sol Cells 2015;133:8–14. https://doi.org/10.1016/j.solmat.2014.10.036.

[40] Wu T, Qin Z, Wang Y, Wu Y, Chen W, Zhang S, et al. The Main Progress of Perovskite Solar Cells in 2020–2021. Nano-Micro Lett 2021;13:152. https://doi.org/10.1007/s40820-021-00672-w.

[41] Liu W, Liu Y, Yang Z, Xu C, Li X, Huang S, et al. Flexible solar cells based on foldable silicon wafers with blunted edges. Nature 2023;617:717–23. https://doi.org/10.1038/s41586-023-05921-z.

[42] Islam R, Saraswat K. Limitation of Optical Enhancement in Ultra-thin Solar Cells Imposed by Contact Selectivity. Sci Rep 2018;8:8863. https://doi.org/10.1038/s41598-018-27155-0.

[43] El-Mellouhi F, Marzouk A, Bentria ET, Rashkeev SN, Kais S, Alharbi FH. Hydrogen Bonding and Stability of Hybrid Organic–Inorganic Perovskites. ChemSusChem 2016;9:2648–55. https://doi.org/10.1002/cssc.201600864.

[44] Niu G, Li W, Meng F, Wang L, Dong H, Qiu Y. Study on the stability of CH 3 NH 3 PbI 3 films and the effect of post-modification by aluminum oxide in all-solid-state hybrid solar cells. J Mater Chem A 2014;2:705–10. https://doi.org/10.1039/C3TA13606J.

[45] Mazumdar S, Du B, Huang C, Lin P, Zhao J, Zeng X, et al. Designing electron transporting layer for efficient perovskite solar cell by deliberating over nano-electrical conductivity. Sol Energy Mater Sol Cells 2019;200:109995. https://doi.org/10.1016/j.solmat.2019.109995.

[46] Aftab S, Koyyada G, Ali Z, Assiri MA, Kim JH, Rubab N, et al. Flexible perovskite solar cells: A revolutionary approach for wearable electronics and sensors. Mater Today Energy 2025;51:101872. https://doi.org/10.1016/j.mtener.2025.101872.

[47] Peng Z, Wei Q, Chen H, Liu Y, Wang F, Jiang X, et al. Cs0.15FA0.85PbI3/CsxFA1-xPbI3 Core/Shell Heterostructure for Highly Stable and Efficient Perovskite Solar Cells. Cell Reports Phys Sci 2020;1:100224. https://doi.org/10.1016/j.xcrp.2020.100224.

[48] Wang S, Li X, Tong T, Han J, Zhang Y, Zhu J, et al. Sequential Processing: Spontaneous Improvements in Film Quality and Interfacial Engineering for Efficient Perovskite Solar Cells. Sol RRL 2018;2. https://doi.org/10.1002/solr.201800027.

[49] Prasanna R, Gold-Parker A, Leijtens T, Conings B, Babayigit A, Boyen H-G, et al. Band Gap Tuning via Lattice Contraction and Octahedral Tilting in Perovskite Materials for Photovoltaics. J Am Chem Soc 2017;139:11117–24. https://doi.org/10.1021/jacs.7b04981.

[50] Sadeghi D, Eslami A, Eslami S, Rahbar K, Kari R. Enhancing PV system modeling accuracy: Comparative analysis of radiation models and data sources. Next Res 2025;2:100165. https://doi.org/10.1016/j.nexres.2025.100165.

[51] Prince KJ, Mirletz HM, Gaulding EA, Wheeler LM, Kerner RA, Zheng X, et al. Sustainability pathways for perovskite photovoltaics. Nat Mater 2025;24:22–33. https://doi.org/10.1038/s41563-024-01945-6.

[52] Jäger-Waldau A. Solar Energy and Photovoltaics. Encycl. Inorg. Bioinorg. Chem., Wiley; 2015, p. 1–10. https://doi.org/10.1002/9781119951438.eibc2312.

[53] Yang WS, Noh JH, Jeon NJ, Kim YC, Ryu S, Seo J, et al. High-performance photovoltaic perovskite layers fabricated through intramolecular exchange. Science (80- ) 2015;348:1234–7. https://doi.org/10.1126/science.aaa9272.

[54] Burdick J, Schmidt P. Install your own solar panels: designing and installing a photovoltaic system to power your home. Storey Publishing; 2017.

[55] A Beginners Guide to Solar Panels for Home - SolarReviews n.d. https://www.solarreviews.com/blog/solar-panels-for-home (accessed February 21, 2025).

[56] Zhang Q, Liu W. A review of enhancement strategies for pyrocatalysis of perovskite oxides and their applications. J Mater Chem A 2025;13:7601–33. https://doi.org/10.1039/D4TA07831D.

[57] Park N. Research Direction toward Scalable, Stable, and High Efficiency Perovskite Solar Cells. Adv Energy Mater 2020;10. https://doi.org/10.1002/aenm.201903106.

[58] Mohammad A, Bari Shovon R, Manzurul Hasan M, Das R, Muhammad Abdul Munayem N, Ahsan Arif M. Perovskite Solar Cell Materials Development for Enhanced Efficiency and Stability 2024;48.

[59] Mahmoudi T, Wang Y, Im YH, Hahn Y-B. Improvement of stability and efficiency of tin-based perovskite solar cells with inclusion of Cu-Sn-graphene oxide composites in interfacial and active layers. Mater Today Energy 2025;50:101866. https://doi.org/10.1016/j.mtener.2025.101866.

[60] Han Y, Meyer S, Dkhissi Y, Weber K, Pringle JM, Bach U, et al. Degradation observations of encapsulated planar CH 3 NH 3 PbI 3 perovskite solar cells at high temperatures and humidity. J Mater Chem A 2015;3:8139–47. https://doi.org/10.1039/C5TA00358J.

[61] Lee J, Kim D, Kim H, Seo S, Cho SM, Park N. Formamidinium and Cesium Hybridization for Photo- and Moisture-Stable Perovskite Solar Cell. Adv Energy Mater 2015;5. https://doi.org/10.1002/aenm.201501310.

[62] Kim IS, Cao DH, Buchholz DB, Emery JD, Farha OK, Hupp JT, et al. Liquid Water- and Heat-Resistant Hybrid Perovskite Photovoltaics via an Inverted ALD Oxide Electron Extraction Layer Design. Nano Lett 2016;16:7786–90. https://doi.org/10.1021/acs.nanolett.6b03989.

[63] Nam JK, Chai SU, Cha W, Choi YJ, Kim W, Jung MS, et al. Potassium Incorporation for Enhanced Performance and Stability of Fully Inorganic Cesium Lead Halide Perovskite Solar Cells. Nano Lett 2017;17:2028–33. https://doi.org/10.1021/acs.nanolett.7b00050.

[64] Saliba M, Matsui T, Domanski K, Seo J-Y, Ummadisingu A, Zakeeruddin SM, et al. Incorporation of rubidium cations into perovskite solar cells improves photovoltaic performance. Science (80- ) 2016;354:206–9. https://doi.org/10.1126/science.aah5557.

[65] Boyd CC, Cheacharoen R, Leijtens T, McGehee MD. Understanding Degradation Mechanisms and Improving Stability of Perovskite Photovoltaics. Chem Rev 2019;119:3418–51. https://doi.org/10.1021/acs.chemrev.8b00336.

[66] Dunfield SP, Bliss L, Zhang F, Luther JM, Zhu K, van Hest MFAM, et al. From Defects to Degradation: A Mechanistic Understanding of Degradation in Perovskite Solar Cell Devices and Modules. Adv Energy Mater 2020;10. https://doi.org/10.1002/aenm.201904054.

[67] Ono LK, Schulz P, Endres JJ, Nikiforov GO, Kato Y, Kahn A, et al. Air-exposure-induced gas-molecule incorporation into spiro-MeOTAD films. J Phys Chem Lett 2014;5:1374–9. https://doi.org/10.1021/jz500414m.

[68] Chen D, Yang F, Yang M, Chai W, Zhu W, Zhang C, et al. A stepwise solvent-annealing strategy for high-efficiency four-terminal Perovskite/Cu(InGa)Se2 tandem solar cells. Mater Today Energy 2025;49:101816. https://doi.org/10.1016/j.mtener.2025.101816.

[69] Zhang F, Yi C, Wei P, Bi X, Luo J, Jacopin G, et al. A Novel Dopant-Free Triphenylamine Based Molecular "Butterfly" Hole-Transport Material for Highly Efficient and Stable Perovskite Solar Cells. Adv Energy Mater 2016;6. https://doi.org/10.1002/aenm.201600401.

[70] Misra RK, Aharon S, Li B, Mogilyansky D, Visoly-Fisher I, Etgar L, et al. Temperature- and Component-Dependent Degradation of Perovskite Photovoltaic Materials under Concentrated Sunlight. J Phys Chem Lett 2015;6:326–30. https://doi.org/10.1021/jz502642b.

[71] Sun M, Zhang F, Liu H, Li X, Xiao Y, Wang S. Tuning the crystal growth of perovskite thin-films by adding the 2-pyridylthiourea additive for highly efficient and stable solar cells prepared in ambient air. J Mater Chem A 2017;5:13448–56. https://doi.org/10.1039/C7TA00894E.

[72] Bai S, Da P, Li C, Wang Z, Yuan Z, Fu F, et al. Planar perovskite solar cells with long-term stability using ionic liquid additives. Nature 2019;571:245–50. https://doi.org/10.1038/s41586-019-1357-2.

[73] Wang R, Xue J, Meng L, Lee JW, Zhao Z, Sun P, et al. Caffeine Improves the Performance and Thermal Stability of Perovskite Solar Cells. Joule 2019;3:1464–77. https://doi.org/10.1016/j.joule.2019.04.005.

[74] Schloemer TH, Raiford JA, Gehan TS, Moot T, Nanayakkara S, Harvey SP, et al. The Molybdenum Oxide Interface Limits the High-Temperature Operational Stability of Unencapsulated Perovskite Solar Cells. ACS Energy Lett 2020;5:2349–60. https://doi.org/10.1021/acsenergylett.0c01023.

[75] Tsotetsi D, Idisi DO, Fakayode O, Mbule P, Dhlamini M. Stability of perovskite active layers induced by caffeine: A mini review. Next Res 2024;1:100022. https://doi.org/10.1016/j.nexres.2024.100022.

[76] Boix PP, Raga SR, Mathews N. Working Principles of Perovskite Solar Cells. Halide

Perovskites, 2018, p. 81–99. https://doi.org/10.1002/9783527800766.ch2_01.

[77] Sheehan TJ, Saris S, Tisdale WA. Exciton Transport in Perovskite Materials. Adv Mater 2024. https://doi.org/10.1002/adma.202415757.

[78] Akel S, Kulkarni A, Rau U, Kirchartz T. Relevance of Long Diffusion Lengths for Efficient Halide Perovskite Solar Cells. PRX Energy 2023;2:013004. https://doi.org/10.1103/PRXEnergy.2.013004.

[79] Afre RA, Pugliese D. Perovskite Solar Cells: A Review of the Latest Advances in Materials, Fabrication Techniques, and Stability Enhancement Strategies. Micromachines 2024;15:192. https://doi.org/10.3390/mi15020192.

[80] Njema GG, Kibet JK, Ngari SM. Advancements in the photovoltaic optimization of a high performance perovskite solar cell based on graphene oxide (GO) hole transport layer. Next Res 2024;1:100055. https://doi.org/10.1016/j.nexres.2024.100055.

[81] Krückemeier L, Liu Z, Kirchartz T, Rau U. Quantifying Charge Extraction and Recombination Using the Rise and Decay of the Transient Photovoltage of Perovskite Solar Cells. Adv Mater 2023;35. https://doi.org/10.1002/adma.202300872.

[82] Liang Z, Ren Z, Wang Z, Yan N, Li L, Zhang B, et al. Boosting charge extraction and efficiency of inverted perovskite solar cells through coordinating group modification at the buffer layer/cathode interface. Nano Res 2025;18:94907075. https://doi.org/10.26599/NR.2025.94907075.

[83] Ndi FE, Perabi SN, Ndjakomo SE, Ondoua Abessolo G, Mengounou Mengata G. Estimation of single-diode and two diode solar cell parameters by equilibrium optimizer method. Energy Reports 2021;7:4761–8. https://doi.org/10.1016/j.egyr.2021.07.025.

[84] Kumar C, Raj TD, Premkumar M, Raj TD. A new stochastic slime mould optimization algorithm for the estimation of solar photovoltaic cell parameters. Optik (Stuttg) 2020;223:165277. https://doi.org/10.1016/j.ijleo.2020.165277.

[85] Jiao S, Chong G, Huang C, Hu H, Wang M, Heidari AA, et al. Orthogonally adapted Harris hawks optimization for parameter estimation of photovoltaic models. Energy 2020;203:117804. https://doi.org/10.1016/j.energy.2020.117804.

[86] Yu K, Qu B, Yue C, Ge S, Chen X, Liang J. A performance-guided JAYA algorithm for parameters identification of photovoltaic cell and module. Appl Energy 2019;237:241–57. https://doi.org/10.1016/j.apenergy.2019.01.008.

[87] Lin P, Cheng S, Yeh W, Chen Z, Wu L. Parameters extraction of solar cell models using a modified simplified swarm optimization algorithm. Sol Energy 2017;144:594–603. https://doi.org/10.1016/j.solener.2017.01.064.

[88] Derick M, Rani C, Rajesh M, Farrag ME, Wang Y, Busawon K. An improved optimization technique for estimation of solar photovoltaic parameters. Sol Energy 2017;157:116–24. https://doi.org/10.1016/j.solener.2017.08.006.

[89] Yu K, Liang JJ, Qu BY, Chen X, Wang H. Parameters identification of photovoltaic models using an improved JAYA optimization algorithm. Energy Convers Manag 2017;150:742–53. https://doi.org/10.1016/j.enconman.2017.08.063.

[90] Liu Y, Heidari AA, Ye X, Liang G, Chen H, He C. Boosting slime mould algorithm for parameter identification of photovoltaic models. Energy 2021;234:121164. https://doi.org/10.1016/j.energy.2021.121164.

[91] Rawa M, Abusorrah A, Al-Turki Y, Calasan M, Micev M, Ali ZM, et al. Estimation of Parameters of Different Equivalent Circuit Models of Solar Cells and Various Photovoltaic

Modules Using Hybrid Variants of Honey Badger Algorithm and Artificial Gorilla Troops Optimizer. Mathematics 2022;10. https://doi.org/10.3390/math10071057.

[92] Luque A, Hegedus S. Photovoltaic Science Handbook of Photovoltaic Science. 2011.

[93] Tao M. Physics of solar cells. 2014. https://doi.org/10.1007/978-1-4471-5643-7_3.

[94] Yadav P, Yadav S, Atri S, Tomar R. A Brief Review on Key Role of Perovskite Oxides as Catalyst. ChemistrySelect 2021;6:12947–59. https://doi.org/10.1002/slct.202102292.

[95] Wu Y, Islam A, Yang X, Qin C, Liu J, Zhang K, et al. Retarding the crystallization of PbI2 for highly reproducible planar-structured perovskite solar cells via sequential deposition. Energy Environ Sci 2014;7:2934–8. https://doi.org/10.1039/c4ee01624f.

[96] Sarukura N, Murakami H, Estacio E, Ono S, El Ouenzerfi R, Cadatal M, et al. Proposed design principle of fluoride-based materials for deep ultraviolet light emitting devices. Opt Mater (Amst) 2007;30:15–7. https://doi.org/10.1016/j.optmat.2006.11.031.

[97] Jeon NJ, Noh JH, Kim YC, Yang WS, Ryu S, Seok S Il. Solvent engineering for high-performance inorganic-organic hybrid perovskite solar cells. Nat Mater 2014;13:897–903. https://doi.org/10.1038/nmat4014.

[98] Xie J, Jin X, Cao J, Shu Y, Yan W, Han L, et al. Synergistic Treatments of Bulk and Surface in Tin-Lead Mixed Perovskite for Efficient All-Perovskite Tandem Solar Cells. Adv Funct Mater 2025;35. https://doi.org/10.1002/adfm.202414819.

[99] Armstrong PJ, Chapagain S, Chacko E, Druffel T, Grapperhaus CA. Evaluation of imidazole blocking layers for perovskite stability. Next Res 2024;1:100036. https://doi.org/10.1016/j.nexres.2024.100036.

[100] Petrović M, Chellappan V, Ramakrishna S. Perovskites: Solar cells & engineering applications - materials and device developments. Sol Energy 2015;122:678–99. https://doi.org/10.1016/j.solener.2015.09.041.

[101] Saikia D, Betal A, Bera J, Sahu S. Progress and challenges of halide perovskite-based solar cell- a brief review. Mater Sci Semicond Process 2022;150. https://doi.org/10.1016/j.mssp.2022.106953.

[102] Li C, Soh KCK, Wu P. Formability of ABO3 perovskites. J Alloys Compd 2004;372:40–8. https://doi.org/10.1016/j.jallcom.2003.10.017.

[103] Li C, Lu X, Ding W, Feng L, Gao Y, Guo Z. Formability of ABX 3 (X = F, Cl, Br, I) halide perovskites. Acta Crystallogr Sect B Struct Sci 2008;64:702–7. https://doi.org/10.1107/S0108768108032734.

[104] Mitzi DB, Chondroudis K, Kagan CR. Design, structure, and optical properties of organic-inorganic perovskites containing an oligothiophene chromophore. Inorg Chem 1999;38:6246–56. https://doi.org/10.1021/ic991048k.

[105] Travis W, Glover ENK, Bronstein H, Scanlon DO, Palgrave RG. On the application of the tolerance factor to inorganic and hybrid halide perovskites: A revised system. Chem Sci 2016;7:4548–56. https://doi.org/10.1039/c5sc04845a.

[106] Yang C, Yang C. Perovskite single crystals for solar cell and photodetector Dissertation by 2022.

[107] Xing J, Zou Y, Zhao C, Yu Z, Shan Y, Kong W, et al. Thickness-dependent carrier lifetime and mobility for MAPbBr3 single crystals. Mater Today Phys 2020;14:1–7. https://doi.org/10.1016/j.mtphys.2020.100240.

[108] Rajagopal A, Stoddard RJ, Jo SB, Hillhouse HW, Jen AKY. Overcoming the Photovoltage

Plateau in Large Bandgap Perovskite Photovoltaics. Nano Lett 2018;18:3985–93. https://doi.org/10.1021/acs.nanolett.8b01480.

[109] Song W, Cao G. Surface-defect passivation through complexation with organic molecules leads to enhanced power conversion efficiency and long term stability of perovskite photovoltaics. Sci China Mater 2020;63:479–80. https://doi.org/10.1007/s40843-020-1249-3.

[110] Ahmed MT, Islam S, Ahmed F. Comparative Study on the Crystallite Size and Bandgap of Perovskite by Diverse Methods. Adv Condens Matter Phys 2022;2022. https://doi.org/10.1155/2022/9535932.

[111] Hamada M, Rana S, Jokar E, Awasthi K, Diau EWG, Ohta N. Temperature-Dependent Electroabsorption Spectra and Exciton Binding Energy in a Perovskite CH3NH3PbI3Nanocrystalline Film. ACS Appl Energy Mater 2020;3:11830–40. https://doi.org/10.1021/acsaem.0c01983.

[112] Qian XY, Tang YY, Zhou W, Shen Y, Guo ML, Li YQ, et al. Strategies to Improve Luminescence Efficiency and Stability of Blue Perovskite Light-Emitting Devices. Small Sci 2021;1:1–15. https://doi.org/10.1002/smsc.202000048.

[113] Baldo MA, Lamansky S, Burrows PE, Thompson ME, Forrest SR. Very High-Efficiency Green Organic Light-Emitting Devices Based on Electrophosphorescence. Electrophosphorescent Mater Devices 2023:59–68. https://doi.org/10.1201/9781003088721-5.

[114] Xu J, Castriotta LA, Skafi Z, Chakraborty A, Di Carlo A, Brown TM. Lead-free solar cells and modules with antimony-based perovskite inspired materials for indoor photovoltaics. Mater Today Energy 2025;49:101823. https://doi.org/10.1016/j.mtener.2025.101823.

[115] Zhou D, Zhou T, Tian Y, Zhu X, Tu Y. Perovskite-Based Solar Cells: Materials, Methods, and Future Perspectives. J Nanomater 2018;2018. https://doi.org/10.1155/2018/8148072.

[116] Lekesi LP, Koao LF, Motloung S V., Motaung TE, Malevu T. Developments on Perovskite Solar Cells (PSCs): A Critical Review. Appl Sci 2022;12. https://doi.org/10.3390/app12020672.

[117] Jiang T, Chen Z, Chen X, Chen X, Xu X, Liu T, et al. Power Conversion Efficiency Enhancement of Low-Bandgap Mixed Pb-Sn Perovskite Solar Cells by Improved Interfacial Charge Transfer. ACS Energy Lett 2019;4:1784–90. https://doi.org/10.1021/acsenergylett.9b00880.

[118] Inamul Hasan Z, Joshi S, Subbaya KM. Halide-based CH3NH3PbI3 hybrid perovskite thin films structural studies using synchrotron source X-ray diffraction. J Mater Sci Mater Electron 2022;33:16369–82. https://doi.org/10.1007/s10854-022-08528-8.

[119] Zimmermann E, Wong KK, Müller M, Hu H, Ehrenreich P, Kohlstädt M, et al. Characterization of perovskite solar cells: Towards a reliable measurement protocol. APL Mater 2016;4. https://doi.org/10.1063/1.4960759.

[120] Joshi S, Hasan ZI, Pruthvi M, Tulsiram MP. X-ray diffraction - A simplistic approach for perovskite based solar cells degradation studies. Mater Today Proc 2019;35:31–4. https://doi.org/10.1016/j.matpr.2019.05.390.

[121] Elloh VW, Yaya A, Abavare EKK. Modelling of (N-vinylcarbazole)/fullerene nanoheterojunction for organic solar cells and photovoltaics applications – A DFT approach. Next Res 2024;1:100042. https://doi.org/10.1016/j.nexres.2024.100042.

[122] Tong G, Song Z, Li C, Zhao Y, Yu L, Xu J, et al. Cadmium-doped flexible perovskite solar cells with a low-cost and low-temperature-processed CdS electron transport layer. RSC Adv 2017;7:19457–63. https://doi.org/10.1039/c7ra01110e.

[123] Abulikemu M, Barbé J, El Labban A, Eid J, Del Gobbo S. Planar heterojunction perovskite solar cell based on CdS electron transport layer. Thin Solid Films 2017;636:512–8. https://doi.org/10.1016/j.tsf.2017.07.003.

[124] Mohamadkhani F, Javadpour S, Taghavinia N. Improvement of planar perovskite solar cells by using solution processed SnO2/CdS as electron transport layer. Sol Energy 2019;191:647–53. https://doi.org/10.1016/j.solener.2019.08.067.

[125] Wang JTW, Ball JM, Barea EM, Abate A, Alexander-Webber JA, Huang J, et al. Low-temperature processed electron collection layers of graphene/TiO 2 nanocomposites in thin film perovskite solar cells. Nano Lett 2014;14:724–30. https://doi.org/10.1021/nl403997a.

[126] Patil P, Mann DS, Nakate UT, Hahn YB, Kwon SN, Na SI. Hybrid interfacial ETL engineering using PCBM-SnS2 for High-Performance p-i-n structured planar perovskite solar cells. Chem Eng J 2020;397:125504. https://doi.org/10.1016/j.cej.2020.125504.

[127] Wang Y, Cai G, Chen Z, Peng H, Zhang Q, Gao L, et al. Transparent Conducting TiO2 Thin Film Induced by Electric-Field Controlled Hydrogen Ion Intercalation. Adv Electron Mater 2024;2400029:1–9. https://doi.org/10.1002/aelm.202400029.

[128] Hossain MK, Toki GFI, Kuddus A, Rubel MHK, Hossain MM, Bencherif H, et al. An extensive study on multiple ETL and HTL layers to design and simulation of high-performance lead-free CsSnCl3-based perovskite solar cells. Sci Rep 2023;13:1–24. https://doi.org/10.1038/s41598-023-28506-2.

[129] Targhi FF, Jalili YS, Kanjouri F. MAPbI3 and FAPbI3 perovskites as solar cells: Case study on structural, electrical and optical properties. Results Phys 2018;10:616–27. https://doi.org/10.1016/j.rinp.2018.07.007.

[130] Li S, Cao YL, Li WH, Bo ZS. A brief review of hole transporting materials commonly used in perovskite solar cells. Rare Met 2021;40:2712–29. https://doi.org/10.1007/s12598-020-01691-z.

[131] Assi AA, Saleh WR, Mohajerani E. Effect of Metals ( Au, Ag, and Ni) as Cathode Electrode on Perovskite Solar Cells. IOP Conf Ser Earth Environ Sci 2021;722. https://doi.org/10.1088/1755-1315/722/1/012019.

[132] Kim H-S, Lee C-R, Im J-H, Lee K-B, Moehl T, Marchioro A, et al. Lead Iodide Perovskite Sensitized All-Solid-State Submicron Thin Film Mesoscopic Solar Cell with Efficiency Exceeding 9%. Sci Rep 2012;2:591. https://doi.org/10.1038/srep00591.

[133] Heo JH, Im SH, Noh JH, Mandal TN, Lim CS, Chang JA, et al. Efficient inorganic-organic hybrid heterojunction solar cells containing perovskite compound and polymeric hole conductors. Nat Photonics 2013;7:486–91. https://doi.org/10.1038/nphoton.2013.80.

[134] Roy P, Kumar Sinha N, Tiwari S, Khare A. A review on perovskite solar cells: Evolution of architecture, fabrication techniques, commercialization issues and status. Sol Energy 2020;198:665–88. https://doi.org/10.1016/j.solener.2020.01.080.

[135] Tang J, Jiao D, Zhang L, Zhang X, Xu X, Yao C, et al. High-performance inverted planar perovskite solar cells based on efficient hole-transporting layers from well-crystalline NiO nanocrystals. Sol Energy 2018;161:100–8. https://doi.org/10.1016/j.solener.2017.12.045.

[136] Im JH, Lee CR, Lee JW, Park SW, Park NG. 6.5% Efficient Perovskite Quantum-Dot-Sensitized Solar Cell. Nanoscale 2011;3:4088–93. https://doi.org/10.1039/c1nr10867k.

[137] Ball JM, Lee MM, Hey A, Snaith HJ. Low-temperature processed meso-superstructured to thin-film perovskite solar cells. Energy Environ Sci 2013;6:1739–43. https://doi.org/10.1039/c3ee40810h.

[138] Liu M, Johnston MB, Snaith HJ. Efficient planar heterojunction perovskite solar cells by vapour deposition. Nature 2013;501:395–8. https://doi.org/10.1038/nature12509.

[139] Yang D, Yang R, Wang K, Wu C, Zhu X, Feng J, et al. High efficiency planar-type perovskite solar cells with negligible hysteresis using EDTA-complexed SnO2. Nat Commun 2018;9. https://doi.org/10.1038/s41467-018-05760-x.

[140] Haider M, Zhen C, Wu T, Liu G, Cheng HM. Boosting efficiency and stability of perovskite solar cells with nickel phthalocyanine as a low-cost hole transporting layer material. J Mater Sci Technol 2018;34:1474–80. https://doi.org/10.1016/j.jmst.2018.03.005.

[141] Li W, Zhang W, Van Reenen S, Sutton RJ, Fan J, Haghighirad AA, et al. Enhanced UV-light stability of planar heterojunction perovskite solar cells with caesium bromide interface modification. Energy Environ Sci 2016;9:490–8. https://doi.org/10.1039/c5ee03522h.

[142] Bryant D, Aristidou N, Pont S, Sanchez-Molina I, Chotchunangatchaval T, Wheeler S, et al. Light and oxygen induced degradation limits the operational stability of methylammonium lead triiodide perovskite solar cells. Energy Environ Sci 2016;9:1655–60. https://doi.org/10.1039/c6ee00409a.

[143] Delugas P, Caddeo C, Filippetti A, Mattoni A. Thermally Activated Point Defect Diffusion in Methylammonium Lead Trihalide: Anisotropic and Ultrahigh Mobility of Iodine. J Phys Chem Lett 2016;7:2356–61. https://doi.org/10.1021/acs.jpclett.6b00963.

[144] Su H, Wu T, Cui D, Lin X, Luo X, Wang Y, et al. The Application of Graphene Derivatives in Perovskite Solar Cells. Small Methods 2020;4:1–12. https://doi.org/10.1002/smtd.202000507.

[145] Saidaminov MI, Kim J, Jain A, Quintero-Bermudez R, Tan H, Long G, et al. Suppression of atomic vacancies via incorporation of isovalent small ions to increase the stability of halide perovskite solar cells in ambient air. Nat Energy 2018;3:648–54. https://doi.org/10.1038/s41560-018-0192-2.

[146] Yuan Y, Huang J. Ion Migration in Organometal Trihalide Perovskite and Its Impact on Photovoltaic Efficiency and Stability. Acc Chem Res 2016;49:286–93. https://doi.org/10.1021/acs.accounts.5b00420.

[147] Yu W, Li F, Wang H, Alarousu E, Chen Y, Lin B, et al. Ultrathin Cu2O as an efficient inorganic hole transporting material for perovskite solar cells. Nanoscale 2016;8:6173–9. https://doi.org/10.1039/c5nr07758c.

[148] Park JH, Seo J, Park S, Shin SS, Kim YC, Jeon NJ, et al. Efficient CH3NH3PbI3 Perovskite Solar Cells Employing Nanostructured p-Type NiO Electrode Formed by a Pulsed Laser Deposition. Adv Mater 2015;27:4013–9. https://doi.org/10.1002/adma.201500523.

[149] Kung PK, Li MH, Lin PY, Chiang YH, Chan CR, Guo TF, et al. A Review of Inorganic Hole Transport Materials for Perovskite Solar Cells. Adv Mater Interfaces 2018;5:1–35. https://doi.org/10.1002/admi.201800882.

[150] Rao H, Ye S, Sun W, Yan W, Li Y, Peng H, et al. A 19.0% efficiency achieved in CuOx-based inverted CH3NH3PbI3-xClx solar cells by an effective Cl doping method. Nano Energy 2016;27:51–7. https://doi.org/10.1016/j.nanoen.2016.06.044.

[151] Jung M, Kim YC, Jeon NJ, Yang WS, Seo J, Noh JH, et al. Thermal Stability of CuSCN Hole Conductor-Based Perovskite Solar Cells. ChemSusChem 2016;9:2592–6. https://doi.org/10.1002/cssc.201600957.

[152] Xia J, Sohail M, Nazeeruddin MK. Efficient and Stable Perovskite Solar Cells by Tailoring of Interfaces. Adv Mater 2023;35. https://doi.org/10.1002/adma.202211324.

[153] Zakaria Y, Aïssa B, Fix T, Ahzi S, Mansour S, Slaoui A. Moderate temperature deposition of

RF magnetron sputtered SnO2-based electron transporting layer for triple cation perovskite solar cells. Sci Rep 2023;13:1–14. https://doi.org/10.1038/s41598-023-35651-1.

[154] Miah MH, Rahman MB, Nur-E-Alam M, Islam MA, Shahinuzzaman M, Rahman MR, et al. Key degradation mechanisms of perovskite solar cells and strategies for enhanced stability: issues and prospects. RSC Adv 2025;15:628–54. https://doi.org/10.1039/d4ra07942f.

[155] Witteck R, Minh DN, Paul G, Harvey SP, Zheng X, Jiang Q, et al. Reducing Thermal Degradation of Perovskite Solar Cells during Vacuum Lamination by Internal Diffusion Barriers. ACS Appl Energy Mater 2024. https://doi.org/10.1021/acsaem.4c02567.

[156] Boyd CC, Cheacharoen R, Bush KA, Prasanna R, Leijtens T, McGehee MD. Barrier Design to Prevent Metal-Induced Degradation and Improve Thermal Stability in Perovskite Solar Cells. ACS Energy Lett 2018;3:1772–8. https://doi.org/10.1021/acsenergylett.8b00926.

[157] Reddy SH, Di Giacomo F, Matteocci F, Castriotta LA, Di Carlo A. Holistic Approach toward a Damage-Less Sputtered Indium Tin Oxide Barrier Layer for High-Stability Inverted Perovskite Solar Cells and Modules. ACS Appl Mater Interfaces 2022;14:51438–48. https://doi.org/10.1021/acsami.2c10251.

[158] Yang G, Wang C, Lei H, Zheng X, Qin P, Xiong L, et al. Interface engineering in planar perovskite solar cells: Energy level alignment, perovskite morphology control and high performance achievement. J Mater Chem A 2017;5:1658–66. https://doi.org/10.1039/c6ta08783c.

[159] Emery Q, Remec M, Paramasivam G, Janke S, Dagar J, Ulbrich C, et al. Encapsulation and Outdoor Testing of Perovskite Solar Cells: Comparing Industrially Relevant Process with a Simplified Lab Procedure. ACS Appl Mater Interfaces 2022;14:5159–67. https://doi.org/10.1021/acsami.1c14720.

[160] Cells S. Διαφορετικές γενιές των ηλιακών κυττάρων Δεύτερη Γενιά Solar Cells : Τρίτης γενιάς Solar Cells : Nat Energy 2014;5:2–3.

[161] Wang T, Yang J, Cao Q, Pu X, Li Y, Chen H, et al. Room temperature nondestructive encapsulation via self-crosslinked fluorosilicone polymer enables damp heat-stable sustainable perovskite solar cells. Nat Commun 2023;14:1–10. https://doi.org/10.1038/s41467-023-36918-x.

[162] Cao M, Ji W, Chao C, Li J, Dai F, Fan X. Recent Advances in UV-Cured Encapsulation for Stable and Durable Perovskite Solar Cell Devices. Polymers (Basel) 2023;15. https://doi.org/10.3390/polym15193911.

[163] Singh P, Ravindra NM. Temperature dependence of solar cell performance - An analysis. Sol Energy Mater Sol Cells 2012;101:36–45. https://doi.org/10.1016/j.solmat.2012.02.019.

[164] Pisoni A, Baris OS, Spina M, Gaa R. Ultra-Low Thermal Conductivity in Organic − Inorganic Hybrid 2014.

[165] Conings B, Drijkoningen J, Gauquelin N, Babayigit A, D’Haen J, D’Olieslaeger L, et al. Intrinsic Thermal Instability of Methylammonium Lead Trihalide Perovskite. Adv Energy Mater 2015;5:1–8. https://doi.org/10.1002/aenm.201500477.

[166] Yusoff ARBM, Nazeeruddin MK. Organohalide Lead Perovskites for Photovoltaic Applications. J Phys Chem Lett 2016;7:851–66. https://doi.org/10.1021/acs.jpclett.5b02893.

[167] Kim NK, Min YH, Noh S, Cho E, Jeong G, Joo M, et al. Investigation of Thermally Induced Degradation in CH3NH3PbI3 Perovskite Solar Cells using In-situ Synchrotron Radiation Analysis. Sci Rep 2017;7. https://doi.org/10.1038/s41598-017-04690-w.

[168] Abdelmageed G, Mackeen C, Hellier K, Jewell L, Seymour L, Tingwald M, et al. Effect of

temperature on light induced degradation in methylammonium lead iodide perovskite thin films and solar cells. Sol Energy Mater Sol Cells 2018;174:566–71. https://doi.org/10.1016/j.solmat.2017.09.053.

[169] Brunetti B, Cavallo C, Ciccioli A, Gigli G, Latini A. On the Thermal and Thermodynamic (In)Stability of Methylammonium Lead Halide Perovskites. Sci Rep 2016;6. https://doi.org/10.1038/srep31896.

[170] Eperon GE, Stranks SD, Menelaou C, Johnston MB, Herz LM, Snaith HJ. Formamidinium lead trihalide: A broadly tunable perovskite for efficient planar heterojunction solar cells. Energy Environ Sci 2014;7:982–8. https://doi.org/10.1039/c3ee43822h.

[171] Niu G, Li W, Li J, Liang X, Wang L. Enhancement of thermal stability for perovskite solar cells through cesium doping. RSC Adv 2017;7:17473–9. https://doi.org/10.1039/c6ra28501e.

[172] Yang J, Liu X, Zhang Y, Zheng X, He X, Wang H, et al. Comprehensive understanding of heat-induced degradation of triple-cation mixed halide perovskite for a robust solar cell. Nano Energy 2018;54:218–26. https://doi.org/10.1016/j.nanoen.2018.10.011.

[173] Tan W, Bowring AR, Meng AC, McGehee MD, McIntyre PC. Thermal Stability of Mixed Cation Metal Halide Perovskites in Air. ACS Appl Mater Interfaces 2018;10:5485–91. https://doi.org/10.1021/acsami.7b15263.

[174] Gaonkar H, Zhu J, Kottokkaran R, Bhageri B, Noack M, Dalal V. Thermally Stable, Efficient, Vapor Deposited Inorganic Perovskite Solar Cells. ACS Appl Energy Mater 2020;3:3497–503. https://doi.org/10.1021/acsaem.0c00010.

[175] Kasparavicius E, Franckevičius M, Malinauskiene V, Genevičius K, Getautis V, Malinauskas T. Oxidized Spiro-OMeTAD: Investigation of Stability in Contact with Various Perovskite Compositions. ACS Appl Energy Mater 2021;4:13696–705. https://doi.org/10.1021/acsaem.1c02375.

[176] Rombach FM, Haque SA, Macdonald TJ. Lessons learned from spiro-OMeTAD and PTAA in perovskite solar cells. Energy Environ Sci 2021;14:5161–90. https://doi.org/10.1039/d1ee02095a.

[177] Kim S, Sabury S, Perini CAR, Hossain T, Yusuf AO, Xiao X, et al. Enhancing Thermal Stability of Perovskite Solar Cells through Thermal Transition and Thin Film Crystallization Engineering of Polymeric Hole Transport Layers. ACS Energy Lett 2024:4501–8. https://doi.org/10.1021/acsenergylett.4c01546.

[178] Mazumdar S, Zhao Y, Zhang X. Stability of Perovskite Solar Cells: Degradation Mechanisms and Remedies. Front Electron 2021;2:1–34. https://doi.org/10.3389/felec.2021.712785.

[179] Zheng H, Liu G, Zhang C, Zhu L, Alsaedi A, Hayat T, et al. The influence of perovskite layer and hole transport material on the temperature stability about perovskite solar cells. Sol Energy 2018;159:914–9. https://doi.org/10.1016/j.solener.2017.09.039.

[180] Jena AK, Numata Y, Ikegami M, Miyasaka T. Role of spiro-OMeTAD in performance deterioration of perovskite solar cells at high temperature and reuse of the perovskite films to avoid Pb-waste. J Mater Chem A 2018;6:2219–30. https://doi.org/10.1039/c7ta07674f.

[181] Tumen-Ulzii G, Qin C, Matsushima T, Leyden MR, Balijipalli U, Klotz D, et al. Understanding the Degradation of Spiro-OMeTAD-Based Perovskite Solar Cells at High Temperature. Sol RRL 2020;4. https://doi.org/10.1002/solr.202000305.

[182] Song W, Rakocevic L, Thiruvallur Eachambadi R, Qiu W, Bastos JP, Gehlhaar R, et al. Improving the Morphology Stability of Spiro-OMeTAD Films for Enhanced Thermal Stability of Perovskite Solar Cells. ACS Appl Mater Interfaces 2021;13:44294–301. https://doi.org/10.1021/acsami.1c11227.

[183] a ESS. Adsorption of pyridine at the Au ( 111 ) -solution. Potentials 1991;307:241–62.

[184] Stolberg L, Richer J, Lipkowski J, Irish DE. Adsorption of pyridine at the polycrystalline gold-solution interface. J Electroanal Chem 1986;207:213–34. https://doi.org/10.1016/0022-0728(86)87073-5.

[185] Snaith HJ, Grätzel M. Enhanced charge mobility in a molecular hole transporter via addition of redox inactive ionic dopant: Implication to dye-sensitized solar cells. Appl Phys Lett 2006;89. https://doi.org/10.1063/1.2424552.

[186] Krüger J, Plass R, Cevey L, Piccirelli M, Grätzel M, Bach U. High efficiency solid-state photovoltaic device due to inhibition of interface charge recombination. Appl Phys Lett 2001;79:2085–7. https://doi.org/10.1063/1.1406148.

[187] Juarez-Perez EJ, Leyden MR, Wang S, Ono LK, Hawash Z, Qi Y. Role of the Dopants on the Morphological and Transport Properties of Spiro-MeOTAD Hole Transport Layer. Chem Mater 2016;28:5702–9. https://doi.org/10.1021/acs.chemmater.6b01777.

[188] Pham HD, Jain SM, Li M, Manzhos S, Feron K, Pitchaimuthu S, et al. Dopant-free novel hole-transporting materials based on quinacridone dye for high-performance and humidity-stable mesoporous perovskite solar cells. J Mater Chem A 2019;7:5315–23. https://doi.org/10.1039/c8ta11361k.

[189] Pham HD, Hayasake K, Kim J, Do TT, Matsui H, Manzhos S, et al. One step facile synthesis of a novel anthanthrone dye-based, dopant-free hole transporting material for efficient and stable perovskite solar cells. J Mater Chem C 2018;6:3699–708. https://doi.org/10.1039/c7tc05238c.

[190] Pham HD, Do TT, Kim J, Charbonneau C, Manzhos S, Feron K, et al. Molecular Engineering Using an Anthanthrone Dye for Low-Cost Hole Transport Materials: A Strategy for Dopant-Free, High-Efficiency, and Stable Perovskite Solar Cells. Adv Energy Mater 2018;8:1–13. https://doi.org/10.1002/aenm.201703007.

[191] Luo J, Jia C, Wan Z, Han F, Zhao B, Wang R. The novel dopant for hole-transporting material opens a new processing route to efficiently reduce hysteresis and improve stability of planar perovskite solar cells. J Power Sources 2017;342:886–95. https://doi.org/10.1016/j.jpowsour.2017.01.010.

[192] Bai Y, Fang Y, Deng Y, Wang Q, Zhao J, Zheng X, et al. Low Temperature Solution-Processed Sb:SnO2 Nanocrystals for Efficient Planar Perovskite Solar Cells. ChemSusChem 2016;9:2686–91. https://doi.org/10.1002/cssc.201600944.

[193] Li M, Wang ZK, Yang YG, Hu Y, Feng SL, Wang JM, et al. Copper Salts Doped Spiro-OMeTAD for High-Performance Perovskite Solar Cells. Adv Energy Mater 2016;6. https://doi.org/10.1002/aenm.201601156.

[194] Chen C, Zhang W, Cong J, Cheng M, Zhang B, Chen H, et al. Cu(II) Complexes as p-Type Dopants in Efficient Perovskite Solar Cells. ACS Energy Lett 2017;2:497–503. https://doi.org/10.1021/acsenergylett.6b00691.

[195] Pham HD, Yang TCJ, Jain SM, Wilson GJ, Sonar P. Development of Dopant-Free Organic Hole Transporting Materials for Perovskite Solar Cells. Adv Energy Mater 2020;10. https://doi.org/10.1002/aenm.201903326.

[196] Guo Y, Lei H, Xiong L, Li B, Fang G. An integrated organic-inorganic hole transport layer for efficient and stable perovskite solar cells. J Mater Chem A 2018;6:2157–65. https://doi.org/10.1039/c7ta09946k.

[197] Ito S, Tanaka S, Vahlman H, Nishino H, Manabe K, Lund P. Carbon-double-bond-free printed solar cells from TiO2/CH 3NH3PbI3/CuSCN/Au: Structural control and photoaging effects.

ChemPhysChem 2014;15:1194–200. https://doi.org/10.1002/cphc.201301047.

[198] Seo S, Park IJ, Kim M, Lee S, Bae C, Jung HS, et al. An ultra-thin, un-doped NiO hole transporting layer of highly efficient (16.4%) organic-inorganic hybrid perovskite solar cells. Nanoscale 2016;8:11403–12. https://doi.org/10.1039/c6nr01601d.

[199] You J, Meng L, Song T Bin, Guo TF, Chang WH, Hong Z, et al. Improved air stability of perovskite solar cells via solution-processed metal oxide transport layers. Nat Nanotechnol 2016;11:75–81. https://doi.org/10.1038/nnano.2015.230.

[200] Wu Y, Xie F, Chen H, Yang X, Su H, Cai M, et al. Thermally Stable MAPbI3 Perovskite Solar Cells with Efficiency of 19.19% and Area over 1 cm2 achieved by Additive Engineering. Adv Mater 2017;29:1–8. https://doi.org/10.1002/adma.201701073.

[201] Shi X, Wu Y, Chen J, Cai M, Yang Y, Liu X, et al. Thermally stable perovskite solar cells with efficiency over 21%: Via a bifunctional additive. J Mater Chem A 2020;8:7205–13. https://doi.org/10.1039/d0ta01255f.

[202] Li L, Chen YW, Zheng Y, Hu H, Wu J. Seismic Evidence for Plume-Slab Interaction by High-Resolution Imaging of the 410-km Discontinuity Under Tonga. Geophys Res Lett 2019;46:13687–94. https://doi.org/10.1029/2019GL084164.

[203] Mbumba MT, Malouangou DM, Tsiba JM, Bai L, Yang Y, Guli M. Degradation mechanism and addressing techniques of thermal instability in halide perovskite solar cells. Sol Energy 2021;230:954–78. https://doi.org/10.1016/j.solener.2021.10.070.

[204] Zhou Y, Hu J, Wu Y, Qing R, Zhang C, Xu X, et al. Review on methods for improving the thermal and ambient stability of perovskite solar cells. J Photonics Energy 2019;9:1. https://doi.org/10.1117/1.jpe.9.040901.

[205] Smith IC, Hoke ET, Solis-Ibarra D, McGehee MD, Karunadasa HI. A Layered Hybrid Perovskite Solar-Cell Absorber with Enhanced Moisture Stability. Angew Chemie - Int Ed 2014;53:11232–5. https://doi.org/10.1002/anie.201406466.

[206] Yao K, Wang X, Li F, Zhou L. Mixed perovskite based on methyl-ammonium and polymeric-ammonium for stable and reproducible solar cells. Chem Commun 2015;51:15430–3. https://doi.org/10.1039/c5cc05879a.

[207] Davy MM, Jadel TM, Qin C, Luyun B, Mina G. Recent progress in low dimensional (quasi-2D) and mixed dimensional (2D/3D) tin-based perovskite solar cells. Sustain Energy Fuels 2021;5:34–51. https://doi.org/10.1039/d0se01520b.

[208] Tian Q, Yao W, Wu Z, Liu J, Liu L, Wu W, et al. Full-spectrum-activated Z-scheme photocatalysts based on NaYF4:Yb3+/Er3+, TiO2 and Ag6Si2O7. J Mater Chem A 2017;5:23566–76. https://doi.org/10.1039/c7ta07529d.

[209] Bai Y, Xiao S, Hu C, Zhang T, Meng X, Lin H, et al. Dimensional Engineering of a Graded 3D–2D Halide Perovskite Interface Enables Ultrahigh Voc Enhanced Stability in the p-i-n Photovoltaics. Adv Energy Mater 2017;7:1–8. https://doi.org/10.1002/aenm.201701038.

[210] Grancini G, Roldán-Carmona C, Zimmermann I, Mosconi E, Lee X, Martineau D, et al. One-Year stable perovskite solar cells by 2D/3D interface engineering. Nat Commun 2017;8:1–8. https://doi.org/10.1038/ncomms15684.

[211] Zhang X, Ren X, Liu B, Munir R, Zhu X, Yang D, et al. Stable high efficiency two-dimensional perovskite solar cells via cesium doping. Energy Environ Sci 2017;10:2095–102. https://doi.org/10.1039/c7ee01145h.

[212] Zhang PP, Zhou ZJ, Kou DX, Wu SX. Perovskite Thin Film Solar Cells Based on Inorganic Hole Conducting Materials. Int J Photoenergy 2017;2017.

https://doi.org/10.1155/2017/6109092.

[213] Saliba M, Matsui T, Seo JY, Domanski K, Correa-Baena JP, Nazeeruddin MK, et al. Cesium-containing triple cation perovskite solar cells: Improved stability, reproducibility and high efficiency. Energy Environ Sci 2016;9:1989–97. https://doi.org/10.1039/c5ee03874j.

[214] Liang Z, Zhang Y, Xu H, Chen W, Liu B, Zhang J, et al. Homogenizing out-of-plane cation composition in perovskite solar cells. Nature 2023;624:557–63. https://doi.org/10.1038/s41586-023-06784-0.

[215] Kubicki DJ, Prochowicz D, Hofstetter A, Zakeeruddin SM, Grätzel M, Emsley L. Phase Segregation in Cs-, Rb- and K-Doped Mixed-Cation (MA)x(FA)1-xPbI3 Hybrid Perovskites from Solid-State NMR. J Am Chem Soc 2017;139:14173–80. https://doi.org/10.1021/jacs.7b07223.

[216] Zhang M, Yun JS, Ma Q, Zheng J, Lau CFJ, Deng X, et al. High-Efficiency Rubidium-Incorporated Perovskite Solar Cells by Gas Quenching. ACS Energy Lett 2017;2:438–44. https://doi.org/10.1021/acsenergylett.6b00697.

[217] Qin Z, Pols M, Qin M, Zhang J, Yan H, Tao S, et al. Over-18%-Efficiency Quasi-2D Ruddlesden-Popper Pb-Sn Mixed Perovskite Solar Cells by Compositional Engineering. ACS Energy Lett 2023;8:3188–95. https://doi.org/10.1021/acsenergylett.3c00853.

[218] Yang F, Zhu K. Advances in Mixed Tin-Lead Narrow-Bandgap Perovskites for Single-Junction and All-Perovskite Tandem Solar Cells. Adv Mater 2024;36. https://doi.org/10.1002/adma.202314341.

[219] Lin Z, Chen J, Duan C, Fan K, Li J, Zou S, et al. Self-assembly construction of a homojunction of Sn-Pb perovskite using an antioxidant for all-perovskite tandem solar cells with improved efficiency and stability. Energy Environ Sci 2024;17:6314–22. https://doi.org/10.1039/d4ee01539h.

[220] Zhou J, Fu S, Zhou S, Huang L, Wang C, Guan H, et al. Mixed tin-lead perovskites with balanced crystallization and oxidation barrier for all-perovskite tandem solar cells. Nat Commun 2024;15:1–10. https://doi.org/10.1038/s41467-024-46679-w.

[221] Sun Q, Zhang Z, Yu H, Huang J, Li X, Dai L, et al. Surface charge transfer doping of narrow-bandgap Sn-Pb perovskites for high-performance tandem solar cells. Energy Environ Sci 2024;17:2512–20. https://doi.org/10.1039/d3ee03898j.

[222] Lee H, Kang SB, Lee S, Zhu K, Kim DH. Progress and outlook of Sn–Pb mixed perovskite solar cells. Nano Converg 2023;10. https://doi.org/10.1186/s40580-023-00371-9.

[223] Thiesbrummel J, Shah S, Gutierrez-Partida E, Zu F, Peña-Camargo F, Zeiske S, et al. Ion-induced field screening as a dominant factor in perovskite solar cell operational stability. Nat Energy 2024;9:664–76. https://doi.org/10.1038/s41560-024-01487-w.

[224] Hieulle J, Wang X, Stecker C, Son DY, Qiu L, Ohmann R, et al. Unraveling the Impact of Halide Mixing on Perovskite Stability. J Am Chem Soc 2019;141:3515–23. https://doi.org/10.1021/jacs.8b11210.

[225] Wang C, Nie R, Dai Y, Tai H, Zhu B, Zhao L, et al. Enhancing the inherent stability of perovskite solar cells through chalcogenide-halide combinations. Energy Environ Sci 2024;17:1368–86. https://doi.org/10.1039/d3ee03612j.

[226] Penukula S, Estrada Torrejon R, Rolston N. Quantifying and Reducing Ion Migration in Metal Halide Perovskites through Control of Mobile Ions. Molecules 2023;28. https://doi.org/10.3390/molecules28135026.

[227] Pering SR, Cameron PJ. The effect of multiple ion substitutions on halide ion migration in

perovskite solar cells. Mater Adv 2022;3:7918–24. https://doi.org/10.1039/d2ma00619g.

[228] Ajibade PA, Adeloye AO, Oluwalana AE, Thamae MA. Organolead halide perovskites: Synthetic routes, structural features, and their potential in the development of photovoltaic. Nanotechnol Rev 2023;12:1–41. https://doi.org/10.1515/ntrev-2022-0547.

[229] Gkini K, Martinaiou I, Falaras P. A review on emerging efficient and stable perovskite solar cells based on g-c3n4 nanostructures. Materials (Basel) 2021;14:1–16. https://doi.org/10.3390/ma14071679.

[230] Calió L, Kazim S, Grätzel M, Ahmad S. Lochtransportmaterialien für Perowskit-Solarzellen. Angew Chemie 2016;128:14740–64. https://doi.org/10.1002/ange.201601757.

[231] Liu A, Zhu H, Bai S, Reo Y, Zou T, Kim MG, et al. High-performance inorganic metal halide perovskite transistors. Nat Electron 2022;5:78–83. https://doi.org/10.1038/s41928-022-00712-2.

[232] Li MH, Shen PS, Wang KC, Guo TF, Chen P. Inorganic p-type contact materials for perovskite-based solar cells. J Mater Chem A 2015;3:9011–9. https://doi.org/10.1039/c4ta06425a.

[233] Li M-H, Chiang Y-H, Shen P-S, Juang SS-Y, Chen PC-Y. P-Type and Inorganic Hole Transporting Materials for Perovskite Solar Cells 2017:63–109. https://doi.org/10.1142/9789813222526_0004.

[234] Chen J, Park NG. Inorganic Hole Transporting Materials for Stable and High Efficiency Perovskite Solar Cells. J Phys Chem C 2018;122:14039–63. https://doi.org/10.1021/acs.jpcc.8b01177.

[235] Li MH, Yum JH, Moon SJ, Chen P. Inorganic p-type semiconductors: Their applications and progress in dye-sensitized solar cells and perovskite solar cells. Energies 2016;9:1–28. https://doi.org/10.3390/en9050331.

[236] Xu Z, Guo Z, Li H, Zhou Y, Liu Z, Wang K, et al. Efficient and stable inverted MA/Br-free 2D/3D perovskite solar cells enabled by α -to- δ phase transition inhibition and crystallization modulation. Energy Environ Sci 2025;18:1354–65. https://doi.org/10.1039/D4EE04136D.